\documentclass[usenatbib]{mnras}
\usepackage{graphicx}
\usepackage{float}
\usepackage{epsfig}
\usepackage{times}
\usepackage{longtable}
\usepackage{afterpage}
\usepackage{pdflscape}
\usepackage{amsmath}

\newcommand{\NII}{[N\,\textsc{ii}]~$\lambda$6583}
\newcommand{\SII}{[S\,\textsc{ii}]}
\newcommand{\OI}{[O\,\textsc{i}]}
\newcommand{\NIIHa}{[N\,\textsc{ii}]~$\lambda$6583$/$H$\alpha$}
\newcommand{\NIIHatwo}{[N\,\textsc{ii}]$/$H$\alpha$}
\newcommand{\SIIHa}{[S\,\textsc{ii}]$/$H$\alpha$}
\newcommand{\OIHa}{[O\,\textsc{i}]$/$H$\alpha$}
\newcommand{\OIIIHb}{[O\,\textsc{iii}]$/$H$\beta$}
\newcommand{\OIII}{[O\,\textsc{iii}]~$\lambda$5007}

\newcommand{\OIIIone}{[O\,\textsc{iii}]~$\lambda$4363}

\begin{document}

\title[The SFRS: Activity demographics and host-galaxy properties]{The Star Formation Reference Survey. II. Activity demographics and host-galaxy properties for Infrared-selected galaxies}

\author[A. Maragkoudakis et al.]
{A. Maragkoudakis,$^1$ $^2$ $^3$ A. Zezas,$^1$ $^2$ $^3$ M. L. N. Ashby,$^2$ S. P. Willner$^2$ \\
$^1$University of Crete, Department of Physics, Heraklion 71003, Greece \\
$^2$Harvard-Smithsonian Center for Astrophysics, Cambridge, MA 02138 \\
$^3$Foundation for Research and Technology - Hellas (FORTH), Heraklion 71003, Greece}

\maketitle
\begin{abstract}
  We present activity demographics and host-galaxy properties of
  infrared-selected galaxies in the local Universe, using the
  representative Star Formation Reference Survey (SFRS). Our
  classification scheme is based on a combination of optical
  emission-line diagrams (BPT) and IR-color diagnostics. Using the
  weights assigned to the SFRS galaxies based on its parent
    sample, a far-infrared-selected sample comprises 71\%
  H\,\textsc{ii} galaxies, 13\% Seyferts, 3\% Transition Objects
  (TOs), and 13\% Low-Ionization Nuclear Emission-Line Regions
  (LINERs). For the SFRS H\,\textsc{ii} galaxies we derive nuclear
  star-formation rates and gas-phase metallicities. We measure
  host-galaxy metallicities for all galaxies with available long-slit
  spectroscopy and abundance gradients for a subset of 12 face-on
  galaxies. The majority of H\,\textsc{ii} galaxies show a narrow
  range of metallicities, close to solar, and flat metallicity
  profiles. Based on their host-galaxy and nuclear properties, the
  dominant ionizing source in the far-infrared selected TOs is
  star-forming activity. LINERs are found mostly in massive hosts
  (median of $10^{10.5}$ M$ _{\sun} $), median $L(60\micron) =
  10^{9}$ L$_{\sun}$, median dust temperatures of $ F60/F100 = 0.36
  $, and median $L_{\textrm{H}\alpha}$ surface density of $ 10^{40.2}
  $ erg s$ ^{-1} $kpc$ ^{-2} $, indicating older stellar populations
  as their main ionizing source rather than AGN activity. 
\end{abstract}


\section{INTRODUCTION}

Distinguishing between the processes of star-formation and active galactic nuclei (AGN) is crucial in the study of energy production in galaxies. The ratio of these processes and their interplay is important on both whole-galaxy and galaxy nucleus scales for understanding their part in galaxy evolution. Several scaling relations between black hole mass ($M_{\textrm{BH}}$), bulge luminosity ($L_{\textrm{bulge}}$), bulge mass ($M_{\textrm{bulge}}$), velocity dispersion ($ \sigma $) e.g., $M_{\textrm{BH}}-L_{\textrm{bulge}}$: \citep{Kormendy95}, $M_{\textrm{BH}}-M_{\textrm{bulge}}$ \citep{Magorrian98}, $M_{\textrm{BH}}-\sigma$ \citep{Ferrarese00}, average black hole growth rate, and mean total stellar mass \citep{Mullaney12} describe empirical relations between the central regions of galaxies and their hosts. However, the origin and degree of influence of such relations are still debated. (For a review see \citealt{KH13}).

The close similarities between the cosmic evolution of the star-formation rate density and the  supermassive black hole (SMBH) mass accretion rate density (e.g., \citealt{Silverman08}; \citealt{Aird10}) point towards a connection between the evolution of star formation and AGN activity. \cite{Davies07} provided strong evidence for a starburst--AGN connection by examining the nuclei of nine Seyfert galaxies and showing that the peak of AGN activity occurs 50--100 Myr after the onset of star formation. The delay between star formation and AGN activity indicates that the starburst had a significant impact on fueling the central black hole. Hydrodynamical simulations (e.g., \citealt{Hobbs11}) support this scenario by showing that stellar winds and supernovae enhance the black hole mass accretion by injecting turbulence into the surrounding gas disc. However, a complete picture of these processes has not yet been established, and many questions remain open regarding the ways star formation and AGN feedback mechanisms impact and regulate one another considering the different physical conditions at different cosmic epochs. 

A complete study of the starburst--AGN connection requires a representative sample of galaxies probing the wide range of different physical conditions that these phenomena engage. The nearby Universe offers the opportunity to study the above processes and overall galaxy properties in observationally resolved conditions. A number of nearby galaxy surveys (e.g., \textit{Spitzer} Infrared Nearby Galaxies Sample$ - $SINGS; \citealt{Kenn03}, Great Observatories All-sky LIRG Survey$ - $GOALS; \citealt{GOALS}, \textit{Herschel} Reference Survey$ - $HRS; \citealt{Boselli10}) have approached the subject by focusing mainly on specific aspects of the star formation or AGN processes. For example, SINGS galaxies constitute a relatively small number of nearby and extended objects at the low-luminosity end of FIR luminosity function and therefore do not depict star-formation in its whole extent. Furthermore, powerful AGNs were intentionally excluded from the SINGS sample, thus limiting the diversity of AGN systems available for a comprehensive study of the starburst--AGN connection. GOALS on the other hand consists predominantly of Luminous Infrared Galaxies (LIRGs, $10^{11}L_{\sun} < L < 10^{12}L_{\sun} $) and a few Ultra-Luminous Infrared Galaxies (ULIRGs $L \geq 10^{12}L_{\sun}$), sampling the higher ends of star formation and galactic activity in general, while the HRS is particularly oriented toward high-density environments.

For a comprehensive study of the star formation and AGN phenomenon in a wide range of environments and physical conditions in the local universe, it is imperative to employ a well defined and representative sample of galaxies. To this cause the ``Star Formation Reference Survey" (SFRS; \citealt{SFRS}) was established. An initial result from the SFRS sample indicating a connection between the central regions of star-forming galaxies and their hosts \citep{Maragkoudakis17} showed a correlation between the nuclear star-formation rate (SFR) and the total stellar mass of the galaxies. This was referred to as the Nuclear Main Sequence of star-forming galaxies.

The current paper presents the activity demographics and the host galaxy properties for the SFRS galaxies and is organized as follows. Section \ref{SFRS} describes the SFRS and gives basic information on the sample properties. Section \ref{Observations} describes the observations, the initial reduction of the spectra, and the measurement of the fluxes of the strong optical emission lines. The activity classification is presented in Section \ref{Classification}. Gas-phase metallicities, metallicity gradients, and nuclear SFR measurements are presented in Section \ref{Z-SFR}. Section \ref{Discussion} provides a detailed discussion of the activity demographics in the local Universe, the different classification methods used, the galaxy properties based on activity type, and the dominant source of ionization in LINERs and transition objects. Finally, we summarize our results in Section \ref{Conclusions}. Throughout the paper the activity class H\,\textsc{ii} (or H\,\textsc{ii} galaxies) is used to describe pure star-forming galaxies (SFGs) that exhibit no fraction of AGN activity.

\section{The SFRS sample} \label{SFRS}

The SFRS galaxies were chosen in order to build a representative sample spanning the full range of properties of local galaxies hosting primarily star-formation activity. The parent sample from which SFRS was drawn is the PSC$ z $ \citep{Saunders00}, a database of 15,411 nearby star-forming galaxies brighter than 0.6 Jy at 60$\mu m$ across 84 per cent of the sky at redshifts $ z \leq 0.2 $. Selection from the PSC$ z $ into the SFRS was based on observational proxies for SFR, specific SFR (sSFR), and dust temperature. Specifically, \cite{SFRS} used the IRAS 60 $\mu$m luminosity as a SFR tracer, the flux ratio of $F_{60}$ to 2MASS $K_{S}$ ($F_{60}/K_{S}$) as sSFR proxy, and the far-IR flux density ratio of $F_{100}/F_{60}$ as a measure of dust temperature to create a sample of galaxies spanning this three-dimensional parameter space. 369\footnote{Quasar 3C~273 and blazar OJ~287 are excluded from the spectroscopic studies related to star formation but are included in the rest of the SFRS analysis.} galaxies were selected \citep{SFRS} to explore all existing combinations of these parameters over their entire range. Each SFRS galaxy was assigned a weight reflecting its relative prevalence with respect to the parent population from which the SFRS was defined \citep{SFRS}. The overall distribution of SFRS's total infrared (TIR) luminosities is $10^{7.95}-10^{12.25} \textrm{L}_{\sun} $, the $ F_{60}/F_{100} $ flux-density ratio distribution is between 0.14--1.64, and the stellar masses range between $10^{7.79}-10^{12.14} \textrm{M}_{\sun} $ \citep{Maragkoudakis17}. Thus, the parameter space covered by the SFRS captures the full range of star-forming and AGN conditions and environments, making it an ideal survey for the study of these processes in the local Universe.

Exploring all conditions of star formation requires an uncensored sample of galaxies, and therefore galaxies hosting AGN activity were not discarded from the SFRS. This also provides the opportunity to examine the AGN contribution to the overall energy output of galaxies and simultaneously investigate the interplay between star formation and AGN mechanisms (i.e., quenching or boosting of star formation). Therefore, we initiated a long-slit spectroscopic campaign to acquire spectra for the SFRS galaxies lacking SDSS spectroscopic coverage and used a combination of optical emission-line and IR diagnostics to identify AGN hosts.

The SFRS has an impressive range of multiband photometric data from UV to radio bands: GALEX FUV--NUV, SDSS $ugriz$, 2MASS $JHKs$, \textit{Spitzer}/IRAC $ 3.6, 4.5, 5.8, 8.0\,\micron$, \textit{Spitzer}/MIPS $ 24\,\micron$, \textit{AKARI FIR All-Sky Survey}, \textit{IRAS}, VLA/NVSS. Furthermore, there is an ongoing H$ \alpha $ imaging campaign at the 1.3-m Ritchey-Cr\'{e}tien telescope at the Skinakas Observatory in Crete. This complete set of multi-wavelength photometry combined with the spectroscopically defined activity classification will allow us to derive fundamental galaxy properties (i.e., SFRs, stellar masses, dust-luminosities) for each galaxy activity class using both luminosity-dependent calibrations as well as spectral energy distribution (SED) fitting techniques.

\section{OBSERVATIONS AND ANALYSIS} \label{Observations}

\subsection{SDSS spectra and aperture effects}

210 SFRS galaxies had available SDSS spectra as of data release 7 (DR7; \citealt{SDSSDR7}). These cover a spectral range from 3800 \AA~ to 9200 \AA~ with a resolution $R = 1500$ at 3800 \AA~ and $R = 2500$ at 9000 \AA. The spectra refer to a 3\arcsec-diameter region of the nucleus of each galaxy. Taking into account that at the mean SFRS redshift of 0.024 the fiber diameter correspond to a physical size of approximately 1.5 kpc, the obtained SDSS spectra provide a reasonable representation of the nuclear region of the SFRS galaxies. Because the SDSS fibers are located on the nucleus, and the region of spectral extraction can not be adjusted, we refer to all SDSS classification as nuclear region classification. Thirty six SFRS galaxies are located at redshifts between $0.05<z<0.2$, 21 of which have SDSS fiber spectra. The projected aperture diameter at these redshifts is between 3 kpc and 12 kpc, which are closer to global spectra of galaxies rather than nuclear. \cite{Maragkoudakis14} showed that line ratios and consequently activity classification obtained from SDSS $3\arcsec$ fibers can be affected by aperture size. Besides AGN spectral features being veiled by the host-galaxy starlight, extranuclear star-forming activity encompassed in the fiber can influence spectral lines, often increasing lower ionization emission-line intensities. Although starlight removal techniques can mitigate these effects they do not fully remove them, especially the contamination by emission lines from the ionized interstellar medium (ISM) in the host galaxy (\citealt{Maragkoudakis14}). However, these galaxies constitute only a very small fraction of the SFRS sample, and they do not change the conclusions of the analysis.

\subsection{Long-slit spectra} \label{spectra}
Long-slit spectra were acquired at the Fred Lawrence Whipple Observatory in Arizona using the 60-inch Tillinghast telescope with the FAST spectrograph \citep{FAST}. The target list included 159 galaxies without SDSS spectra plus 19 galaxies observed by SDSS but with insufficient signal-to-noise ratio (S/N) or bad fiber centering. Each galaxy was observed with a $3\arcsec$ wide $\times6\arcmin$ long slit visually centered on the nucleus and oriented along the galaxy major axis. The data were collected using the 2688 $\times$ 512 pixel FAST3 CCD camera. We used two configurations: a) a 600 l mm$^{-1}$ grating giving a spectral coverage of the 3800 -- 5700 \AA\ or 5500 -- 7800\AA\ wavelength range in the blue and red part of the spectrum respectively with a dispersion of 0.75\AA\ pixel$^{-1}$ b) a 300 l mm$^{-1}$ grating covering the spectral range of 3400 -- 7200 \AA\ with a dispersion of 1.47 \AA\ pixel$ ^{-1} $. The first configuration was used for the observation of 39 galaxies of which 23 also have SDSS spectra. The second configuration provided simultaneous coverage of all diagnostic lines of interest with sufficient resolution and was used for most of the galaxies. The resolution of the 300 l mm$^{-1}$ grating was 5.993 \AA\ as measured from the full width at half maximum (FWHM) of the [O\,\textsc{i}]~5577 \AA\ sky line. Two to three exposures were acquired for each object to correct for the presence of cosmic rays on the images. The individual exposure times varied between 600 and 1800 seconds in order to achieve a S/N of 40 at the H$\alpha$ emission line. For large galaxies covering a major portion of the slit, separate sky exposures were obtained as close as possible to the location of the galaxies in the sky to allow proper sky subtraction. Spectrophotometric standard-star exposures from the list of \cite{Massey88} were also obtained during each night and used for flux calibration of the spectra. 

The initial reduction of the long-slit spectra was performed with \textsc{iraf} (Image Reduction and Analysis Facility). Standard procedures of bias subtraction, flat fielding, and wavelength calibration were applied to the two-dimensional CCD images. The different exposures of each galaxy were combined in order to remove cosmic ray imprints on the data. Two types of one-dimensional spectra were extracted with \textsc{iraf}'s \textsc{apall} task: integrated spectra with aperture size matched to the point where the galaxy's flux reached the level of the background and nuclear spectra, extracted from a $3.5\arcsec \times 3\arcsec$ aperture, determined from the average seeing at the time of the observations. The standard star spectra were processed in the same way as the galactic spectra. We identified the strongest lines on the extracted spectra and measured their location with \textsc{iraf}'s \textsc{splot} task. The measured emission-line centroids were used to establish galaxy redshifts and shift all spectra to the rest frame. This step was also performed on the SDSS spectra. During the SFRS long-slit campaign (2010$-$2014), 45 galaxies with no previous SDSS data were added to the SDSS data releases. We have analyzed these galaxies to examine the agreement, or possible deviations, between long-slit and fiber spectroscopy.

\subsection{Starlight Subtraction}

The stellar continuum was removed from both the long-slit (integrated and nuclear) and the SDSS spectra using the \textsc{starlight} v.04 code \citep{STARLIGHT} to model the galaxy spectra. \textsc{starlight} fits the stellar continuum on emission-line-free regions of the spectrum using linear combinations of simple stellar populations (SSPs) from the Bruzual \& Charlot (\citeyear{BC03}) (hereafter BC03) libraries. We used a base of 138 SSPs with 23 ages ranging between 1 Myr and 13 Gyr and 6 metallicities from 0.005 to 2.5$Z\sun$. The extinction ($A_{V}$) was also fitted by \textsc{starlight}, assuming the \cite{CCM89} reddening law with the ratio of total to selective extinction $R_{V} \equiv A_{V}/E(B-V) = 3.1 $. In order to handle any age mixture biases in the fitting process for galaxies having an excessive blueness when $A_{V}$ is close to 0, we allowed $A_{V}$ to take negative values as discussed by \cite{STARLIGHT_TEST}. Negative $A_{V}$ was also allowed in studies using different fitting methods (e.g., \cite{Kauffmann03}). A statistical justification for this unphysical condition is that when $A_{V}$ is truly = 0, unbiased estimates should oscillate around 0, including both negative and positive values. There are only 14 cases showing negative extinctions but very close to zero, with a median of $-0.08$.

\begin{figure}
	\begin{center}
		\includegraphics[keepaspectratio=true, scale=.6]{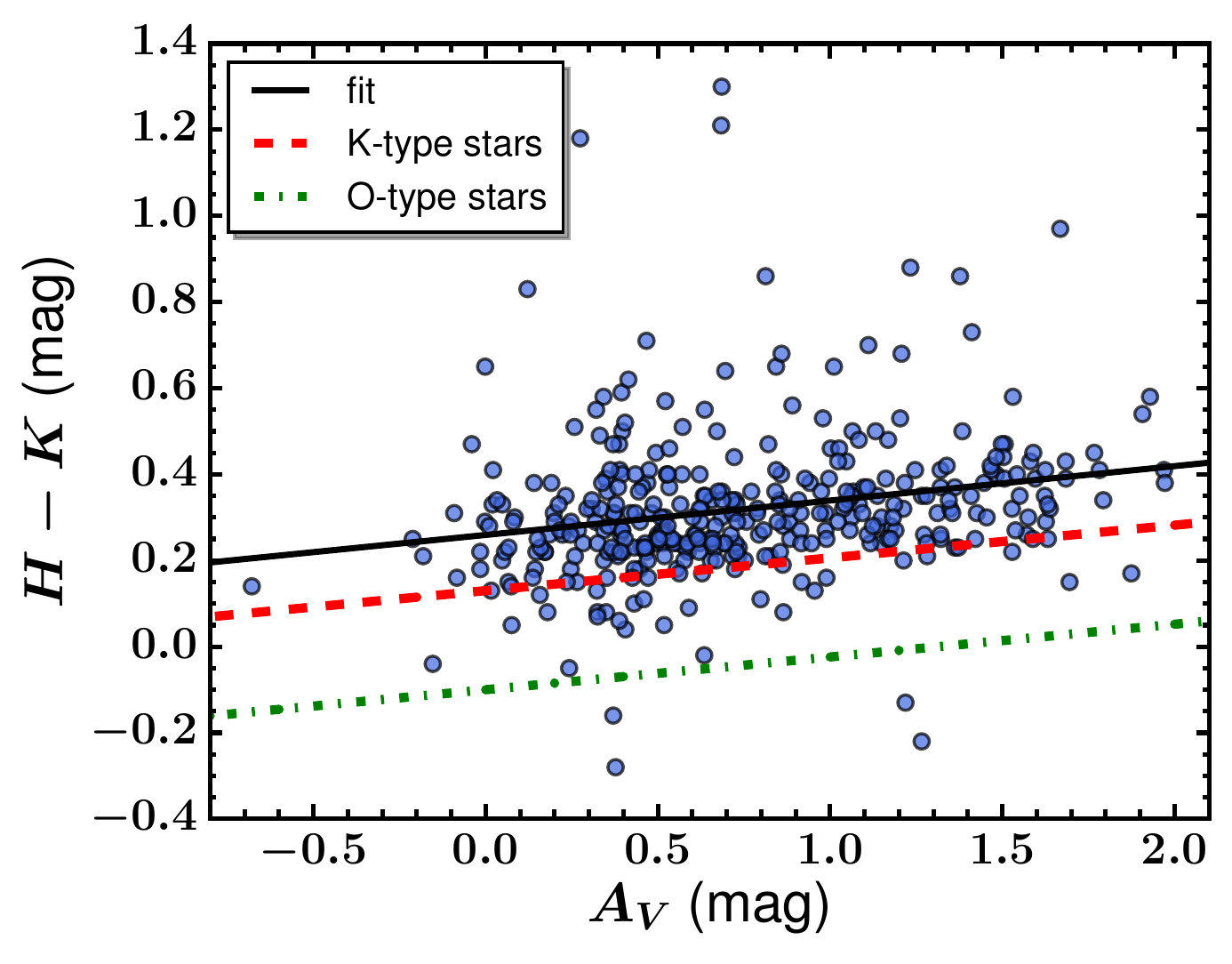}
		\caption{Comparison between the $A_{V}$ extinction calculated by \textsc{starlight} and the 2MASS $H-K$ colors in Vega magnitudes. The black solid line is the fit to the data. The red dashed line is the theoretical $H-K$ reddening curve \protect\citep{CCM89}, assuming intrinsic $ (H-K)_{o} = 0.1 $ for K-type stars \protect\citep{Blum00}. Similarly, the green dash-dot line is the theoretical $H-K$ reddening curve, assuming intrinsic $ (H-K)_{o} = -0.1 $ based on the O-type stars calibrations by \protect\cite{Martins06}, representing the bluest possible stellar populations. NGC~4688 is the galaxy with $A_{V}=-0.68$ also reported with $E(B-V)=0$ by \protect\cite{Ho97c}.}
		\label{Ext1}
	\end{center}
\end{figure}

An initial assessment of starlight subtraction results can be done by comparing the estimated $A_{V}$ extinction with the Two-Micron All-Sky Survey (2MASS) $H-K$ colors. It has been recognized for some time that infrared colors could be used to directly measure extinction (e.g., \citealt{Hyland80}; \citealt{Lada94}). Fig. \ref{Ext1} shows the theoretical $H-K$ reddening curve for K-type and O-type stars and the fit to \textsc{starlight} estimated extinction. The observed correlation between the two quantities indicates reasonably good fitting results. Based on the fitted spectra, the range of equivalent widths of H$\alpha$ and H$\beta$ absorption lines are 2.49\AA\ $<$ EW(H$\beta$) $<$ 4.70\AA\ and 2.05\AA\ $<$ EW(H$\alpha$) $<$ 3.04\AA. Appendix~\ref{Balmer-test} shows that the uncertainties in the calculated EWs are $\la$0.2~\AA\ and should have negligible effect on the results.


\textsc{starlight} also provides physical parameters such as the dominant stellar populations in the host galaxies, their corresponding metallicities, stellar ages, and masses. However adopting these parameters as the actual or most representative parameters of the input galaxies is not a trivial matter, as many degeneracies are involved in the fitting process and cannot be neglected. (See Appendix \ref{ST-SIMs}.) Despite the uncertainties in the parameters of the stellar populations, \textsc{starlight} appears to be robust in measuring, fitting, and removing the stellar features (continuum and absorption lines) in our spectra. We reached this result by fitting the same spectra using different sets of stellar population.

\begin{figure}
	\begin{center}
		\includegraphics[keepaspectratio=true, scale=.42]{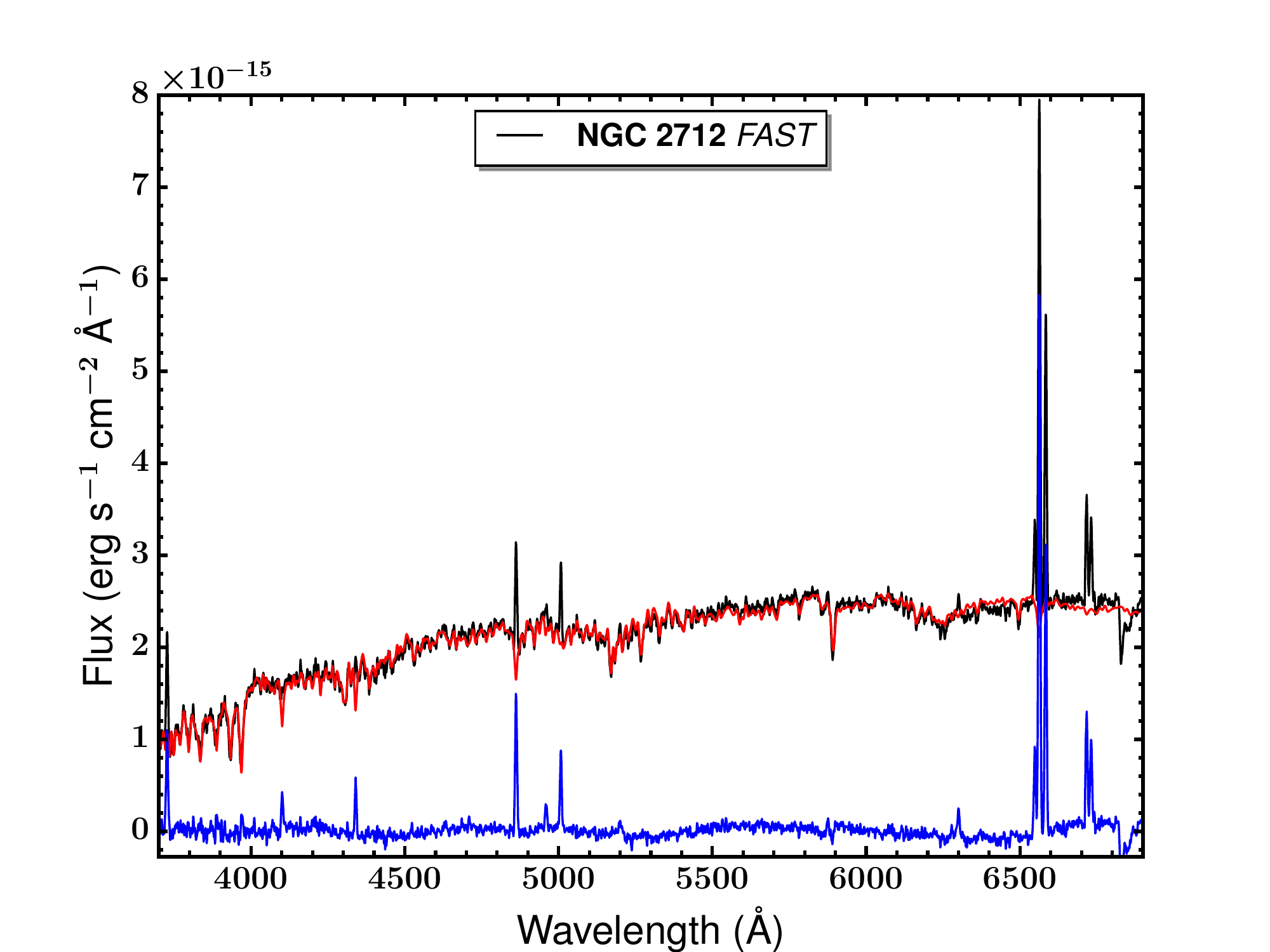}
		\caption{The observed (black line), \textsc{starlight}-fitted (red line), and starlight-subtracted (blue line) spectra of NGC~2712 observed with the FAST spectrograph.}
		\label{spectrum}
	\end{center}
\end{figure}

Figure \ref{spectrum} shows an example of the observed, \textsc{starlight}-fitted, and starlight-subtracted spectra of NGC~2712. All the SFRS galaxy spectra are available online from MNRAS.

\subsection{Emission Line Measurements} \label{emission-lines}

\begin{figure}
	\begin{center}
		\hspace*{-0.4cm}\includegraphics[keepaspectratio=true, scale=.45]{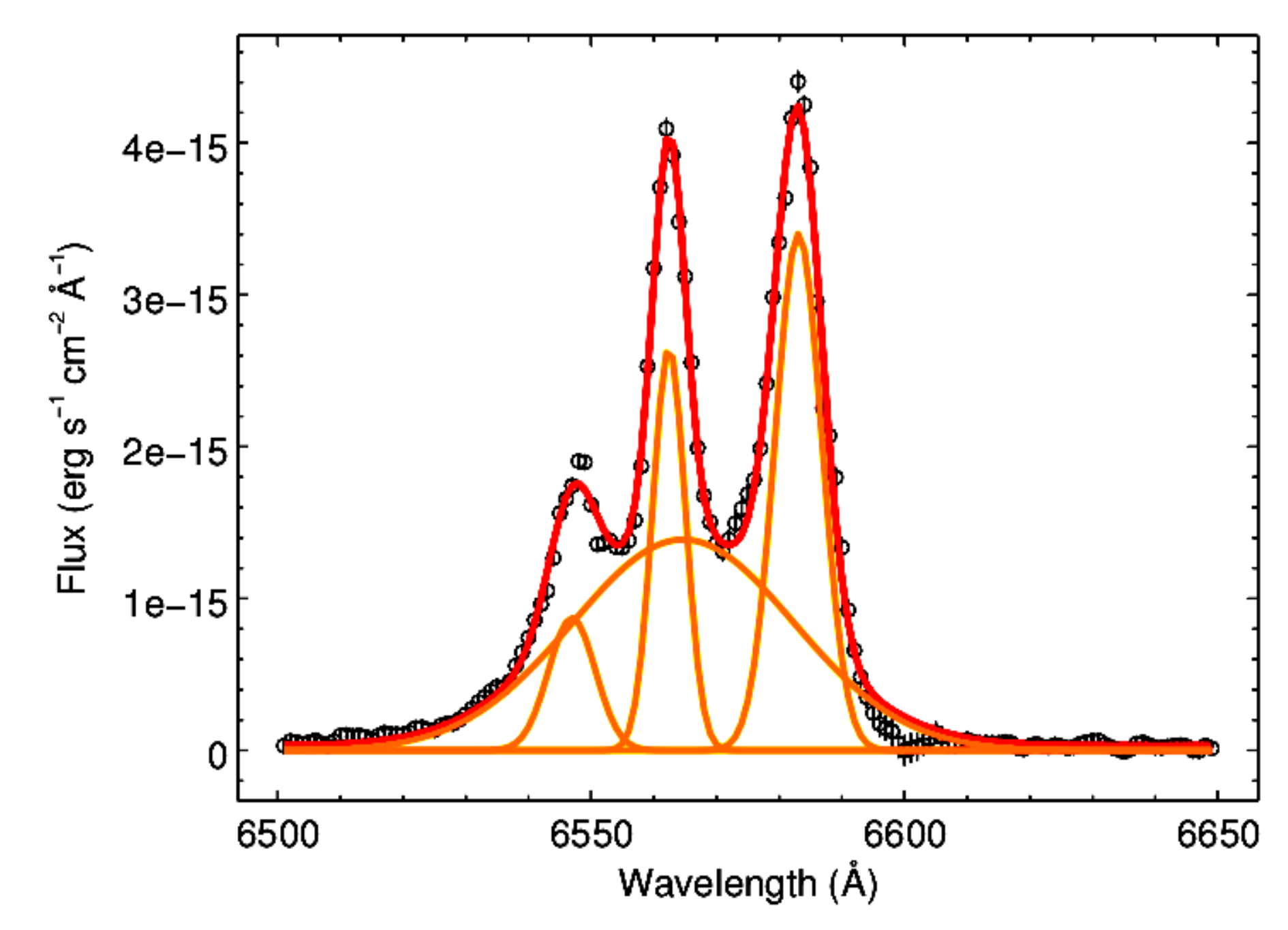}
		\caption{Example of a multi-component emission-line fit with \textsc{sherpa} of the H$ \alpha $ and [N\,\textsc{ii}]~$\lambda\lambda$6548,6583 doublet in the nuclear spectrum of Seyfert IC~910. An additional broad component is used in this case. Points show starlight-subtracted data. Orange lines show four individual emission line components, and red line shows the sum which corresponds to the best fit.}
		\label{sherpa}
	\end{center}
\end{figure}

The emission lines of interest were fitted with \textsc{sherpa} \citep{SHERPA}, a general-purpose fitting environment designed for CIAO, the Chandra (X-Ray Observatory) Interactive Analysis of Observations software package \citep{CIAO}. \textsc{sherpa} can handle multi-component models (Fig. \ref{sherpa}) such as emission lines with broad components (e.g., Seyfert 1 galaxies) or doubled-peaked line profiles resulting from galaxy rotation. One of the most important advantages of \textsc{sherpa} is its capability of fitting spectra accounting for the flux density uncertainties. We fitted the intensity, full width at half-maximum (FWHM), and central wavelength of each emission line using Gaussian profiles and a constant for the local continuum. The latter is a good approximation given that we are fitting the starlight-subtracted spectra. Uncertainties for the parameters were calculated at the 1$\sigma$ confidence intervals. This was done by varying a parameter's value in a grid of values while at the same time the values of the other model parameters were allowed to float to new best-fit values. 

\begin{figure}
	\begin{center}
		\includegraphics[keepaspectratio=true, scale=.42]{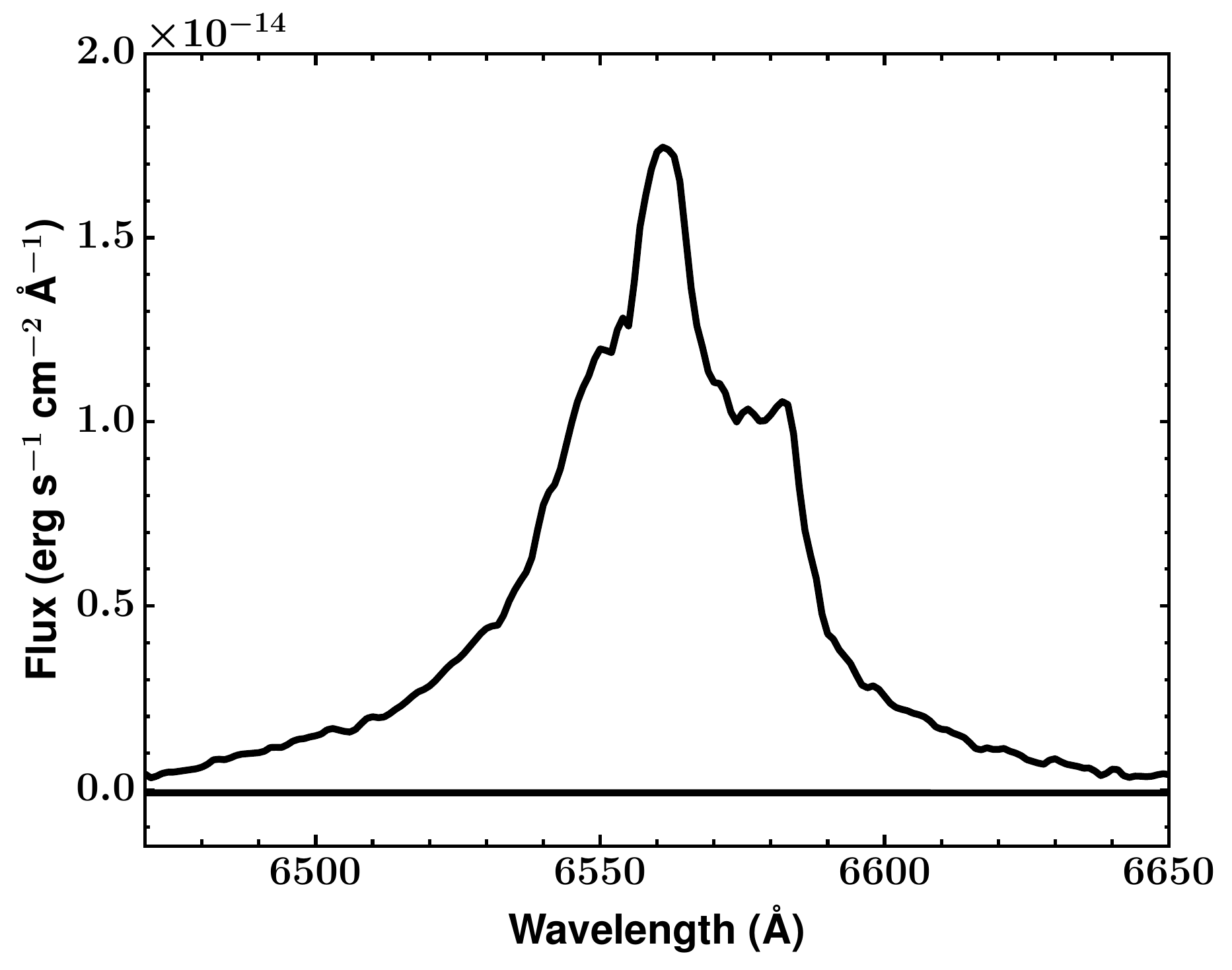}
		\caption{An example of a Type-1 AGN spectrum (NGC~3758) at the H$ \alpha $ line wavelength region, where no emission line fitting was performed. For five such galaxies a Sy-1 classification was automatically assigned.}
		\label{Type1AGN}
	\end{center}
\end{figure}

We measured the intensities of the forbidden lines [O\,\textsc{iii}]~$\lambda$5007, [N\,\textsc{ii}]~$\lambda$6583, [S\,\textsc{ii}]~$\lambda\lambda$6716, 6731, [O\,\textsc{i}]~$\lambda$6300, and the Balmer H$\beta$ and H$\alpha$ lines. For five galaxies (NGC~3758, UGC~8058, IRAS~13218+0552, MK~268, and UGC~9412) having spectra with extremely broad lines (Fig. \ref{Type1AGN}) no emission line fitting was performed, since from the profile of the lines they could be clearly classified as AGN. Therefore, they were assigned a Sy-1 classification (see Section \ref{Final_Class}). The emission-line intensity ratios with respect to the H$\alpha$ or H$\beta$ lines used in the three activity classification diagnostic diagrams (see Section \ref{sec:BPT}), along with the intensity of the H$\alpha$ and the \textsc{starlight} fitted extinction, are presented in Table \ref{table1} along with the corresponding H$\alpha$/H$\beta$ ratio.

\subsection{H$ \alpha $ emission line extinction correction} \label{ext-cor}
Because the emission-line ratios used in our analysis are very close in wavelength, they are almost completely insensitive to reddening and therefore we do not perform extinction correction to the respective emission lines. However, because we measured nuclear SFR (Section \ref{Z-SFR}) based on L$_{\textrm{H}\alpha} $, we did correct the H$ \alpha $ flux. The extinction was derived from the relative observed strengths of the H$ \alpha $ and H$ \beta $ Balmer lines. The intrinsic Balmer decrement remains roughly constant for typical gas conditions, and in our analysis we assumed an intrinsic flux ratio value (FH$ \alpha/$FH$ \beta $)$ _{0} $ of 2.86 for H\,\textsc{ii} regions, corresponding to a temperature T = 10$ ^{4} $ K and an electron density n$ _{e} $ = 10$ ^{2} $ cm$ ^{-3} $ for Case B recombination \citep{Osterbrock89}. For the narrow line region of AGN a value of 3.1 is generally adopted, where H$ \alpha $ emission is slightly enhanced by collisional excitation due to higher density gas and a much higher ionizing continuum. 

We adopted the \cite{CCM89} extinction law and calculated the gas extinction A$ _{V}$ based on the following relation:
\begin{equation} \label{ext-gas}
A_{V} = \frac{\log_{10}[(FH\alpha/FH\beta)/(FH\alpha/FH\beta)_{0}]}{0.4 \times (A_{H\beta} - A_{H\alpha})},
\end{equation}
where $A_{H\beta}$ and $A_{H\alpha}$ are the magnitude attenuations at the H$ \alpha $ and H$ \beta $ wavelengths. The extinction-corrected H$ \alpha $ fluxes and the corresponding gas reddening E(B-V) values calculated from Eq. \ref{ext-gas} and $R_{V} \equiv A_{V}/E(B-V) = 3.1 $, are given in Table \ref{table1}.


\begin{table*}
\begin{scriptsize}

	\hspace*{-0.9cm}\begin{minipage}{194mm}
	\caption{Nuclear fluxes and flux ratios of optical diagnostic emission lines. Column (1) SFRS ID (\citealt{SFRS}); Column (2) Galaxy name; Column (3) H$\alpha$ emission line flux, normalized to 10$^{-14}$ erg cm$^{-2}$ s$^{-1}$; Column (4) Ratio of H$\alpha$ and H$\beta$ Balmer line fluxes. Columns (5) $-$ (8) BPT line ratio measurements and uncertainties; Column (9) Extinction corrected H$\alpha$ emission line flux normalized to 10$^{-14}$ erg cm$^{-2}$ s$^{-1}$; Columns (10) and (11) Integrated and nuclear extinction as calculated by $\textsc{starlight}$ code; Column (12) The nuclear E(B-V) color excess of the gas as calculated from the Balmer decrement; Column (13) Spectral observation method. Emission line fluxes and ratios for the FAST observations come from the nuclear aperture spectrum. F(H$\alpha$)$_{cor}$ and E(B-V)$_{\textrm{gas}} $ values flagged with a * symbol indicate uncertain measurements due to differences in the flux calibration between the blue and red spectral regions of galaxies observed with the 600 l mm$^{-1}$ grating configuration (Section \ref{FAST-SDSS}). The full version of Table \ref{table1} is available online from MNRAS.} 
	\label{table1}
	{\renewcommand{\arraystretch}{1.35}
	\begin{tabular}{@{}ccccccccccccc}
 \hline
SFRS & Galaxy & F(H$\alpha$) & H$\alpha$/H$\beta$ & [O\,{\sc iii}]/H$\beta$ & [N\,{\sc ii}]/H$\alpha$ & [S\,{\sc ii}]/H$\alpha$ & [O\,{\sc i}]/H$\alpha$ & F(H$\alpha$)$_{cor}$ & A$_{V}$ & A$_{V}\textrm{nuc}$ & E(B-V)$ _{\textrm{gas}} $ & Observation \\
 (1) & (2) & (3) & (4) & (5) & (6) & (7) & (8) & (9) & (10) & (11) & (12) & (13) \\
 \hline

1 & IC~486 & 2.430$\substack{+0.037 \\ -0.026}$ & 5.830$\substack{+0.214 \\ -0.205}$ & 1.058$\substack{+0.015 \\ -0.015}$ & 0.071$\substack{+0.006 \\ -0.008}$ & -0.105$\substack{+0.005 \\ -0.007}$ & -0.727$\substack{+0.019 \\ -0.020}$ & 10.788$\substack{+0.164 \\ -0.115}$ & \ldots & 0.68 & 0.64 & SDSS \\ 
2 & IC~2217 & 7.660$\substack{+0.045 \\ -0.045}$ & 4.935$\substack{+0.085 \\ -0.087}$ & -0.552$\substack{+0.029 \\ -0.029}$ & -0.422$\substack{+0.006 \\ -0.006}$ & -0.474$\substack{+0.003 \\ -0.003}$ & -1.588$\substack{+0.052 \\ -0.054}$ & 27.755$\substack{+0.163 \\ -0.163}$ & 0.50 & 0.82 & 0.55 & FAST \\ 
3 & NGC~2500 & 0.151$\substack{+0.007 \\ -0.007}$ & 3.096$\substack{+0.497 \\ -0.482}$ & 0.132$\substack{+0.080 \\ -0.083}$ & -0.446$\substack{+0.057 \\ -0.058}$ & -0.268$\substack{+0.022 \\ -0.019}$ & 0.127$\substack{+0.031 \\ -0.029}$ & 0.182$\substack{+0.008 \\ -0.008}$ & 0.81 & 0.68 & 0.08 & FAST \\ 
4 & NGC~2512 & 10.800$\substack{+0.008 \\ -0.008}$ & 4.713$\substack{+0.017 \\ -0.017}$ & -0.677$\substack{+0.008 \\ -0.008}$ & -0.310$\substack{+0.001 \\ -0.001}$ & -0.564$\substack{+0.002 \\ -0.002}$ & -1.535$\substack{+0.014 \\ -0.014}$ & 35.104$\substack{+0.026 \\ -0.026}$ & 0.52 & 0.46 & 0.50 & FAST \\ 
5 & MCG~6-18-009 & 6.230$\substack{+0.041 \\ -0.041}$ & 5.253$\substack{+0.075 \\ -0.072}$ & -0.517$\substack{+0.015 \\ -0.014}$ & -0.289$\substack{+0.004 \\ -0.004}$ & -0.564$\substack{+0.003 \\ -0.003}$ & -1.441$\substack{+0.023 \\ -0.024}$ & 26.157$\substack{+0.172 \\ -0.172}$ & \ldots & 0.63 & 0.61 & SDSS \\ 
6 & MK~1212 & 0.403$\substack{+0.005 \\ -0.004}$ & 6.165$\substack{+0.323 \\ -0.318}$ & -0.316$\substack{+0.054 \\ -0.049}$ & -0.285$\substack{+0.009 \\ -0.008}$ & -0.562$\substack{+0.005 \\ -0.005}$ & -1.441$\substack{+0.053 \\ -0.072}$ & 2.469$\substack{+0.031 \\ -0.025}$ & \ldots & 0.94 & 0.78 & SDSS \\ 
7 & IRAS~08072+1847 & 1.790$\substack{+0.013 \\ -0.013}$ & 8.168$\substack{+0.165 \\ -0.160}$ & -0.239$\substack{+0.020 \\ -0.018}$ & -0.196$\substack{+0.005 \\ -0.005}$ & -0.469$\substack{+0.003 \\ -0.003}$ & -1.187$\substack{+0.023 \\ -0.021}$ & 17.611$\substack{+0.128 \\ -0.128}$ & \ldots & 1.51 & 0.98 & SDSS \\ 
8 & NGC~2532 & 2.300$\substack{+0.060 \\ -0.056}$ & 7.691$\substack{+1.296 \\ -1.221}$ & -0.519$\substack{+0.068 \\ -0.072}$ & -0.415$\substack{+0.028 \\ -0.030}$ & -0.492$\substack{+0.010 \\ -0.011}$ & -1.693$\substack{+0.010 \\ -0.011}$ & 23.745$\substack{+0.619 \\ -0.578}$ & 0.53 & 0.79 & 1.00 & FAST \\ 
9 & UGC~4261 & 8.290$\substack{+0.043 \\ -0.043}$ & 4.195$\substack{+0.039 \\ -0.040}$ & 0.014$\substack{+0.005 \\ -0.005}$ & -0.597$\substack{+0.004 \\ -0.004}$ & -0.591$\substack{+0.002 \\ -0.002}$ & -1.690$\substack{+0.012 \\ -0.012}$ & 20.472$\substack{+0.106 \\ -0.106}$ & \ldots & 0.49 & 0.39 & SDSS \\ 
10 & NGC~2535 & 2.310$\substack{+0.013 \\ -0.013}$ & 4.137$\substack{+0.089 \\ -0.092}$ & -0.885$\substack{+0.077 \\ -0.072}$ & -0.434$\substack{+0.007 \\ -0.007}$ & -0.614$\substack{+0.002 \\ -0.002}$ & -1.741$\substack{+0.163 \\ -0.257}$ & 5.520$\substack{+0.031 \\ -0.031}$ & 0.65 & 0.48 & 0.37 & FAST \\ 

\hline
	\end{tabular}
	}

\end{minipage}
\end{scriptsize}
\end{table*}


\subsection{Fiber and nuclear long-slit spectra emission-line comparison} \label{FAST-SDSS}

\begin{figure*}
	\begin{center}
		\includegraphics[keepaspectratio=true, scale=.55]{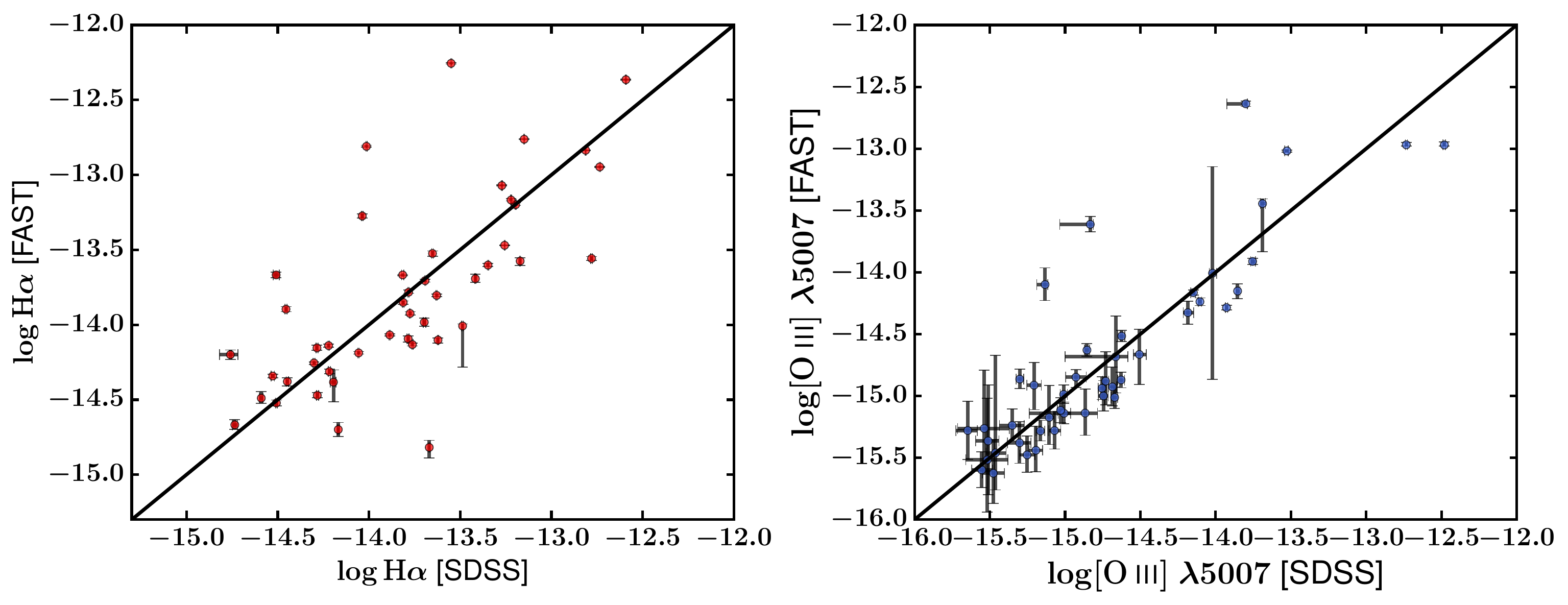}
		\caption{Comparison between the H$ \alpha $ and \OIII{} fluxes measured from the nuclear long-slit and fiber spectroscopy. One-to-one lines are shown. Units are in erg s$ ^{-1} $ cm$ ^{-2} $.}
		\label{F-S-Flux}
	\end{center}
\end{figure*}

\begin{figure}
	\begin{center}
		\hspace*{-1.2cm}\includegraphics[keepaspectratio=true, scale=.43]{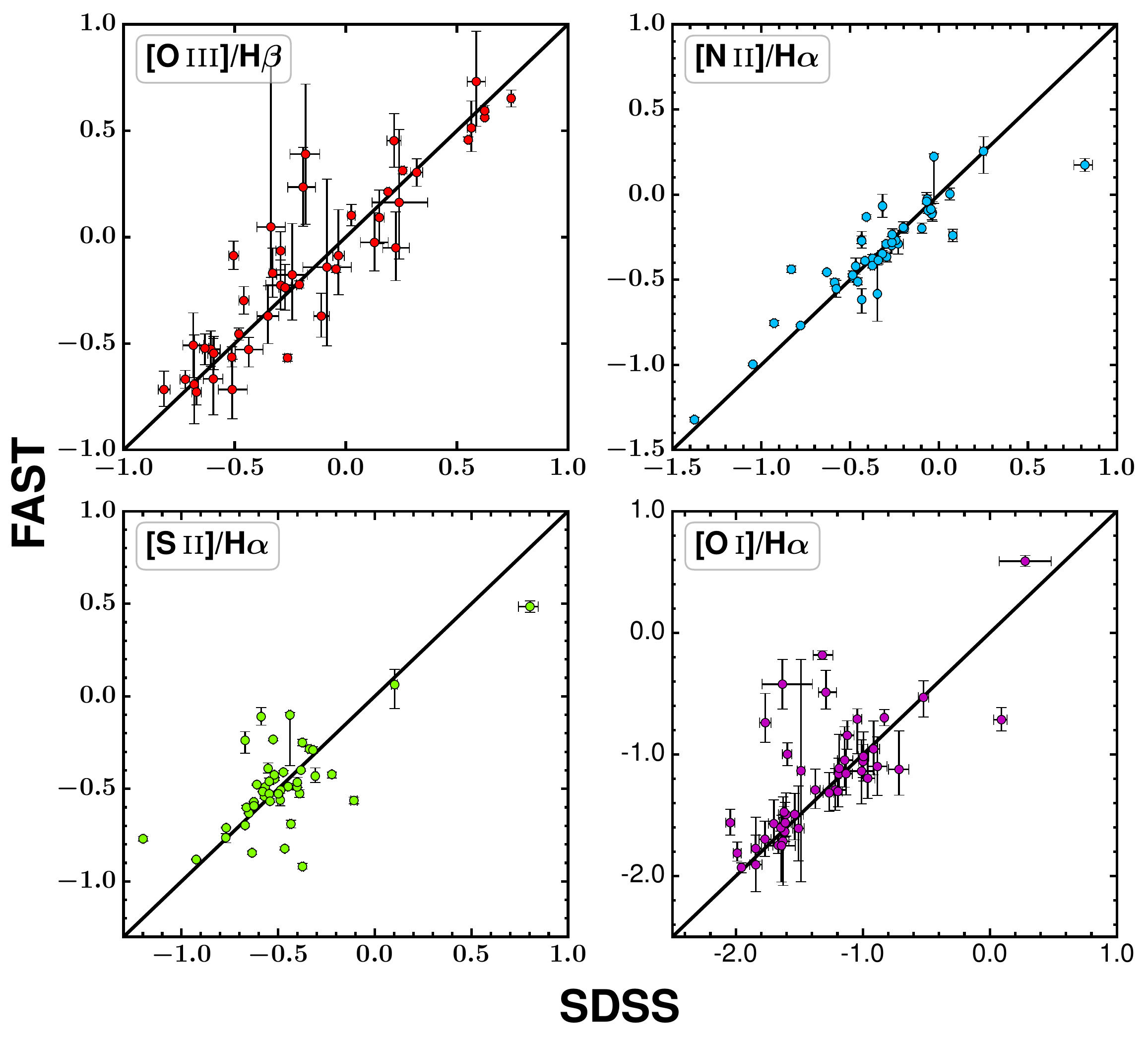}
		\caption{Comparison between the diagnostic line ratios derived from the nuclear long-slit and fiber spectroscopy. FAST measurements are shown on the ordinate and SDSS measurements on the abscissa. All measurements represent the logarithmic values of the line ratios. One-to-one lines are shown.}
		\label{F-S-Lines}
	\end{center}
\end{figure}

Fig. \ref{F-S-Flux} shows a comparison between the nuclear fluxes
measured from long-slit spectroscopy and SDSS fibers for the 45
galaxies observed with both methods. The measured scatter between the
two methods is 0.45~dex for H$\alpha $ and 0.37~dex for
[O\,\textsc{iii}]~$\lambda5007$, and the largest deviations appear
for galaxies in interacting groups. Depending on distance, the
$3.5\arcsec \times 3\arcsec$ long-slit aperture and the
3\arcsec-radius circular aperture may encompass different fractions
of galaxy light as well as companion galaxy
light. Fig. \ref{F-S-Lines} compares the four line ratios used in the
three BPT classification diagrams. There is an overall good agreement
between the two observing methods especially in the cases of the most
commonly used \OIIIHb{} and \NIIHatwo{} ratios. Part of the observed
scatter can be attributed to sensitivity to the exact co-alignment of
the fiber and the long slit, which translates to a difference between
the areas covered by the two methods. In addition, in the case of the
\OIHa{} and \SIIHa{} ratios, the scatter is also attributed to the
generally weaker \OI{} and \SII{} emission lines. Indeed, the largest
deviations are observed in the few cases of low S/N long-slit
spectra, where the fitting of the emission-line profiles is more
uncertain. In certain cases, galaxies observed with the
  600~l~mm$^{-1}$ grating configuration had the blue and red part of their
  spectrum observed on different nights. As a result, differences in
  the flux calibration of the two spectral regions can occur
  depending on the observational conditions and/or the (mis)alignment
  of the long slit at the exact same location on the galaxy between
  observing nights. For these cases, we used the emission lines
  measured from the long-slit observations for the nuclear activity
  classification because the line ratios used in the optical
  diagrams are very close in wavelength and in the same spectrum
  (blue or red). However, for the nuclear SFR measurements
  (Section~\ref{Z-SFR}), we used the SDSS spectra, where available, to
  calculate the Balmer decrement and obtain extinction corrected H$
  \alpha $ fluxes. Seven star-forming galaxies observed with the
  600~l~mm$^{-1}$ grating configuration and having mismatches between their
  red and blue  continua have no available SDSS spectra, and
  Table~\ref{Nuc-Host-Z-SFR} flags their SFR measurements as
  uncertain.

\section{Activity Classification} \label{Classification}

\subsection{Optical Diagnostics}
\label{sec:BPT}
\begin{figure*}
	\begin{center}
		\hspace*{-0.4cm}\includegraphics[keepaspectratio=true, scale=.52]{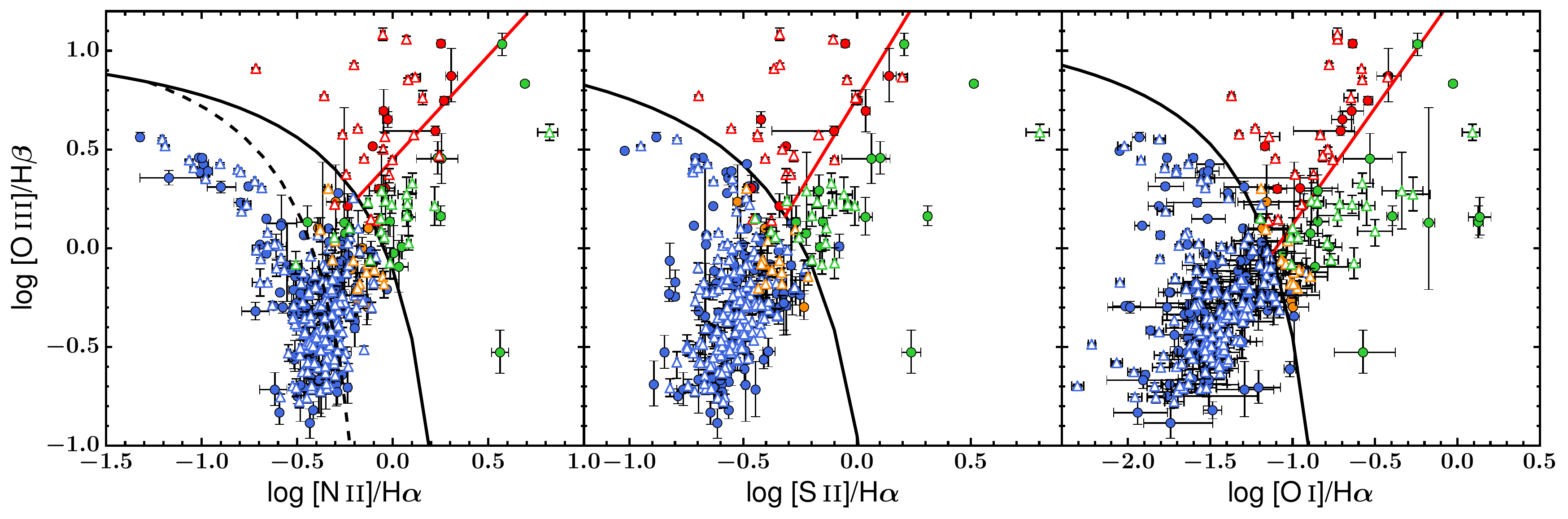}
		\caption{The three BPT diagnostic diagram for the SFRS galaxies. From left to right the [O\,\textsc{iii}]$/$H$\beta$ versus [N\,\textsc{ii}]$/$H$\alpha$, [O\,\textsc{iii}]$/$H$\beta$ versus [S\,\textsc{ii}]$/$H$\alpha$, and [O\,\textsc{iii}]$/$H$\beta$ versus [O\,\textsc{i}]$/$H$\alpha$ are plotted. The dashed line in the [N\,\textsc{ii}]$/$H$\alpha$ diagram (left-hand panel) is the empirical star-forming line of \protect\cite{Kauffmann03}, while the solid black line in all three diagrams is the theoretical maximum starburst line from \protect\cite{Kewley01}. The red lines in all diagrams separate Seyfert nuclei from LINERs, and for the [N\,\textsc{ii}]$/$H$\alpha$ diagram it is defined by \protect\cite{Schawinski07} while for the [S\,\textsc{ii}]$/$H$\alpha$ and [O\,\textsc{iii}]$/$H$\beta$ diagrams they are defined by \protect\cite{Kewley06}. Different color points correspond to the final activity classification obtained using all three diagnostic diagrams. The color code is as follows; Blue: star-forming galaxies, red: Seyfert, orange: TO, green: LINER. Solid circles represent galaxies observed with long-slit spectroscopy and open triangles represent SDSS fiber spectra. Different demarcation lines are described in detail in Section \ref{sec:BPT}.}
		\label{ALL_BPT}
	\end{center}
\end{figure*}

The standard optical activity diagnostic diagrams of [O\,\textsc{iii}]~$\lambda$5007$/$H$\beta$ versus [N\,\textsc{ii}]~$\lambda$6583$/$H$\alpha$, [S\,\textsc{ii}]~$\lambda\lambda$6716, 6731$/$H$\alpha$, and [O\,\textsc{i}]~$\lambda$6300$/$H$\alpha$ (\citealt*{BPT} (hereafter BPT); \citealt{VO87}) are the primary classifiers for the galactic nuclear activity in our sample (Fig. \ref{ALL_BPT}). The standard BPT diagrams are probably the most widely used and best calibrated activity classification diagnostics for galaxies in the nearby universe ($z < 0.4$). These diagrams are able to distinguish star formation from AGN powered galaxies because the relative intensity of spectral lines with different excitation energies depends on the hardness of the ionizing continuum. The diagrams were refined by Kewley et al. (\citeyear{Kewley01}) and Kauffmann et al. (\citeyear{Kauffmann03}), with the former defining a theoretical upper bound to the location of the SFGs using H\textsc{ii} region model spectra and the latter empirically separating the pure star forming galaxies in the [O\,\textsc{iii}]$/$H$\beta$ versus [N\,\textsc{ii}]$/$H$\alpha$ diagnostic based on a sample of 122,808 galaxies from the SDSS. Objects that host both starburst and AGN activity lie between these lines and are known as composite or transition objects (TO; e.g., \citealt{Ho93}). Another refinement on the [S\,\textsc{ii}]$/$H$\alpha$ and [O\,\textsc{i}]$/$H$\alpha$ diagrams was introduced by \cite{Kewley06}, who calculated an empirical separating line in the AGN plane distinguishing the Seyfert populations from the Low-Ionization Nuclear Emission-Line Regions (LINER) \citep{Heckman80}. Lastly, \cite{Schawinski07} defined an empirical line in the [N\,\textsc{ii}]$/$H$\alpha$ BPT diagnostic separating Seyfert from LINER classes, similarly to the \cite{Kewley06} empirical lines.

\begin{table*}
      \begin{minipage}{142mm}
\caption{Activity classification based on different diagnostics. Column (1) SFRS ID (\citealt{SFRS}); Column (2) Galaxy name; Columns (3) - (5) present the nuclear activity classification of the three diagnostic BPT diagrams. Column (6) shows the IRAC color classification based on both the nuclear and integrated colors. Columns (7) and (8) present the MEx and CEx diagrams classification, and Column (9) presents the final adopted classification based on the combination of the three BPT diagnostics and the IRAC color--color diagram. Broad-line (Type-1) Sy and LINERs are reported as Sy-1 and LINER-1 respectively, otherwise Type-2 AGNs are implied. The full version of Table \ref{table2} is available online from MNRAS.} 
	\label{table2}
	\begin{tabular}{@{}ccccccccc}
 \hline
 
 SFRS & Galaxy & \multicolumn{3}{c}{Nuclear Spectra BPT} & IRAC Colors & MEx & CEx & Final Classification \\
\multicolumn{2}{c}{} & [N\,{\sc ii}]/H$\alpha$ & [S\,{\sc ii}]/H$\alpha$ & [O\,{\sc i}]/H$\alpha$ & & & \\
(1) & (2) & (3) & (4) & (5) & (6) & (7) & (8) & (9) \\
 \hline
  1 & IC~486 & Sy & Sy & Sy & Sy & Sy & Sy & Sy\\
2 & IC~2217 & HII & HII & HII & HII & HII & HII & HII\\
3 & NGC~2500 & TO & LINER & LINER & HII & HII & \ldots & LINER\\
4 & NGC~2512 & HII & HII & HII & HII & HII & \ldots & HII\\
5 & MCG~6-18-009 & HII & HII & HII & HII & HII & HII & HII\\
6 & MK~1212 & TO & HII & HII & HII & HII & HII & HII\\
7 & IRAS~08072+1847 & TO & HII & HII & Sy & HII & TO & Sy\\
8 & NGC~2532 & HII & HII & HII & \ldots & HII & HII & HII\\
9 & UGC~4261 & HII & HII & HII & HII & TO & HII & HII\\
10 & NGC~2535 & HII & HII & HII & HII & HII & \ldots & HII\\
 \hline
	\end{tabular}

\end{minipage}
\end{table*}

\subsection{Infrared Diagnostics} \label{IR-diagnostics}

While the BPT method performs well in distinguishing the different type of energy mechanisms in the majority of nearby galaxies, it is insensitive to faint or highly obscured AGN where the characteristic diagnostic lines may be faint or veiled by the gaseous and dusty torus. For these cases, diagnostics employing infrared lines and/or continuum measurements appear more sensitive than optical methods. Stern et al. \citeyearpar{Stern05} used mid-infrared photometry to distinguish AGN dominated from star-forming galaxy SEDs. Galaxies that lie within the following region in the IRAC color--color space are defined as AGN: 
\begin{equation} \label{stern-eq}
\begin{split}
([5.8]-[8.0])&>0.6, \\
([3.6]-[4.5])&>0.2\cdot([5.8]-[8.0])+0.18 , ~ \textrm{and} \\
([3.6]-[4.5])&>2.5\cdot([5.8]-[8.0])-3.5
\end{split}
\end{equation}
where [3.6], [4.5], [5.8], [8.0] are the IRAC magnitudes at 3.6, 4.5, 5.8, and 8.0\,\micron\ correspondingly in the Vega system.

\begin{figure*}
	\begin{center}
		\includegraphics[keepaspectratio=true, scale=.52]{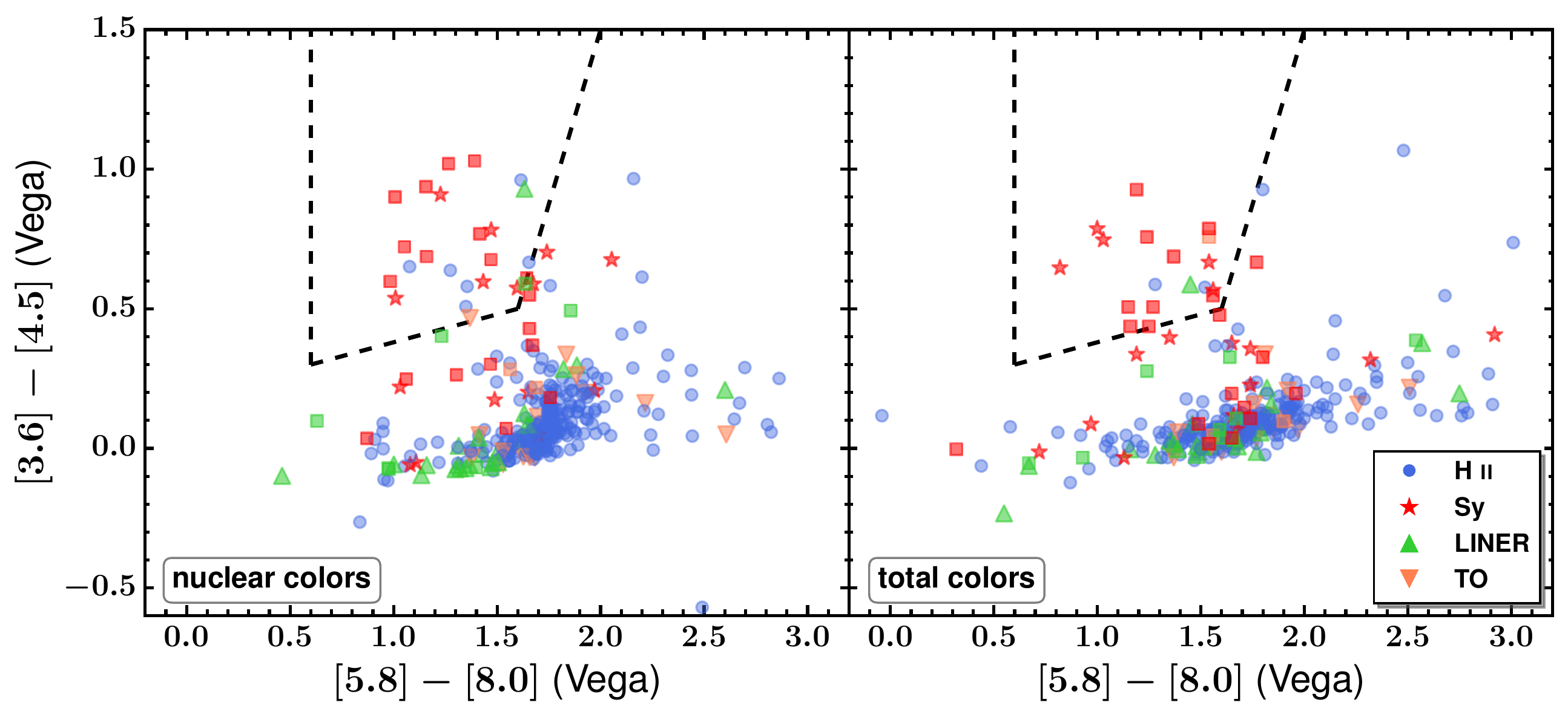}
		\caption{IRAC color--color diagram in the Vega magnitude system using nuclear (left panel) and integrated colors (right panel). The color code is based on the combined BPT classification scheme shown in Fig. \ref{ALL_BPT}. Specifically, blue circles represent H\,\textsc{ii} galaxies, red stars show Sy galaxies, green triangles show LINERs, and upside-down orange triangles show TOs. Squares denote galaxies with broad H$ \alpha $ profiles color-coded based on the combined BPT classification. The dashed lines are the empirically defined lines \citep{Stern05} described in Eq \ref{stern-eq}, that separate AGN withing the wedge from SFGs outside of it.}
		\label{Stern-plot}
	\end{center}
\end{figure*}

In an effort to uncover even the most ``hidden" cases of galaxies hosting AGN activity, we applied the \cite{Stern05} criteria to the SFRS galaxies, using both integrated IRAC colors as well as colors from their nuclear region. The use of nuclear colors has the advantage of decreasing host-galaxy light contamination that can potentially mask the presence of AGN. In order to measure nuclear colors in a consistent way avoiding aperture effects due to varying galaxy distances, we performed matched aperture photometry from physical sizes of $1 \times 1$ kpc$^{2}$ to all galaxies having robust IRAC data in all bands. The apertures were centered on the nucleus, based on the source coordinates (\citealt{SFRS}), and the angular aperture sizes in each galaxy were adjusted according to galaxy distance to ensure consistent sampling of linear scales, following the method described by \cite{Maragkoudakis17}. Fig. \ref{Stern-plot} shows the comparison of the nuclear and integrated color--color plots. Galaxies that are BPT-classified as H\,\textsc{ii}, TO, or LINERs and lie within the AGN region defined by Eq \ref{stern-eq} in both nuclear and total IRAC colors are considered obscured AGN and assigned a Sy classification.

Another criterion to diagnose AGN is based on IR continuum measurements such as the flux-density ratio of the 25 and 60\,\micron\ IRAS bandpasses: ``Warm" sources with $F25 /F60 > 0.2$ are indicative of AGN signatures \citep{Sanders88}. Because many SFRS galaxies are not reliably detected by IRAS in the 25\,\micron\ band, we used MIPS 24\,\micron\ flux measurements, which on average closely track the IRAS 25\,\micron\ \citep{Dale09}. There are 6 galaxies above the $F25 /F60 > 0.2$ threshold that are not recovered by the other diagnostic methods (3 BPT diagrams or the IRAC color-color plot). However, half of them were reported as starbursts by \cite{Balzano83}, and therefore we do not adopt those 6 galaxies as AGN. Ashby et al. \citeyearpar{SFRS} used the \cite{Stern05} integrated IRAC color diagram, the $F_{24}/F_{60}$ ratio criterion, along with a preliminary application of the BPT method (using the \NIIHa{} diagram) for the available at the time SDSS spectra to obtain an initial sense of the AGN content for the SFRS sample. Of the 165 SDSS sources, 30 were identified as AGN with the BPT method, while 19 and 22 galaxies from the entire sample where assigned an AGN classification using the \cite{Stern05} wedge and the $F_{24}/F_{60}$ ratio respectively, giving a total of 52 AGN-powered galaxies when accounting for overlaps among the methods.

\subsection{Final Activity Classification} \label{Final_Class}

The final adopted activity classification for the SFRS galaxies -- H\,\textsc{ii}, Sy, LINER, TO -- was based on a combination of the three variations of the optical emission-line diagrams (BPT), supplemented by the \cite{Stern05} IR color--color diagnostics to account for obscured AGN. Specifically, galaxies classified as H\,\textsc{ii} in all three diagrams or classified as TO in the \NIIHatwo{} BPT and H\,\textsc{ii} in the remaining two diagrams where assigned H\,\textsc{ii} class. In general, the TO class is identified explicitly on the \NIIHatwo{} diagram. However, by definition TOs are considered to have a composite contribution from both star forming and AGN activity. Therefore, we assigned TO classification when all of the following three conditions apply: (i) defined as TO in the \NIIHatwo{} diagnostic; (ii) have Sy or LINER classification in one of the other BPTs; and (iii) have H\,\textsc{ii} classification in the remaining diagram. Sy and LINERs were defined when presenting an AGN or TO class in the \NIIHatwo{} BPT and Sy or LINER classification accordingly in both \SIIHa{} and \OIHa{} diagrams. Cases of non-unanimous classification were visually inspected in all BPT diagrams (see Appendix \ref{Notes}), based on both their positions and their uncertainties in the BPTs, to derive their final classification. To quantify the BPT activity classification uncertainties, especially for the ambiguous cases, we performed Monte Carlo sampling to all 4 diagnostic emission line ratios based on their corresponding ratio uncertainties and derived the probability of a galaxy to belong to a certain class in each BPT diagram (see Appendix \ref{Probabilistic-class}). As mentioned in Section \ref{emission-lines}, five galaxies were classified Sy-1 on the basis of extremely broad lines. Finally, 6 H\,\textsc{ii} galaxies, 1 LINER, and 3 TOs that were IR-diagnosed as AGN under the nuclear or total IRAC-color criteria were reassigned a Sy classification. As a result, based on our combined analysis the SFRS sample consists of: 269 (73\%) star-forming galaxies, 50 (13\%) Seyferts (including 3C~273 and OJ~287), 33 (9\%) LINERs, and 17 (5\%) TOs.

\subsection{MEx - CEx Diagrams} \label{MEx-CEx}

Recently, several other diagnostics were proposed to distinguish between the different activity types for intermediate or high redshift galaxies. These include the Mass-Excitation (MEx) diagram \citep{MEx}, which substitutes for the standard \NIIHatwo{} ratio of the BPT diagram and plots the [O\,\textsc{iii}]~$\lambda$5007/H$\beta$ against stellar mass, and the Color-Excitation (CEx) diagram \citep{CEx} of [O\,\textsc{iii}]~$\lambda$5007$/$H$\beta$ versus the rest-frame $U - B$ color. These can prove beneficial in the case of high-redshift galaxies, as \NII{} and H$\alpha$ can only be observed in the optical out to $z \la$ 0.45. These methods also avoid blending of lines in low-resolution spectra. In the MEx diagram, two empirically determined dividing lines \citep{MEx2} separate the star-forming, TO, and AGN regions. In the CEx diagram, \cite{CEx} provided a demarcation line to distinguish the star-forming and AGN populations, and \cite{MEx} further added a line to mark the region of TO populations. 

We used the MEx and CEx diagrams to compare the activity classification derived with the other diagnostics. The SFRS stellar masses were calculated using the asymptotic $K-$band fluxes measured from fitting the galaxy profiles (Bonfini et al. in preparation) and using the $ M/L $ calibrations described by \cite{Bell03}: \begin{equation} \label{eq:MassKs} \frac{M_{\star}}{M_{\sun}} = 10^{-0.273+(0.091)(u-r)} \times \frac{L_{K_{s}}}{L_{K_{\sun}}}\end{equation} where $u-r$ are the Petrosian SDSS colors in AB magnitudes. Because the MEx diagram was calibrated based on a Chabrier \citeyearpar{Chabrier03} initial mass function (IMF) and the \cite{Bell03} mass-to-light ratio calibration assumes a Salpeter IMF, we converted the stellar masses to the Chabrier IMF following \cite{Longhetti09}.

\begin{figure}
	\begin{center}
		\includegraphics[keepaspectratio=true, scale=.42]{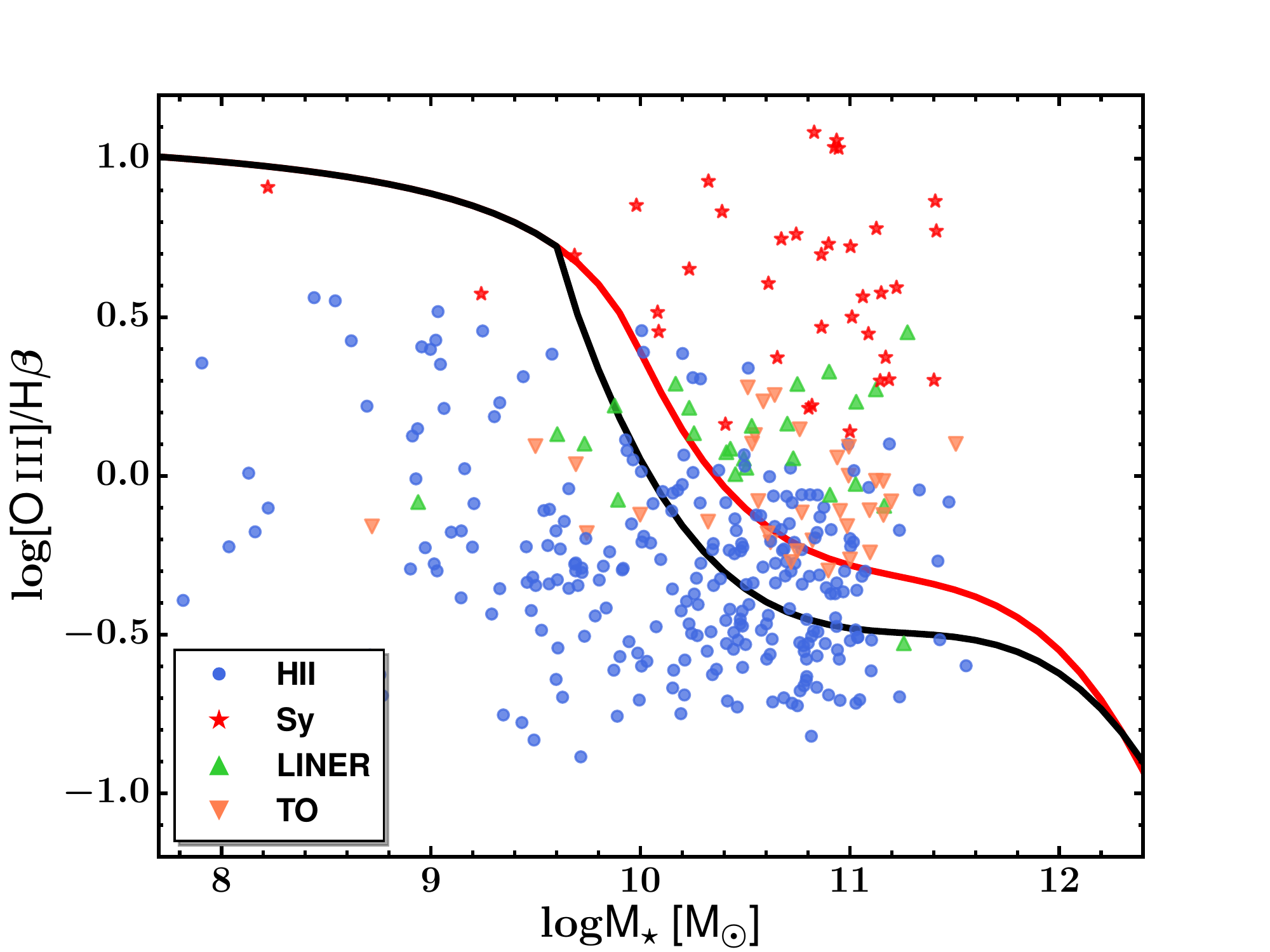}
		\caption{Mass-excitation \citep{MEx} diagram for the SFRS galaxies. Lines \citep{MEx2} divide the classes with AGN to the upper right, SFGs to the lower left, and TO in the narrow middle region. The point colors are based on the BPT-only classification; blue: star-forming galaxies, red: Seyfert, orange: TO, green: LINER.}
		\label{MEx}
	\end{center}
\end{figure}

\begin{figure}
	\begin{center}
		\includegraphics[keepaspectratio=true, scale=.42]{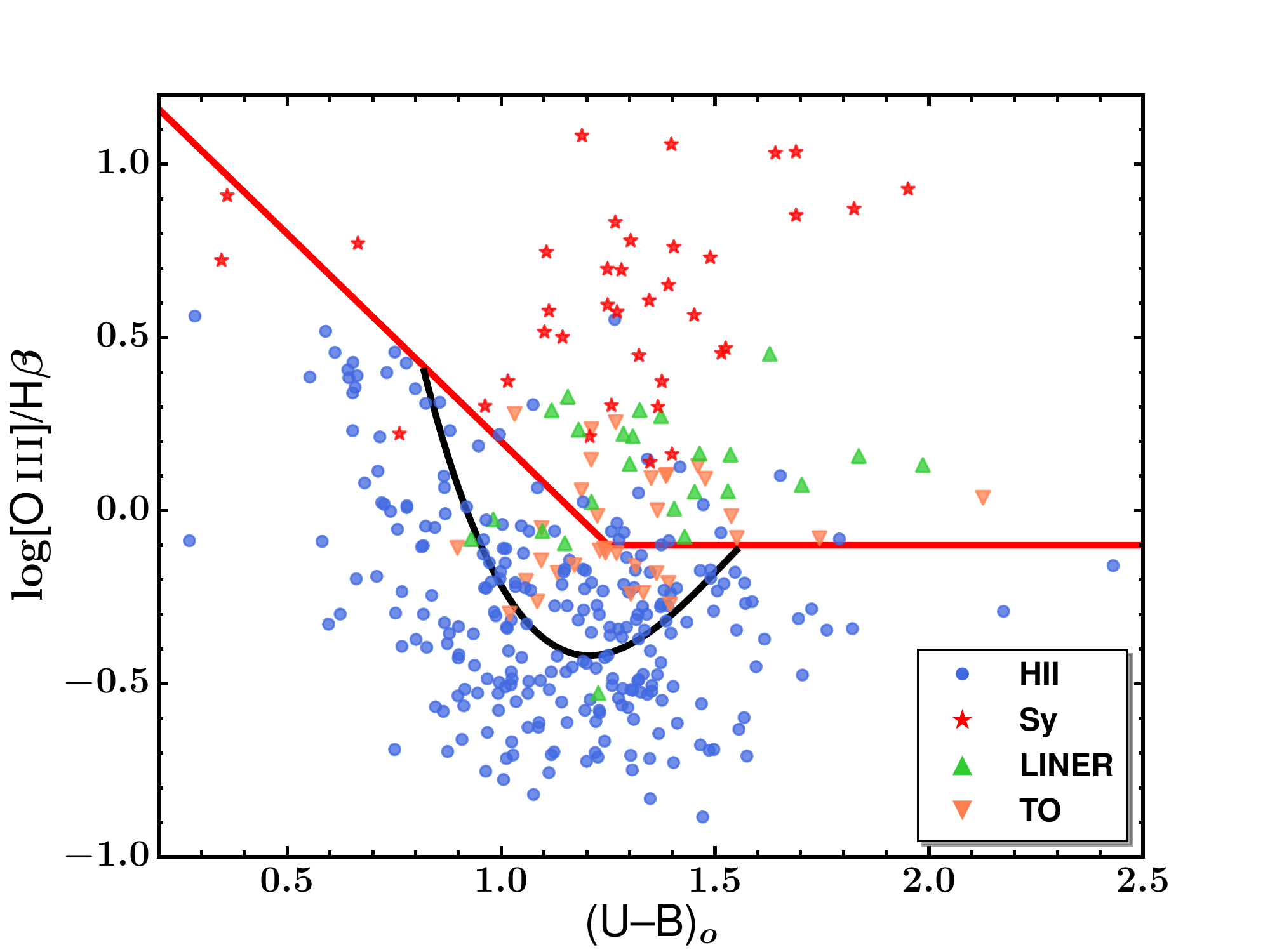}
		\caption{Color-excitation diagram \citep{CEx} for the SFRS galaxies. The red line separates AGN (above the line) from SFGs. The black curve was added by \protect\cite{MEx} and follows the transition where the AGN probability is P(AGN) $> 30\%$. The new region between this curve and the straight lines is analogous to the MEx-intermediate region of the MEx diagram. The point colors are based on the BPT-only classification; blue: star-forming galaxies, red: Seyfert, orange: TO, green: LINER.}
		\label{CEx}
	\end{center}
\end{figure}

Fig. \ref{MEx} and Fig. \ref{CEx} show the SFRS galaxies on the MEx and CEx diagram respectively. While the plots visualize the comparison between the BPT and MEx-CEx results, the classification from MEx and CEx methods shown in Table \ref{table2} was obtained using the IDL routines provided by \citeauthor{MEx} (\citeyear{MEx}; \citeyear{MEx2}). These routines give the probabilities for a given object to belong to each activity class depending on its position on the MEx and CEx diagrams and were calibrated based on the bivariate distribution of a $ z \sim 0 $ SDSS DR7 sample of galaxies with known activity types.

\section{Gas-Phase Metallicities And Nuclear Star Formation Rates} \label{Z-SFR}

The most direct way of measuring gas-phase metallicities is based on temperature-sensitive line ratios such as the [O\,\textsc{iii}]~$\lambda\lambda$4959,5007$/$\OIIIone{} because the flux ratio of lines originating from different excitation states is temperature-dependent, and metals are the primary coolants of H~\textsc{ii} regions. Higher metallicity increases the rate of collisional excitation followed by radiative decay resulting in lower nebular temperatures. Unfortunately, this method cannot be used when \OIIIone{} is undetectable or extremely weak as it often is. To overcome this difficulty, several empirical or theoretical calibrations involving strong emission lines were developed (e.g. \citealt{Pagel}, \citealt{KD02}). Here we derive metallicities using two different calibrations as provided by Pettini \& Pagel (\citeyear{PP04}, hereafter PPO4):
\begin{equation} \label{eq:N2l}
12 + \log(\textrm{O/H}) = 8.90 + 0.57 \times \textrm{N2}
\end{equation}
\begin{equation} \label{eq:O3N2}
12 + \log(\textrm{O/H}) = 8.73 - 0.32 \times \textrm{O3N2}
\end{equation} where $\textrm{N2}\equiv\log([\textrm{N}\,\textsc{ii}]\,\lambda6583/\textrm{H}\alpha)$, and $\textrm{O3N2}\equiv\log(($\OIII{}$/\textrm{H}\beta)/($\NIIHa{}$))$. Equation (\ref{eq:O3N2}) is valid in the range of $-1 < \textrm{O3N2} < 1.9$. These relations were calibrated using the electron temperature (T$_{e}$)-based metallicity for a sample of 137 H\,\textsc{ii} regions. Because the [O\,\textsc{iii}]~$\lambda$4363 line was not available in all of our spectra, we used these calibrations to derive the nuclear-region metallicities for the SFGs and the host-galaxy metallicities for all galaxies with available long-slit spectra.

Table \ref{Nuc-Host-Z-SFR} presents the circumnuclear and host-galaxy gas-phase metallicities of the star-forming SFRS galaxies as derived from both N2 and O3N2 calibrations. For galaxies with available long-slit spectra, we also measured host-galaxy elemental abundances, regardless of activity type, by subtracting the nucleus contribution from the integrated spectrum (Table \ref{non-HII-Z}). While in certain cases the long-slit or SDSS fibers don't cover the entire galaxy surface, and the intensity of the individual lines does depend on the area of the galaxy covered by the slit, their ratio on the other hand does not provided the slit or fiber samples a representative section of the galaxy.

\begin{landscape}
	\begin{figure}
		\begin{center}
			\includegraphics[keepaspectratio=true, scale=.65]{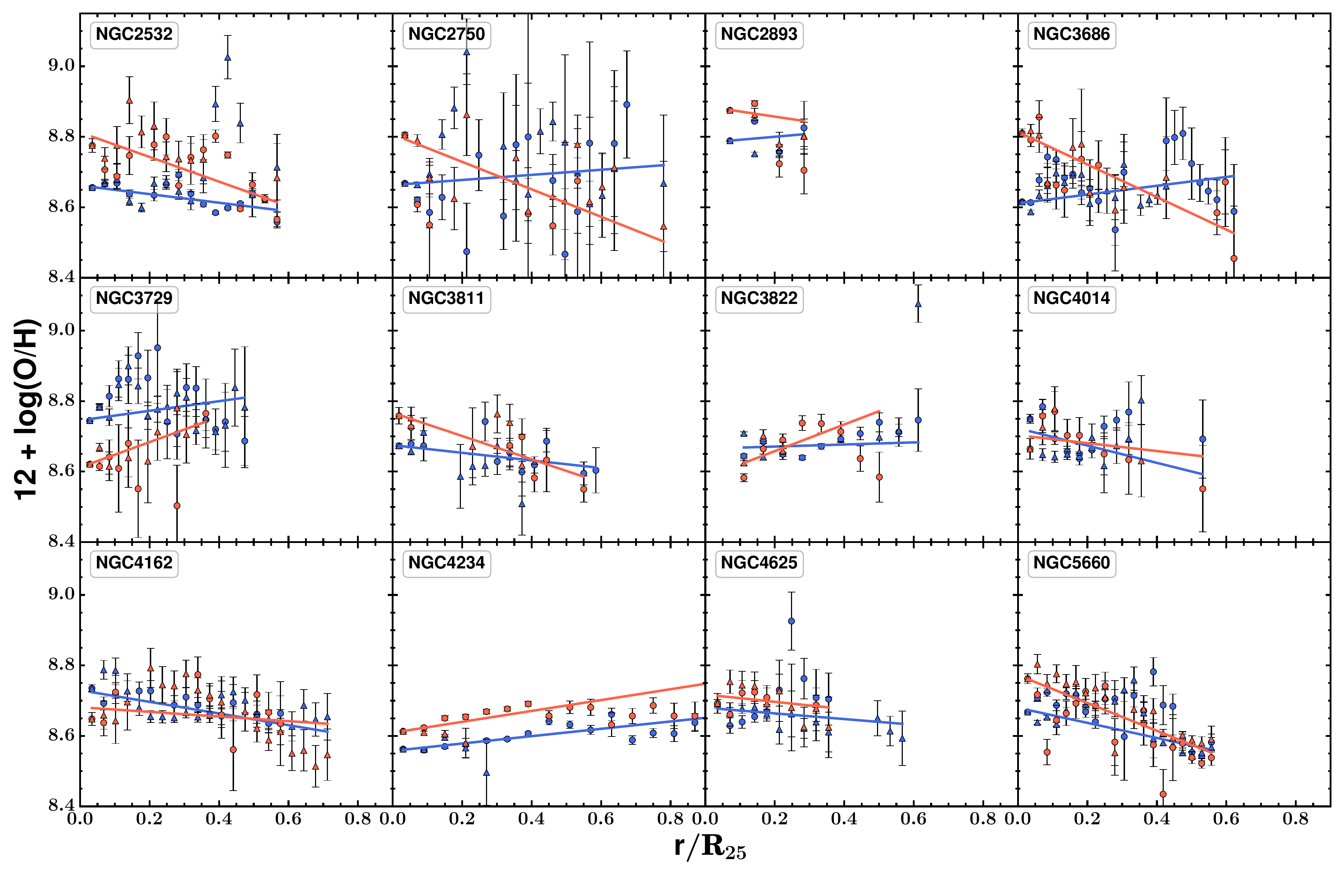}
			\caption{Abundance gradients for the sub-sample of 12 SFRS galaxies. Red points represent metallicities calculated based on the O3N2 calibration, and blue points using the N2 calibration. The red and blue lines are the fit to the O3N2 and N2 calibrations correspondingly. Circles and triangles represent metallicities measured from regions on the left and on the right side of the slit respectively.}
			\label{Z-Gradients}
		\end{center}
	\end{figure}
\end{landscape}

Because the emission line spectrum, characterized in the BPT diagrams, is dependent on both the shape of the ionizing radiation as well as on the metallicity, we derived abundance gradients for 12 large, face-on galaxies observed with the FAST spectrograph in order to measure any systematic effects or variations on the average metallicity measured from the integrated long-slit spectra. We extracted spectra from 3.5\arcsec-wide sub-apertures from successive regions of the galactic disks, starting from the nucleus and moving outwards on both sides along the slit. We performed the standard analysis and starlight subtraction procedure on each individual sub-spectrum and measured the \NII, \OIII, H$\alpha$, and H$\beta$ emission lines on the corresponding regions. Using the PP04 calibrations, we measured each region's metallicity with respect to the galactocentric radius, normalized to the disk radius ($R_{25}$) at the $B = 25$ mag arcsec$^{-2}$ isophote (Fig. \ref{Z-Gradients}). The abundance gradients were derived using linear regression analysis on the metallicity values calculated with calibrations (\ref{eq:N2l}) and (\ref{eq:O3N2}). The basic parameters of the 12 galaxies used to derive metallicity gradients are summarized in Table \ref{z-grad}, along with the gradient slopes and the abundances at the central and characteristic radius $r = 0.4R_{25}$. 

Nuclear SFRs were calculated for the SFRS SFGs (Table \ref{Nuc-Host-Z-SFR}) based on the extinction-corrected H$\alpha$ emission line luminosity. The H$\alpha$ luminosity was measured from the $3.5\arcsec$-wide spectral apertures in long-slit data or the SDSS 3\arcsec-diameter fibers. These correspond to 1.3 and 1.1 kpc respectively at the median distance of the SFRS galaxies. The SFRs were derived based on the \cite{KE12} calibration: 
\begin{equation} \label{eq:SFRHa}
\frac{\textrm{SFR}_{\textrm{H}\alpha}}{(\textrm{M}_{\sun}/\textrm{yr)}} = \frac{L_{\textrm{H}\alpha}}{\textrm{(erg/s)}} \times 10^{-41.27}
\end{equation}
The nuclear SFRs range between $10^{-5}$ and 6.15 M$_{\sun}$ yr$^{-1}$ with mean 0.26 M$_{\sun}$ yr$^{-1}$ and median 0.08 M$_{\sun}$ yr$^{-1}$.

\begin{landscape}

\begin{table}
	\begin{minipage}{136mm}
		\begin{center}
			
	\caption{Host galaxy and nuclear metallicities and SFRs for the SFRS SFGs. Host-galaxy metallicities are given only for galaxies observed with long-slit spectroscopy, with their nuclear contribution subtracted. SFR values flagged with a * symbol indicate uncertain measurements due to differences in the flux calibration between the blue and red spectral regions of galaxies observed with the 600 l mm$^{-1}$ grating configuration (Section \ref{FAST-SDSS}). The full version of Table \ref{Nuc-Host-Z-SFR} is available online from MNRAS.} 
	\label{Nuc-Host-Z-SFR}
	\begin{tabular}{@{}ccccccc}

 \hline
 SFRS & Galaxy & \multicolumn{2}{c}{Nuclear Metallicity} & \multicolumn{2}{c}{Host Galaxy Metallicity} & $\log\textrm{SFR}$\\
\multicolumn{2}{c}{} & O3N2 & N2 & O3N2 & N2 & (Nuclear) \\ 
 \hline
 2 & IC~2217 & 8.772 $\pm$ 0.009 & 8.659 $\pm$ 0.003 & 8.717 $\pm$ 0.011 & 8.941 $\pm$ 0.004 & 0.014 $\pm$ 0.003\\
 4 & NGC~2512 & 8.847 $\pm$ 0.002 & 8.723 $\pm$ 0.001 & 8.743 $\pm$ 0.013 & 8.909 $\pm$ 0.003 & 0.035 $\pm$ 0.001\\
 5 & MCG~6-18-009 & 8.803 $\pm$ 0.005 & 8.735 $\pm$ 0.002 & \ldots & \ldots & 0.657 $\pm$ 0.003\\
 6 & MK~1212 & 8.740 $\pm$ 0.017 & 8.738 $\pm$ 0.005 & \ldots & \ldots & $-$0.322 $\pm$ 0.005\\
 8 & NGC~2532 & 8.763 $\pm$ 0.024 & 8.663 $\pm$ 0.017 & 8.722 $\pm$ 0.025 & $-$0.084 $\pm$ 0.017 & $-$0.037 $\pm$ 0.011\\
 9 & UGC~4261 & 8.534 $\pm$ 0.002 & 8.560 $\pm$ 0.002 & \ldots & \ldots & 0.058 $\pm$ 0.002\\
 10 & NGC~2535 & 8.874 $\pm$ 0.024 & 8.653 $\pm$ 0.004 & 8.656 $\pm$ 0.027 & 9.004 $\pm$ 0.009 & $-$0.871 $\pm$ 0.002\\
 11 & NGC~2543 & 8.770 $\pm$ 0.006 & 8.701 $\pm$ 0.003 & \ldots & \ldots & $-$0.865 $\pm$ 0.003\\
 12 & NGC~2537 & 8.613 $\pm$ 0.027 & 8.609 $\pm$ 0.023 & 8.714 $\pm$ 0.028 & 8.962 $\pm$ 0.023 & $-$3.264 $\pm$ 0.012\\
 13 & IC~2233 & 8.167 $\pm$ 0.003 & 8.213 $\pm$ 0.005 & \ldots & \ldots & $-$2.397 $\pm$ 0.003\\
\hline
	\end{tabular}
		\end{center}
	\end{minipage}
\hfill
     \begin{minipage}{92mm}
\caption{Host galaxy (non-nuclear) metallicities of the non-star-forming SFRS galaxies with available long-slit spectroscopy The nuclear component is subtracted similarly to the SFG host-galaxy metallicities. The full version of Table \ref{non-HII-Z} is available online from MNRAS.} 
\label{non-HII-Z}
	\begin{tabular}{@{}ccccc}
	
	\hline
	SFRS & Galaxy & \multicolumn{2}{c}{Host Galaxy Metallicity} & Classification \\
	\multicolumn{2}{c}{} & O3N2 & N2 \\ 
	\hline
	3 & NGC~2500 & 8.789 $\pm$ 0.033 & 8.914 $\pm$ 0.033 & LINER \\
	36 & IRAS~08572+3915NW & 8.716 $\pm$ 0.025 & 8.905 $\pm$ 0.013 & TO \\
	42 & IC~2434 & 8.755 $\pm$ 0.024 & 8.891 $\pm$ 0.017 & TO \\
	47 & NGC~2824 & 8.689 $\pm$ 0.032 & 8.924 $\pm$ 0.015 & LINER \\
	51 & IRAS~09197+2210 & 8.719 $\pm$ 0.071 & 8.895 $\pm$ 0.032 & TO \\
	61 & CGCG~181-068 & 8.676 $\pm$ 0.058 & 8.885 $\pm$ 0.021 & TO \\
	62 & NGC~2936 & 8.669 $\pm$ 0.052 & 8.714 $\pm$ 0.033 & LINER \\
	79 & IRAS~10120+1653 & 8.743 $\pm$ 0.064 & 8.891 $\pm$ 0.029 & LINER \\
	117 & IRAS~11102+3026 & 8.523 $\pm$ 0.241 & 8.960 $\pm$ 0.044 & Sy \\
	108 & IRAS~10565+2448W & 8.761 $\pm$ 0.012 & 9.031 $\pm$ 0.003 & Sy \\
	\hline
\end{tabular}

\end{minipage}

\end{table}

\begin{table}
	\vspace{10mm}
	\begin{minipage}{242mm}

			\caption{Basic parameters and metallicity gradients for 12 galaxies of the SFRS. Column (1) SFRS Index; Column (2) Galaxy name; Column (3) Morphological types obtained from NED; Column (4) D25 angular diameters taken from Ashby et al.; Column (5) Distances in Mpc taken from Ashby et al.; Column (6) Activity classifications based on the nuclear spectra of galaxies. Columns (7) - (8) Circumnuclear and integrated slit Oxygen abundances, measured from calibration described in equation (\ref{eq:N2l}); Column (9) Central abundance (at radius r = 0) based on the derived N2 abundance gradient; Column (10) Characteristic abundance at radius r = 0.4$R_{25}$ based on the N2 derived abundance gradient; Column (11) Slope of the N2 abundance gradient. Column (12) Central abundance (at radius r = 0) based on the derived O3N2 abundance gradient; Column (13) Characteristic abundance at radius r = 0.4$R_{25}$ based on the 03N2 derived abundance gradient; Column (14) Slope of the 03N2 abundance gradient. The full version of Table \ref{z-grad} is available online from MNRAS.}
			\label{z-grad}
			\begin{tabular}{@{}cccccccccccccc}
				\hline
				SFRS & Galaxy & Morphology & D25 & Distance & Class & \multicolumn{2}{c}{12 + log(O/H)} & N2 & N2 & N2 Gradient & O3N2 & O3N2 & O3N2 Gradient \\
				&  & & (arcmin) & (Mpc) &  & (Nucleus) & (Integrated) & at r = 0 & at r = 0.4$R_{25}$ & (dex $R_{25}^{-1}$) & at r = 0 & at r = $0.4 R_{25}$ & (dex $R_{25}^{-1}$) \\
				(1) & (2) & (3) & (4) & (5) & (6) & (7) & (8) & (9) & (10) & (11) & (12) & (13) & (14)  \\
				\hline
				8 & NGC~2532 & SAB(rs)c & 2.2 & 77.6 & H\,{\sc ii} & 8.76 (8.66) & 8.72 (8.94) &  8.662 $\pm$ 0.003 & 8.608 $\pm$ 0.004 & $-$0.12 $\pm$ 0.01 & 8.81 $\pm$ 0.01 & 8.66 $\pm$ 0.01 & $-$0.35 $\pm$ 0.03 \\
				41 & NGC~2750 & SABc & 2.2 & 37.0 & H\,{\sc ii} & 8.83 (8.68) & 8.74 (9.04) & 8.663 $\pm$ 0.003 & 8.69 $\pm$ 0.02 & 0.07 $\pm$ 0.04 & 8.81 $\pm$ 0.01 & 8.64 $\pm$ 0.03 & $-$0.39 $\pm$ 0.07 \\
				55 & NGC~2893 & (R)SB0/a & 1.1 & 24.0 & H\,{\sc ii} & 8.85 (8.75) & 8.72 (8.95) & 8.783 $\pm$ 0.003 & 8.80 $\pm$ 0.01 &	0.09 $\pm$ 0.04 & 8.89 $\pm$ 0.01 &	8.85 $\pm$ 0.02 & $-$0.15 $\pm$ 0.08 \\  
				129 & NGC~3686 & SB(s)bc & 3.2 & 21.0 & H\,{\sc ii} & 8.81 (8.62) & 8.75 (9.02) & 8.611 $\pm$ 0.002 & 8.69 $\pm$ 0.02 & 0.12 $\pm$ 0.03 & 8.81 $\pm$ 0.01 &	8.52 $\pm$ 0.05 &	$-$0.46 $\pm$ 0.08 \\  
				137 & NGC~3729 & SB(r)apec & 2.8 & 17.1 & H\,{\sc ii} & 8.63 (8.76) & 8.79 (8.99) & 8.745 $\pm$ 0.002 & 8.82 $\pm$ 0.02 & 0.14 $\pm$ 0.04 & 8.612 $\pm$ 0.004 & 8.81 $\pm$ 0.06 & 0.36 $\pm$ 0.11 \\
				147 & NGC~3811 & SB(r)cd? & 2.2 & 54.2 & H\,{\sc ii} & 8.74 (8.67) & 8.72 (8.93) & 8.675 $\pm$ 0.005 & 8.63 $\pm$ 0.01 & $-$0.11 $\pm$ 0.03 & 8.77 $\pm$ 0.01 & 8.62 $\pm$ 0.03 & $-$0.33 $\pm$ 0.06 \\  
				148 & NGC~3822 & S0? & 1.4 & 94.6 & Sy &  \ldots & 8.86 (8.97) & \ldots & 8.67 $\pm$ 0.01 & 0.03 $\pm$ 0.02 &	\ldots & 8.69 $\pm$ 0.02 & 0.38 $\pm$ 0.06 \\  
				163 & NGC~4014 & S0/a & 2.2 & 62.6 & H\,{\sc ii} & 8.67 (8.76) & 8.79 (8.90) & 8.723 $\pm$ 0.009 & 8.61 $\pm$ 0.03 & $-$0.25 $\pm$ 0.06 & 8.70 $\pm$ 0.02 & 8.65 $\pm$ 0.07 &	$-$0.11 $\pm$ 0.15 \\  
				178 & NGC~4162 & (R)SA(rs)bc & 2.3 & 42.5 & H\,{\sc ii} & 8.67 (8.75) & 8.74 (8.89) & 8.730 $\pm$ 0.007 &	8.65 $\pm$ 0.01 & $-$0.17 $\pm$ 0.02 & 8.68 $\pm$ 0.01 & 8.65 $\pm$ 0.02 &	$-$0.07 $\pm$ 0.03 \\  
				186 & NGC~4234 & (R')SB(s)m & 1.3 & 30.0 & H\,{\sc ii} & 8.61 (8.56) & 8.86 (9.09) & 8.558 $\pm$ 0.002 &	8.585 $\pm$ 0.002 &	0.10 $\pm$ 0.01 & 8.61 $\pm$ 0.002 & 8.649 $\pm$ 0.003 & 0.154 $\pm$ 0.01 \\
				218 & NGC~4625 & SAB(rs)mpec & 2.2 & 9.2 & H\,{\sc ii} & 8.69 (8.69) & 8.75 (8.88) & 8.680 $\pm$ 0.007 &	8.64 $\pm$ 0.02 &	$-$0.08 $\pm$ 0.04 & 8.72 $\pm$ 0.02 & 8.67 $\pm$ 0.05 &	$-$0.10 $\pm$ 0.11 \\
				316 & NGC~5660 & SAB(rs)c & 2.8 & 38.9 & H\,{\sc ii} & 8.75 (8.67) & 8.73 (8.95) & 8.680 $\pm$ 0.002 & 8.56 $\pm$ 0.01 & $-$0.21 $\pm$ 0.01 & 8.77 $\pm$ 0.01 & 8.55 $\pm$ 0.01 & $-$0.39 $\pm$ 0.02 \\  
				\hline
			\end{tabular}

	\end{minipage}

\end{table}

\end{landscape}

\section{Results And Discussion} \label{Discussion}

\subsection{Activity Demographics of IR-selected galaxies}

The weights assigned to the SFRS galaxies project the SFRS sample back to the parent PSC$z$ population. Therefore, given the SFRS activity classifications we are able to deliver activity classification fractions for the parent sample (PSC$z$). Specifically, the PSC$z$ comprises 71\% H\,\textsc{ii}, 13\% Seyferts, 3\% TOs, and 13\% LINERs. The close similarity between the SFRS demographics and its parent sample demonstrates the representative nature of the SFRS with respect to local IR-selected galaxies. 

The fraction of all AGNs (Sy, TO, broad-line LINERs) in the PSC$z$ is 20\%. 13\% of the PSC$z$ galaxies are Seyferts, of which 39\% show broad-line spectral features. The mean FWHM of SFRS galaxies fitted with a broad H$ \alpha $ component is $ \sim$2923 km s$ ^{-1} $. The landmark galaxy demographics study of \cite{Ho97} (H97), which used the optical magnitude-limited ($B_{T} \leq 12.5$ mag) sample of the Palomar survey, found 10\% Seyfert galaxies of which 20\% showed broad lines.The PSC$z$ 39\% broad-line fraction is substantially higher than the 20\% of H97. The PSC$z$ broad-line LINERs constitute 25\%, comparable to the 23\% found by H97. In contrast to the PSC$z$, H97 identified a relatively large number of low-luminosity AGNs (LLAGN) with a higher fraction of TOs and LINERs (13\% and 19\% correspondingly) as opposed to the PSC$z$ (3\% and 13\% correspondingly). In the case of TOs, this can be attributed to the strict conditions we applied to define them based on all three BPT diagnostics (Section \ref{Final_Class}), contrary to the common identification based solely on the \NIIHa{} BPT. Furthermore, the higher fraction of LLAGN in the H97 sample can be attributed to the larger number of less-active and early-type systems present in the H97 sample.

Star-forming galaxies are the dominant galaxy class in both the SFRS (73\%) and PSC$z$ (71\%). Despite the broad range in the three-dimensional parameter space covered by the SFRS (Section \ref{SFRS}), far IR selection is oriented towards selecting star-forming populations, ensuring a small AGN ``contamination". This result is of particular importance considering that IR-selection methods are expected to capture both the star-forming phenomenon, as stars are born embedded in dense concentrations of dust and gas, as well as obscured AGN activity. However, the cosmic decrease in the AGN bolometric energy density (e.g., \citealt{AH12}) indicating a downsizing of AGN activity and a subsequent increase of LLAGN in the low-redshift Universe is in agreement with the demographics of H97 but seems at odds with our results. The discrepancy arises because optical selection criteria such as those of H97, are not biased against passive or elliptical galaxies and are able to capture a fair number of LLAGN systems, while IR-selection criteria show a strong preference for SFGs over LLAGN populations.

\subsection{Classification Comparison}\label{sec:classification-comparison}

The BPT and MEx diagrams provide consistent classification for 245 out of the 355 galaxies which lie within the predefined bound of the parameter space covered by the SDSS calibration sample (Section \ref{MEx-CEx}) and have available stellar-mass measurements. However, 57 BPT-defined SFGs are classified as TOs by MEx, and likewise 6 BPT-SFGs fall within the Sy region of the MEx diagram. These represent 66\% and 15\% of the TO and Sy galaxies defined with the MEx method. In reverse 7 and 6 BPT-classified Sy and TOs have an H\,\textsc{ii} classification in the MEx diagram. For BPT-classified LINERs, 12 have a TO classification, 10 have an H\,\textsc{ii} classification, and 5 are classified as Sy in the MEx diagnostic. An important drawback of the MEx diagram is the fact that it is mass dependent and does not rely on direct observable measurements such as flux ratios. This is important because different prescriptions in the $ M/L $ calibrations or SED fitting parameters used to determine stellar masses will generally yield different classification results.

The CEx method classifies 300 galaxies that fall within the calibration regions of the CEx diagram. The activity types for 190 of those galaxies are in agreement with the BPT method. Similarly to the MEx classification, the CEx method tends to classify more BPT-SFGs as TO (64 galaxies representing 75\% of the CEx TO class). In addition, 10 BPT-Sy and 4 BPT-TOs are classified as LINERs (33\% of the CEx LINER class) with the CEx method. The CEx diagram is expected to be biased against broad-line AGNs where the broadband color is not dominated by the host galaxy. Furthermore, as reported by \cite{CEx}, the emphasis given in the empirical definition of the demarcation line in the CEx diagram was based on limiting contamination of the AGN calibration sample. 

The IRAC color classification scheme does not provide detailed demographics on the different activity sub-classes. The separation is done purely on the basis whether a galaxy is classified as an SFG or contains AGN activity and is classified as AGN, regardless its type (Sy, LINER, TO) or intensity. This is evident in Figure \ref{Stern-plot}, where objects of both Type-1 (broad-line) and Type-2 (narrow-line) classes are found within the AGN ``wedge''. Specifically, taking into account both the nuclear and integrated color approaches of the IRAC color diagrams (Section \ref{IR-diagnostics}), 28 out of the 369 SFRS galaxies are classified as AGN based on the empirical \cite{Stern05} criteria, 10 of which had a non-Sy classification in the BPT diagrams (6 BPT-H\,\textsc{ii}, 3 BPT-TOs, and 1 BPT-LINER). Furthermore, the BPT and IRAC colors diagnostics classify unanimously 72\% (265 objects) of the SFRS galaxies as SFGs and 5\% (17 objects) as AGN (BPT-Sy).

\begin{figure*}
	\begin{center}
		\hspace*{-0.7cm}\includegraphics[keepaspectratio=true, scale=.48]{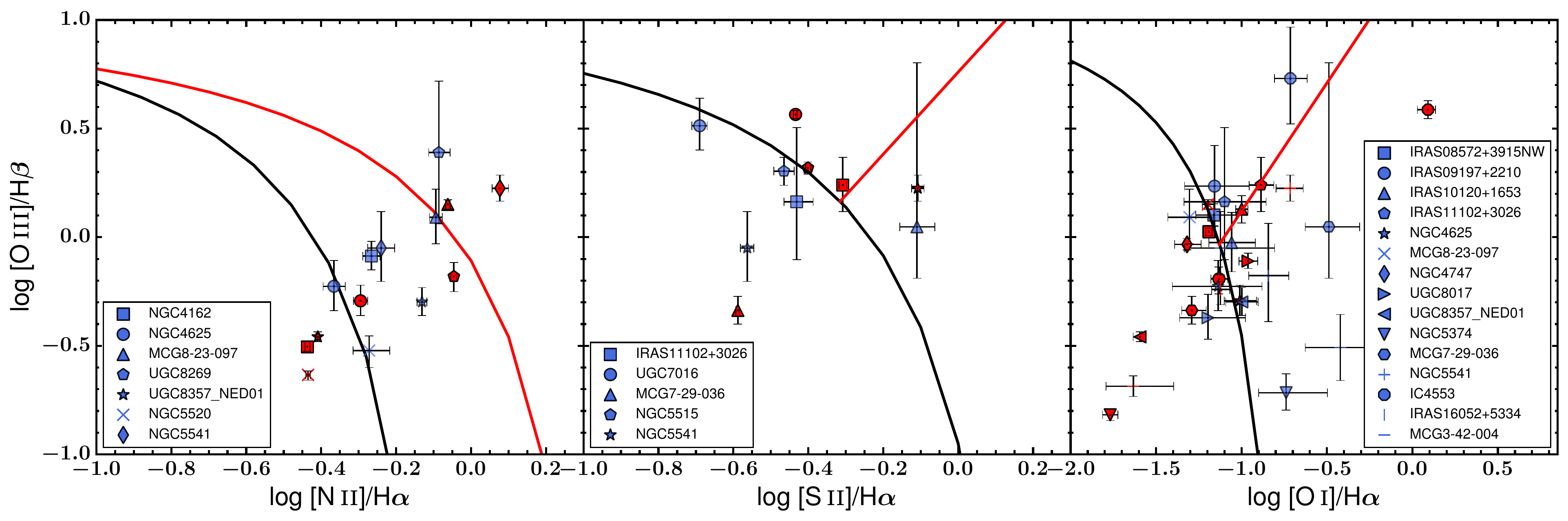}
		\caption{Comparison between the BPT diagrams based on nuclear long-slit and fiber spectroscopy of the SFRS galaxies with inconsistent classifications. Each shape corresponds to a different galaxy; blue shapes show long-slit observation and red shapes SDSS fiber spectra.}
		\label{F-S-Comp}
	\end{center}
\end{figure*}

There are 45 SFRS galaxies which have both nuclear long-slit and SDSS spectra. There is a good agreement between the two observing methods with 38, 39, and 27 galaxies having the same classification in the \NIIHatwo{}, \SIIHa{}, and \OIHa{} BPT diagrams respectively. Fig. \ref{F-S-Comp} show the galaxies with discrepant classifications in one or more BPT diagrams. Most cases presenting a difference between the two methods agree within 3$\sigma$ of the line ratios.

\subsection{Galaxy Properties With Respect To Activity Type} \label{galaxy_properties}

\begin{figure*}
	\begin{center}
		\includegraphics[keepaspectratio=true, scale=.5]{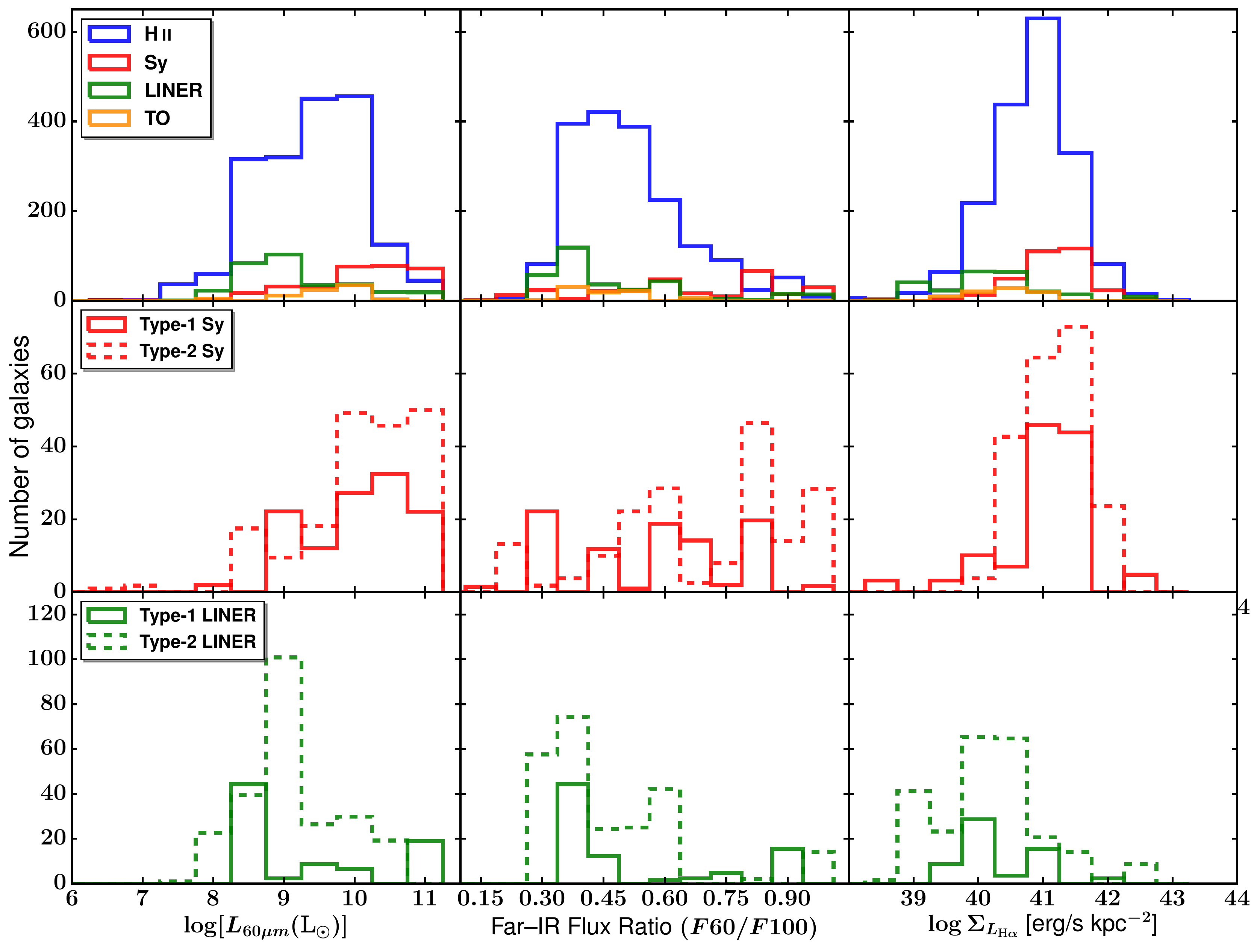}
		\caption{Top panel: Histogram of IRAS 60$\mu$m luminosity (left), $F60/F100$ flux ratio (middle), and circumnuclear $ L_{\textrm{H}\alpha}$ corrected for aperture size (right) for IR-selected galaxies, shown for different activity types. The nuclear area for the $ L_{\textrm{H}\alpha} $ surface density ($ \Sigma_{L_{\textrm{H}\alpha}} $) is calculated based on the $3.5\arcsec \times 3\arcsec$ and 3\arcsec-radius apertures of the FAST and SDSS spectra in each case. Blue lines show SFGs (labeled as H\textsc{ii}), red lines show Seyfert galaxies, orange lines TOs, and green lines LINERs. Middle panel: The corresponding Type-1 and Type-2 Seyfert class distributions. Bottom panel: The corresponding distributions for Type-1 and Type-2 LINERs. The ordinates are based on using SFRS weights to project back to the parent population.}
		\label{SFRSHist}
	\end{center}
\end{figure*}

Being an IR-selected sample, the PSC$z$ contains mostly star-forming galaxies. SFGs have a broad range in 60$\micron$ luminosity between $6.53 < \log( L(60_{\micron})/L_{\sun}) < 11.18$ as seen in Fig. \ref{SFRSHist}, highlighting the fact that the SFRS selection criteria capture all amplitudes of star-forming activity. AGN hosts (Sy and TO) are preferentially found at $L(60\micron) > 10^{9} L_{\sun}$ while LINERs can be seen at lower $L(60\micron)$ with a drop at luminosities higher than $\sim$10$^{9} L_{\sun}$, indicative of hosts without intense star formation. Interestingly, Type-1 and Type-2 Sy and LINERs show similar $L(60\micron)$ distributions with $p$-values of 0.86 and 0.26 respectively returned from a Kolmogorov-Smirnov (K-S) test. 

The $F60/F100$ flux ratio distribution of SFGs which is a proxy for the dust temperature (Fig. \ref{SFRSHist}) peaks around 0.45, which is close to the limit ($\sim$0.5) above which starburst galaxies are typically found (\citealt{RC89}). Seyfert galaxies have a broad distribution of $F60/F100$ and present a peak around $F60/100 \sim 0.85 $ attributed to AGN radiation heating dust to high temperatures. TOs have a dust-temperature distribution with median 0.46, close to the median of SFGs at 0.49, implying that the major contribution to dust heating is from star formation. On the other hand $F60/F100$ in LINERs is confined to lower values, implying both a low radiation density and less dusty environments, the latter typical of early-type galaxies. Again, Type-1 and Type-2 Sy and LINERs have similar $F60/F100$ distributions (K-S $p$-values 0.93 and 0.56 respectively).

The nuclear H$\alpha$ luminosity distribution, serving also as a proxy of SFR for non-AGN hosts, ranges over 7 orders of magnitudes ($10^{36} - 10^{44}$ erg s$ ^{-1} $) in the case of SFGs. Fig. \ref{SFRSHist} shows that Sy show a similar broad range of $L_{\textrm{H}\alpha}/\textrm{kpc}^{2} $ distribution as SFGs ($p$-value = 0.87). In a different manner, TOs and LINERs present lower $ L_{\textrm{H}\alpha}/\textrm{kpc}^{2} $ values as opposed to Seyferts, pointing towards gas-deficient hosts. The opposite view is seen in the case of Seyferts, where the AGN enhances gas ionization, and the $ L_{\textrm{H}\alpha}/\textrm{kpc}^{2} $ distribution peaks at $10^{40.75}$ erg s$ ^{-1} $.

\begin{figure}
	\begin{center}
		\includegraphics[keepaspectratio=true, scale=.52]{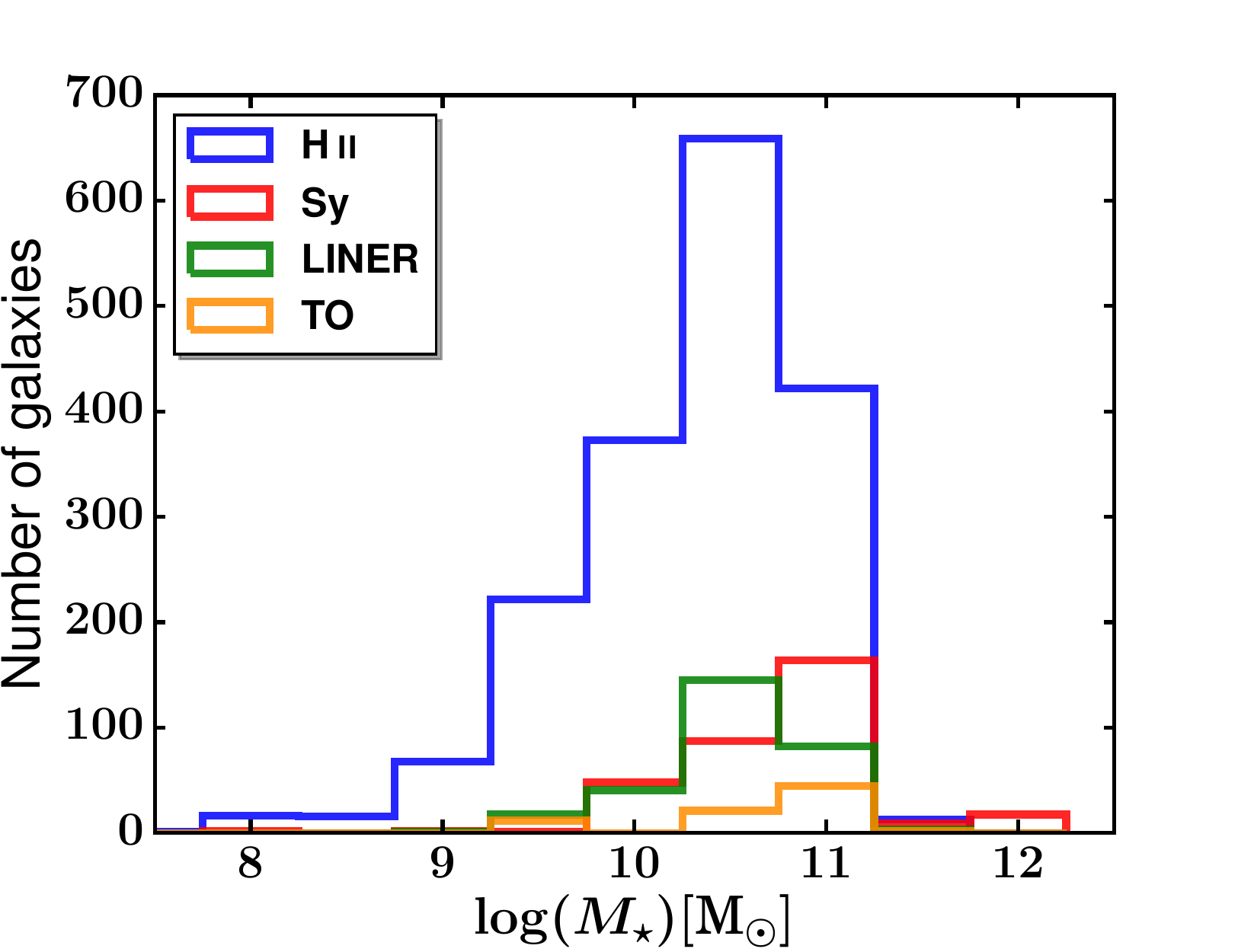}
		\caption{Stellar mass distribution for the different activity types. Blue lines show SFGs (labeled as H\textsc{ii}), red lines show Seyfert galaxies, orange lines TO, and green lines LINERs. Masses refer to Chabrier \citeyearpar{Chabrier03} IMF. Galaxy numbers refer to the SFRS parent sample.}
		\label{MassHist}
	\end{center}
\end{figure}

The SFGs cover a broad range of stellar masses between $10^{7.64}-10^{11.56}$ M$ _{\sun} $ as shown in Fig. \ref{MassHist}. The \cite{Bell03} relations on which mass estimations were based should ideally be applied exclusively to star-forming populations because the presence of non-stellar emission from AGN contaminates the NIR band fluxes, resulting in an overestimation of stellar mass. Therefore, the stellar masses of non-SFG populations should be considered upper limits. Fig. \ref{MassHist} shows that all galaxy types other than SFGs are preferentially located in massive hosts. With the exception of broad-line LINERs that are clearly related to the AGN phenomenon, LINER emission in local galaxies is now considered to result from photoionisation by hot evolved stars and not AGN (\citealt{CidFernandes11}; \citealt{Belfiore16}). Therefore, mass estimates of narrow-line LINERs using $ M/L $ calibrations should be accurate provided correct stellar populations are assumed. LINER nuclei are predominantly found at higher stellar masses with a median of $10^{10.52}$ M$ _{\sun} $. Mass estimates of TOs where the ionization continuum stems from both star-forming and AGN processes should in most cases be treated as upper limits similarly to Sy mass estimations. The median host-galaxy properties of the SFRS sample and the parent population (an unbiased subset of the PSC$z$) from which SFRS was defined are summarized in Table \ref{Median_Prop}.

\begin{table*}
		\begin{center}
		     \begin{minipage}{133mm}
			\caption{Synopsis of the main characteristics of the SFRS and IR-selected galaxies, grouped or in individual activity or luminosity classes. The values in parentheses are calculated based on the weights defined by \protect\cite{SFRS} and correspond to the parent population from which SFRS was defined. Column (1): Description of the activity or luminosity class; Column (2): Number of galaxies in each class; Column (3): $60\micron$ luminosity; Column (4): $F_{60}/F_{100}\,\micron$ flux density ratio; Column (5): Nuclear $ L_{\textrm{H}\alpha} $ surface density; Column (6): Galaxy mass.} 
			\label{Median_Prop}
			\begin{tabular}{@{}lcccccc}
				
				\hline
				&  & $\log L_{60} $ &  & $\log \Sigma_{L_{\textrm{H}\alpha}} $ & $\log M_{\star} $ & Distance \\
				Activity Class & No. & (L$ _{\sun} $) & $ F_{60}/F_{100} $ & (erg s$ ^{-1} $kpc$ ^{-2} $) & (M$ _{\sun} $) & (Mpc)\\
				\hline
				All galaxies & 369 (2560) & 9.55 (9.55) & 0.49 (0.49) & 40.08 (40.10) & 10.49 (10.51) & 77 (71) \\
				SFGs & 269 (1815) & 9.41 (9.47) & 0.48 (0.49) & 40.15 (40.17) & 10.35 (10.43) & 69 (69) \\
				Seyferts & 50 (345) & 10.07 (10.24) & 0.71 (0.80) & 40.24 (40.26) & 10.87 (10.82)  & 117 (105) \\
				Transition Objects & 17 (79) & 9.72 (9.71) & 0.41 (0.46) & 39.62 (39.60) & 10.82 (10.80) & 114 (108) \\
				LINERs & 33 (321) & 9.26 (9.04) & 0.45 (0.36) & 39.50 (39.53) & 10.55 (10.52) & 54 (34) \\
				LIRGs / ULIRGs & 117 / 11 & 10.14 (10.18) & 0.51 (0.55) & 40.08 (40.15) & 11.16 (11.07) & 147 (131) \\
				\hline
			\end{tabular}
		\end{minipage}
		\end{center}
\end{table*}

\subsection{Dominant Central Ionizing Mechanism in LINERs and TOs}

TOs are considered to be galaxies with composite contributions from star-forming and AGN processes in their central energy output. LINERs on the other hand have a much more controversial status in the literature regarding their main ionizing energy source. They are sometimes considered as the low-luminosity, low-accretion-rate branch of Seyfert galaxies (e.g., \citealt{Ho93}, \citealt{Ho03}), which is still a viable assertion at least for LINERs showing broad H$ \alpha $ features. However, this view is disputed in the case of narrow-line LINERs considering that post asymptotic giant branch (post-AGB) stars can produce the required ionizing spectrum necessary to excite LINER emission (e.g., \citealt{Binette94}, \citealt{Stasinska08}, \citealt{Belfiore16}).

The $L(60\micron)$ distribution of TOs follows that of SFGs ($p$-value = 0.08) with a median of $\log L(60\micron) = 9.7$ L$ _{\sun} $ and 9.5 L$ _{\sun} $ respectively, indicative of a similar star-forming activity profile in their host galaxies. Similarly, the median $ F60/F100 $ ratio of both TOs and SFGs is 0.46 and is distinct from the median of Seyferts at 0.80, suggesting star-forming activity as the main source of dust heating in TOs. The $L_{\textrm{H}\alpha}$ surface density distribution of TOs has a smaller range than Sy and SFGs, with median $ L_{\textrm{H}\alpha}/\textrm{kpc}^{2} = 10^{40.30} $ erg s$ ^{-1} $kpc$ ^{-2} $. While the host-galaxy stellar mass is not a property directly linked to the main source of ionization, it gives a sense of the environment where the TOs reside. TOs are allocated predominately at masses above $10^{10.5} $. However, as discussed in Section \ref{galaxy_properties}, Sy-host masses should be considered only as upper limits, and this would apply to TOs depending on their dominant ionizing mechanism. Nevertheless, given the previously discussed similarities of TOs host-galaxy properties with those of SFGs, star-forming activity should be considered as the main source of ionization in the SFRS (PSC$z$) TOs. Given this assertion, TOs reside in massive hosts.

LINERs in the SFRS sample show a sharp drop in their number densities above $L(60_{\micron}) \sim 10^{9} L_{\sun}$, illustrating weak star-forming activity in their hosts. The bulk of LINERs are concentrated mostly at lower values of $ F60/F100 $ with a median of 0.36, with just 12 LINERs lying above the 0.5 starburst threshold as opposed to 118, 38, and 3 SFGs, Sy, and TOs respectively. Furthermore, the $L_{\textrm{H}\alpha}$ surface density distribution of LINERs is confined in the narrow range of $ 10^{38.5} - 10^{40.5} $ erg s$ ^{-1} $kpc$ ^{-2} $ and drops significantly above that, as opposed to Sy and SFGs. Lastly, LINERs are found predominantly in high stellar-mass hosts at a median of $ 10^{10.5}\ \textrm{M}_{\sun}$. Given their generally low H$ \alpha $ and 60\micron\ luminosity it is unlikely that this is the result of bias due to AGN contamination. The above properties demonstrate that the host-galaxy and nuclear properties of LINER bear close resemblance to those of passive early-type systems, hinting that their ionizing continuum stems mostly from older stellar populations (i.e., post-AGB stars) rather than AGN activity. Nevertheless, if LINERs should be considered as AGN powered systems, given the fact that they reside in more massive hosts they should be associated with more massive central black holes.

\subsection{LIRGs and ULIRGs} \label{LIRGs}

The class of LIRGs is characterized by total IR luminosities ($L(\textrm{TIR})$) between $ 10^{11} - 10^{12}$ L$ _{\sun} $, while galaxies with $L(\textrm{TIR}) > 10^{12} $ L$ _{\sun} $ are described as ULIRGs. The SFRS sample contains 117 LIRGs and 11 ULIRGs. Previous studies have demonstrated that a varying fraction of local LIRGs and ULIRGs, ranging from 25\% \citep{Veilleux97} to 70\% \citep{Nardini10}, host an AGN. In the case of SFRS (U)LIRGs, 25\% are Sy hosts, while this percentage increases to 43\% when including LINERs and TOs. \cite{Lee11} using a sample of 115 ULIRGs showed that Type-2 AGNs are more frequently encountered (49 galaxies) compared to Type-1 AGNs (8 galaxies), and the percentage of type 2 ULIRGs increases with infrared luminosity. The SFRS LIRGs consist of 45 AGN, considering all non-SFG classes, out of which 13 are Type-1. Similarly 10 out of 11 SFRS ULIRGs are AGN, with 4 being Type-1, showing an agreement with previous studies. The joined properties of SFRS LIRGs and ULIRGs are summarized in Table \ref{Median_Prop}.

\subsection{Metallicities}\label{sec:Metallicity-Disc}

The SFG's metallicities derived from the O3N2 and N2 calibrations agree for galaxies with sub-solar metallicities but differ for metallicities larger than solar. However, as discussed by PP04, the O3N2 calibration is particularly useful at solar and super-solar metallicities where [N\,\textsc{ii}] saturates, but the strength of [O\,\textsc{iii}] continues to decrease with increasing metallicity. Furthermore, the abundance gradient slopes derived from the O3N2 calibration are generally comparable to the N2 gradients (Fig. \ref{Z-Gradients}). However, the O3N2 calibration indicates in general higher metallicities for the central galactic regions but shows similar values at the characteristic radii of $r = 0.4R_{25}$ (Table \ref{z-grad}). There is also a good agreement between the nuclear abundances measured directly from the central aperture and the ones calculated from the metallicity gradients at $r = 0$ in both calibrations.

The SFG's abundances have a narrow range close to solar (12 + $\log$(O/H) = 8.66), with a median value of 8.72 and 8.67 in the O3N2 and N2 calibrations respectively. The sub-sample of 12 galaxies used to derive metallicity gradients show overall flat metallicity profiles as a function of galactocentric radius (Table \ref{z-grad} and Fig. \ref{Z-Gradients}). The relatively flat profiles derived from our analysis seem to be at odds with similar studies in nearby galaxies (e.g., \citealt{Pilyugin04}; \citealt{Moustakas10}). However, flat average metallicity profiles were also observed by \cite{Moran12} (M12) in a sample of 174 local star-forming galaxies using the O3N2 calibration by PP04. As discussed by M12, galaxies from their local sample with $\log(M_{\star}) > 10.2$  appear to have flatter metallicity profiles and metallicities close to solar. Furthermore, M12 argued that inner metallicity profiles which decline steadily with radius are observed only at the lowest masses. In addition, M12 re-examined the 21 galaxies from the \cite{Moustakas10} sample and estimated the stellar masses of these galaxies, noting that five out of eight galaxies with gradients have $\log(M_{\star}) < 10.2$. Nine out of 12 galaxies SFRS galaxies used to derive metallicity gradients in this paper have $\log(M_{\star}) > 10.2$. Similarly, 178 out of the 261 SFRS SFGs have $\log(M_{\star}) > 10.2$ and metallicities close to solar, indicating that they are consistent with the metallicity gradients observed in local galaxies. Furthermore, the small abundance gradients with respect to galactocentric radius ensures that on average the emission lines measured from the central extraction aperture are not affected by metallicity, regardless the portion of host-galaxy light encompassed, and therefore the resulting classification is robust, capturing the hardness of the central ionizing radiation spectrum.

\section{Conclusions} \label{Conclusions}

Using a fully representative sample of FIR-selected galaxies in the local Universe, we have presented the activity demographics, nuclear metallicities, and host-galaxy properties of the 369 galaxies in the SFRS and derived metallicity gradients for a sub-sample of 12 galaxies. Using the weights assigned to the SFRS galaxies that project the SFRS sample back to the parent PSCz population, we derived the activity demographics and host-galaxy properties for the parent population, summarized as follows.

(i) BPT classification of FIR-selected galaxies gives 71\% SFGs, 13\% Seyferts, 13\% LINERs, and 3\% TOs.

(ii) NIR classification \citep{Stern05} reveals 6 H\,\textsc{ii} galaxies, 1 LINER, and 3 TOs as defined with the BPT method to be AGN galaxies.

(iii) The MEx method is in agreement with the BPT classification for 69\% of galaxies but tends to classify 21\% of BPT-SFGs as TO and 2\% as Sy. 

(iv) The CEx method agrees with BPT classification in 63\% of the cases but classifies 24\% of BPT-SFGs as TO and 21\% as LINER.

(v) The inferred fraction of LLAGN galaxies in the local Universe using IR sample-selection criteria is small compared to SFGs.

(vi) In terms of host-galaxy properties, Sy and TOs are preferentially found at $L(60\micron) > 10^{9}\textrm{L}_{\sun}$ while LINERs have lower $L(60\micron)$ values with few at luminosities higher than $10^{9.5}\textrm{L}_{\sun}$, indicative of hosts with no intense star-formation. SFGs cover a broad range in L(60$\mu$m) between $6.5 < \log( L(60_{\micron})/\textrm{L}_{\sun}) < 11.18$.

(vii) Seyfert galaxies have the highest $F60/F100$ ratios. These can be attributed to AGN radiation heating dust to high temperatures. TOs have the same median dust-temperature as SFGs implying star-formation as the dominant ionizing mechanism, while LINERs are confined to lower dust-temperatures indicative of low-ionization mechanisms and dust-deprived environments. 

(viii) The stellar mass distribution of SFGs covers a broad range ($10^{7.79}-10^{12.14}$ M$ _{\sun} $), while TOs and Seyfert are found exclusively in high stellar-mass hosts, albeit their mass estimations are considered mostly as upper limits. LINERs, which are now strongly considered as galaxies powered by non-AGN processes, are found in high stellar masses galaxies with a median of 10$ ^{10.5} \textrm{M}_{\sun} $.

(ix) SFGs have a wide distribution of nuclear $L_{\textrm{H}\alpha}$ surface densities, ranging between $10^{36.8} < L_{\textrm{H}\alpha}/\textrm{kpc}^{2} < 10^{43.0}$ erg s$ ^{-1} $kpc$ ^{-2} $, and TOs show a similar distribution. Sy nuclear $ L_{\textrm{H}\alpha}/\textrm{kpc}^{2} $ is concentrated above $ 10^{39.5} $ erg s$ ^{-1} $kpc$ ^{-2} $, while LINERs on the other hand are mainly distributed at lower values with median $ L_{\textrm{H}\alpha}/\textrm{kpc}^{2} = 10^{40.18} $ erg s$ ^{-1} $kpc$ ^{-2} $.

(x) Based on their host-galaxy and nuclear properties, the dominant ionizing source in the SFRS (PSC$z$) TOs is star-forming activity. Similarly, LINER host-galaxy characteristics resemble those of passive early-type systems, indicating older stellar populations as their main ionizing source, rather than AGN activity.

(xi) The SFRS SFGs have a narrow range of abundances close to solar and have flat metallicity radial profiles, as found from a sub-sample of 12 galaxies. This is evident for galaxies with $\log(M_{\star}) > 10.2$, as discussed by M12, which is the case for 68\% of the SFRS SFGs.

\section*{ACKNOWLEDGMENTS}
We would like to thank the referee for the constructive comments and suggestions that have improved the clarity of this paper. AM and AZ acknowledge funding from the European Research Council under the European Union’s Seventh Framework Programme (FP/2007-2013) / ERC Grant Agreement n. 617001. Funding for SDSS-III has been provided by the Alfred P. Sloan Foundation, the Participating Institutions, the National Science Foundation, and the U.S. Department of Energy Office of Science. The SDSS-III web site is http://www.sdss3.org/. This research has made use of the NASA/IPAC Extragalactic Database (NED) which is operated by the Jet Propulsion Laboratory, California Institute of Technology, under contract with the National Aeronautics and Space Administration. This publication makes use of data products from the Two Micron All Sky Survey, which is a joint project of the University of Massachusetts and the Infrared Processing and Analysis Center/California Institute of Technology, funded by the National Aeronautics and Space Administration and the National Science Foundation. This work is based [in part] on observations made with the Spitzer Space Telescope, which is operated by the Jet Propulsion Laboratory, California Institute of Technology under a contract with NASA. For this research, we have made extensive use of the Tool for Operations on Catalogues And Tables (TOPCAT) software package (\citealt{TOPCAT}).

\bibliographystyle{mnras}
\bibliography{Bibliography}

\appendix

\section{Probabilistic Activity Classification} \label{Probabilistic-class}

Galaxy activity classification based on the BPT diagrams often stumbles upon the complication of deriving an unambiguous or consistent classification from all three diagnostic diagrams, especially in the occasions where galaxies fall near the demarcation lines. Therefore, obtaining a sense of the uncertainty in a galaxy's classification is important, especially for the ambiguous cases. The uncertainties of the intensities of the diagnostic lines can be translated into an uncertainty in a galaxy's activity classification. We quantified this activity classification uncertainty by calculating the probability that a galaxy falls in the locus of a given class. For each galaxy we generated 1000 samples of their line ratios drawn from a Gaussian distribution, based on their line-ratio uncertainties in all three BPT diagrams, and examined their placement and resulting activity classification on each diagnostic. Thus we derived the probability for each galaxy to be classified as H\,\textsc{ii}, TO, or AGN (Seyfert or LINER) in each of the three diagnostic diagrams. The probabilistic activity classifications of the SFRS sample are presented in Table \ref{Prob-AC}.

\begin{table*}
      \begin{minipage}{140mm}
	\caption{Probabilistic activity classification types based on the uncertainties of line measurements and following the methods outlined in Section \ref{Prob-AC}. The full version of Table \ref{Prob-AC} is available online from MNRAS.}
	\label{Prob-AC}
	\begin{tabular}{@{}ccccccccccc}
 \hline
 SFRS & Galaxy & \multicolumn{3}{c}{[N\,{\sc ii}]/H$\alpha$} & \multicolumn{3}{c}{[S\,{\sc ii}]/H$\alpha$} & \multicolumn{3}{c}{[O\,{\sc i}]/H$\alpha$} \\
\multicolumn{2}{c}{} & H\,{\sc ii} & TO & Sy / LINER & H\,{\sc ii} & LINER & Sy & H\,{\sc ii} & LINER & Sy \\
 \hline  
1 & IC~486 & 0.000 & 0.000 & 1.000 & 0.000 & 0.000 & 1.000 & 0.000 & 0.000 & 1.000 \\ 
2 & IC~2217 & 1.000 & 0.000 & 0.000 & 1.000 & 0.000 & 0.000 & 1.000 & 0.000 & 0.000 \\ 
3 & NGC~2500 & 0.334 & 0.666 & 0.000 & 0.326 & 0.573 & 0.101 & 0.000 & 1.000 & 0.000 \\ 
4 & NGC~2512 & 1.000 & 0.000 & 0.000 & 1.000 & 0.000 & 0.000 & 1.000 & 0.000 & 0.000 \\ 
5 & MCG~6-18-009 & 0.729 & 0.271 & 0.000 & 1.000 & 0.000 & 0.000 & 1.000 & 0.000 & 0.000 \\ 
6 & MK~1212 & 0.000 & 1.000 & 0.000 & 1.000 & 0.000 & 0.000 & 1.000 & 0.000 & 0.000 \\ 
7 & IRAS~08072+1847 & 0.000 & 1.000 & 0.000 & 1.000 & 0.000 & 0.000 & 1.000 & 0.000 & 0.000 \\ 
8 & NGC~2532 & 1.000 & 0.000 & 0.000 & 1.000 & 0.000 & 0.000 & 1.000 & 0.000 & 0.000 \\ 
9 & UGC~4261 & 1.000 & 0.000 & 0.000 & 1.000 & 0.000 & 0.000 & 1.000 & 0.000 & 0.000 \\ 
10 & NGC~2535 & 1.000 & 0.000 & 0.000 & 1.000 & 0.000 & 0.000 & 1.000 & 0.000 & 0.000 \\ 
 \hline
	\end{tabular}

\end{minipage}
\end{table*}

\section{STARLIGHT Code Simulation Study} \label{ST-SIMs}

\subsection{Modeling the input spectrum} 

\subsubsection{Statistical fluctuations} \label{ST-Flux_Dist}
A primary concern when measuring emission lines in galaxy spectra is the efficient subtraction of the underlying stellar continuum and absorption features at the wavelengths where the emission lines of interest reside. While spectral synthesis codes suffer from known degeneracies where different combinations of SSP properties (e.g., ages, metallicities) and galaxy properties (e.g., extinction, AGN component) can fairly reproduce the observed spectrum, to a first level, any combination reproducing the input spectrum is considered adequate for starlight subtraction. However, due to \textsc{starlight}'s complicated architecture where the fit is carried out using a mixture of simulated annealing, Metropolis algorithm, and Markov Chain Monte Carlo techniques, intrinsic uncertainties are introduced in the process, and even when making identical-parameter runs the results can slightly vary.

\begin{figure*}
	\begin{center}
		\includegraphics[keepaspectratio=true, scale=.40]{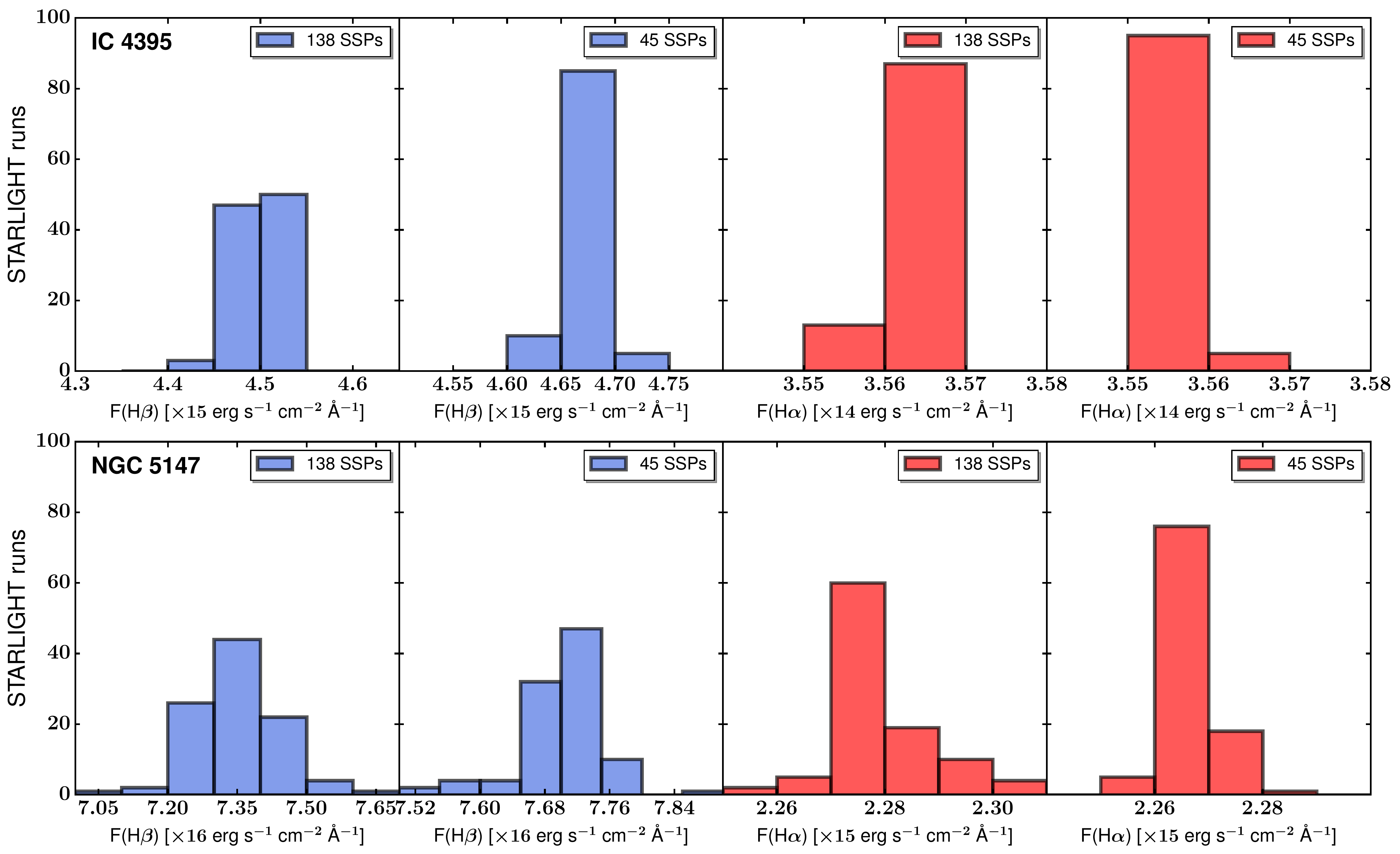}
		\includegraphics[keepaspectratio=true, scale=.40]{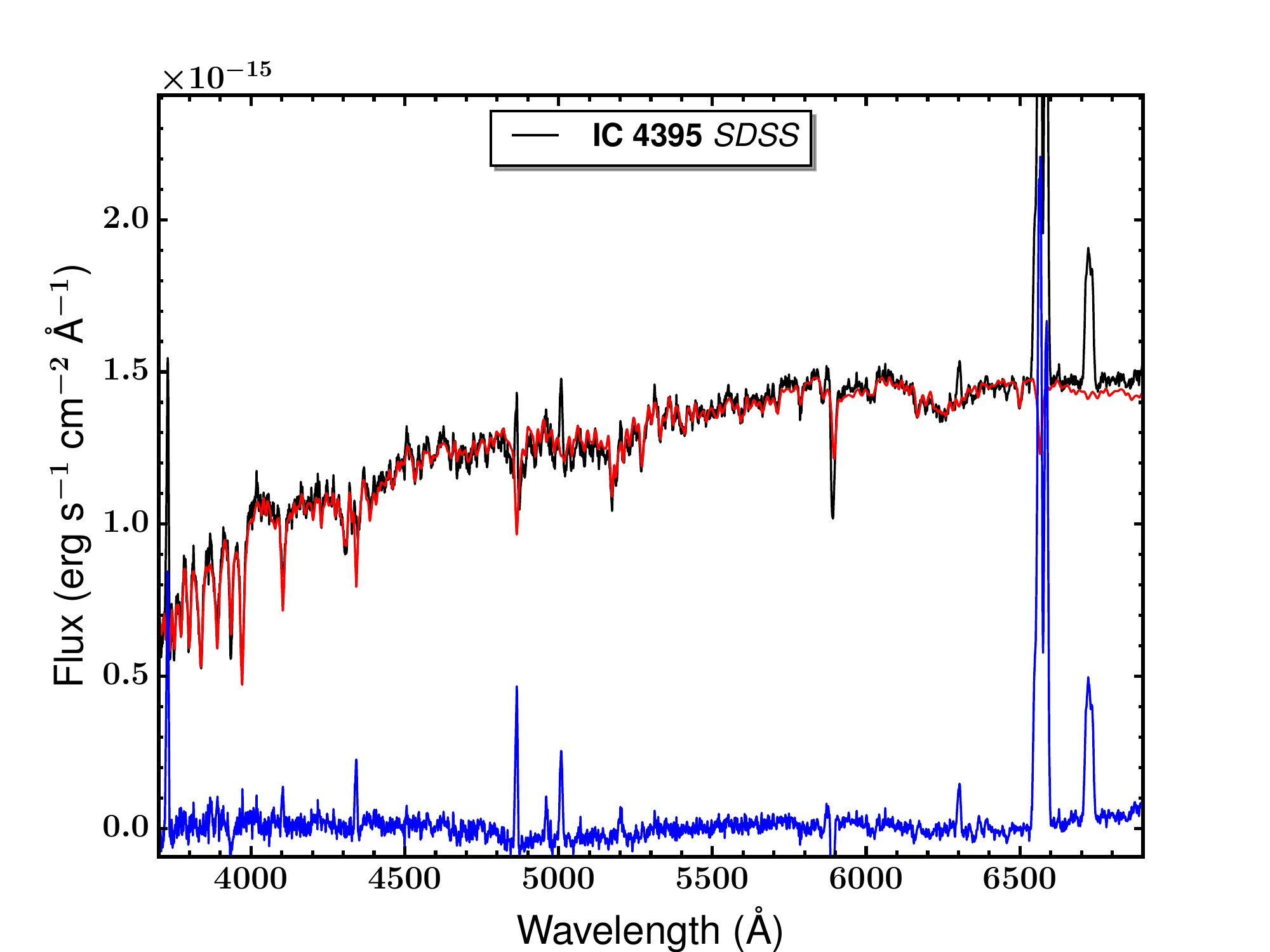}
		\includegraphics[keepaspectratio=true, scale=.40]{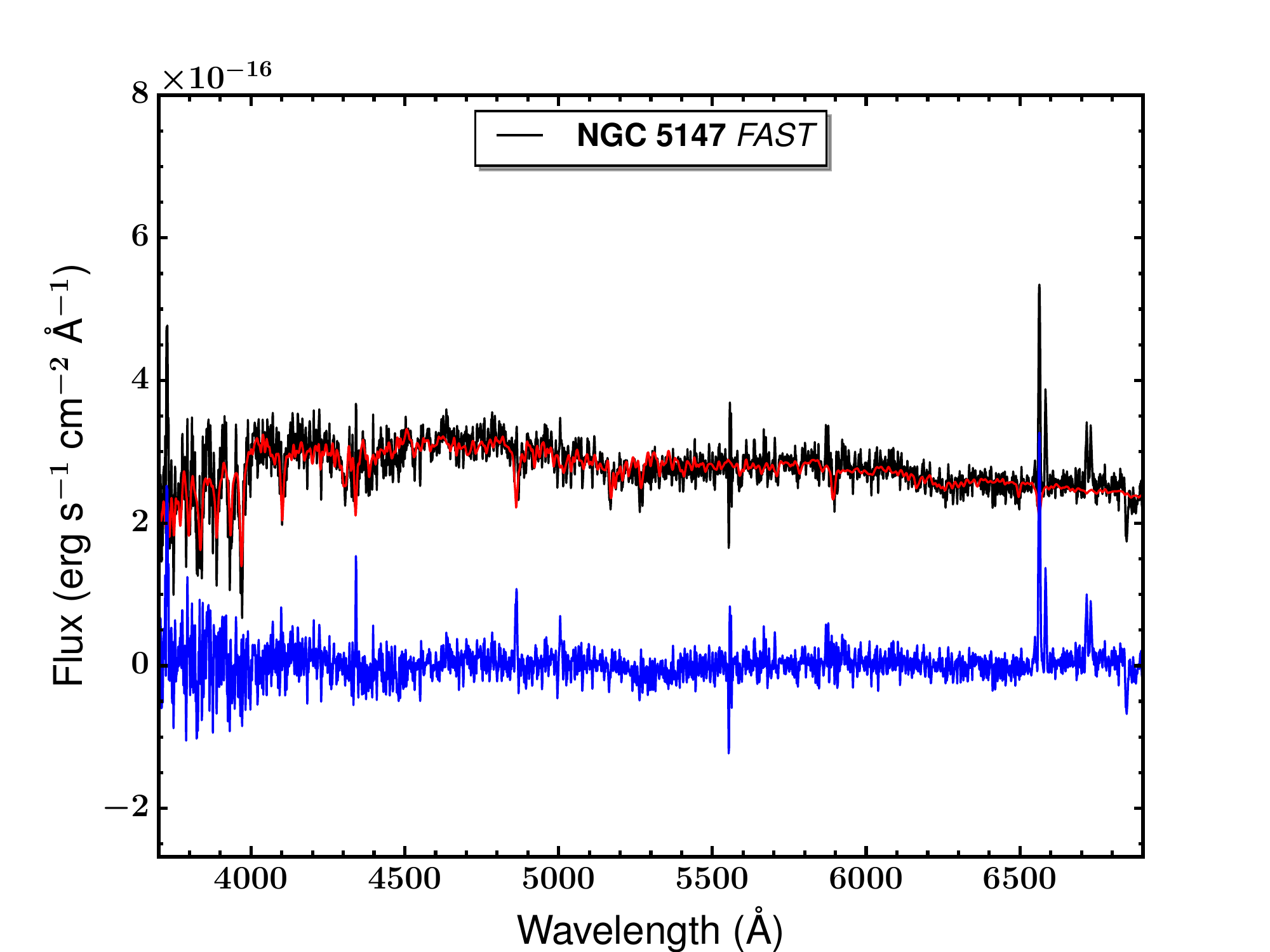}
		\caption{The distribution of F(H$ \beta $) and F(H$ \alpha $) emission lines for galaxies IC~4395 (top row) and NGC~5147 (middle row) measured after performing 100 \textsc{starlight} runs and subtracting the model from the observed spectrum. Two sets of identical-parameter \textsc{starlight} runs were performed using a grid of 138 and 45 SSPs respectively. Their respective nuclear spectra are shown in the bottom row. The observed spectrum is shown with a black line, an example \textsc{starlight} fit with a red line, and the corresponding starlight subtracted spectrum with a blue line.}
		\label{STARLIGHT-Hist}
	\end{center}
\end{figure*}

To quantify the above uncertainties in \textsc{starlight}, we performed 100 identical-parameter runs for two galaxies observed with SDSS (IC~4395) and FAST (NGC~5147), using a two-set grid of base spectra, one with 138 SSPs and one with 45 SSPs. In each run we subtracted the model spectrum from the observed and fitted the H$ \beta $ and H$ \alpha $ emission lines using \textsc{sherpa}. The distributions of the resulting emission line fluxes F(H$ \beta $) and F(H$ \alpha $) for the two sets of 100 runs (using the 138 and 45 SSP grids respectively) are shown in Figure \ref{STARLIGHT-Hist}. While the spread of F(H$ \beta $) is somewhat larger compared to F(H$ \alpha $), especially when using the 45 SSPs grid, the variance between the values is within the measured line-flux uncertainties. Specifically, in all cases the standard deviation of the F(H$ \beta $) and F(H$ \alpha $) distributions is smaller by a factor of 10 compared to the uncertainties of the measured fluxes. These results are independent of the number of runs performed. We verified this by repeating the tests for 500 identical-parameter runs and recovering a factor of 10 difference between the standard deviation of the fluxes and their measured uncertainties. Therefore, the intrinsic uncertainties introduced by \textsc{starlight} have no impact on our results.

\subsubsection{The Balmer regions} \label{Balmer-test}

\textsc{starlight} masks the most notorious optical emission lines in
the spectrum including the Balmer series before performing a
fit. While Balmer lines can be valuable in characterizing the ages of
the underlying stellar populations, the presence of an emission
component at the same wavelength of the absorption feature makes
their usage impractical. Therefore the fit is based on modeling the
emission-free regions of the galaxy spectrum. Assessing the goodness
of the fit in the Balmer regions is not feasible when modeling
emission-line galaxies, but it is important because an inadequate fit
can lead to the over- or under-subtraction of the absorption
components, which will have an impact on the measured emission lines.

In order to determine the accuracy of \textsc{starlight} in modeling
the Balmer absorption lines, we performed simulations using a
modified version of the SED modeling code CIGALE \citep{CIGALE2}. We used the high-resolution BC03 libraries (instead of the default low-resolution BC03 libraries) with a predefined set of
star-formation histories and dust attenuations. Specifically, we
generated galaxies with a single star-formation episode decaying with
an e-folding time fixed at 50~Myr. Onsets were modeled at 50, 100,
200, 400, 500, 1000, 1500, and 2000~Myr before now.  These ages
sample the range of Balmer equivalent widths for both young and old
populations. Each of the 8 base SEDs was reddened by one of three
$E(B-V)$ color excess values: 0.16, 0.32, or 0.65~mag.  CIGALE
applied these reddening values only to stars $<$10~Myr old with older
stars being reddened by 0.001 times as much. No gas or nebular
emission was included to ensure the creation of emission-free galaxy
SEDs. The models gave a total of 24 galaxy spectra referred to as
``reference'' spectra (Table~\ref{Balmer-Table}).  The model results
are as expected with Balmer equivalent widths peaking at ages near
500~Myr.  Because reddening was applied mainly to the youngest
populations, it reduces their contribution to the emerging light and
slightly increases the equivalent width in the resulting light output
at each age.  This effect becomes negligible for populations
$\ga$400~Myr old.

In order to simulate observations, we generated a set of 100 spectra for each of the 24 reference SEDs adding Gaussian noise. The noise
amplitude was the median uncertainty of the measured H$ \alpha $
emission line of the SFRS galaxies.  We  ran \textsc{starlight}
for the 2400 spectra and obtained the best-fit model for each,
referred to as ``model'' spectra. We then computed the mean and
standard deviation of $\rm EW_{mod} - EW_{ref} $ for both H$ \beta $
and H$ \alpha $. Table~\ref{Balmer-Table} summarizes the results.

The models reproduce the absorption EWs reasonably well, showing
standard deviation in H$\beta$ EW $\sim$0.1~\AA\ for the youngest
populations and less than that for populations older than
400~Myr. Differences from the reference values are of the same order
and always $\le$0.18~\AA\ except for the very youngest
populations. H$\beta$ is by far the more important line because the
stellar absorption EW is about twice that of H$ \alpha $, and the
emission line EW is typically 1/5 of that for H$ \alpha $. The lowest
emission EW of H$\beta $ measured in the SFRS sample is 0.19~\AA\ for
a single galaxy, and the second lowest value is 0.45~\AA. Therefore,
\textsc{starlight} over- or under-fitting the absorption lines should
not have a significant impact on the emission line measurements
themselves or on the results for either the classifications or the
H$\alpha$/H$\beta$ extinction estimates (Sec.~\ref{ext-cor}).

\begin{table*}
		        \caption{SED characteristic of the ``reference" galaxies generated with CIGALE, and EW differences between ``reference" and ``model" spectra. Columns (1) - (2) The age of the onset of star formation in Myrs and the E($ B-V $), describing the star-formation history and color excess of the generated ``reference" SEDs; Column (3) - (4) The EW of the H$ \beta $ and H$ \alpha $ absorption lines for the ``reference" spectra; Columns (5) - (6) The mean and standard deviation of the difference of the EW(H$ \beta $) for each of the ``model" spectra from their corresponding ``reference" spectrum (see Section \ref{Balmer-test}); Columns (7) - (8) Similar to Columns (5) and (6) for the EW(H$ \alpha $). All EWs and $\langle \Delta EW\rangle$ values are given in units of \AA.}
		\label{Balmer-Table}
                \hspace*{-1.1cm}\begin{tabular}{@{}cccccccc}
			\hline
Age & $E( B-V $) & $EW_{H \beta}$ & $EW_{H \alpha}$ & $\langle \Delta EW_{H \beta}\rangle$ & $\sigma_{\Delta EW_{H\beta}} $ & $\langle \Delta EW_{H \alpha}\rangle$ & $\sigma_{\Delta EW_{H\alpha}} $ \\
           (Myr) & (mag) & (Ref) & (Ref) & (Mod--Ref) & (Mod--Ref) & (Mod--Ref) & (Mod--Ref)\\            
			(1) & (2) & (3) & (4) & (5) & (6) & (7) & (8) \\
			\hline

  50 & 0.16 & 3.89 & 2.42 & 0.30 & 0.12 & 0.16 & 0.06\\
  50 & 0.32 & 3.97 & 2.45 & 0.23 & 0.11 & 0.12 & 0.05\\
  50 & 0.65 & 4.09 & 2.48 & 0.11 & 0.10 & 0.06 & 0.05\\
  100 & 0.16 & 4.92 & 2.92 & 0.18 & 0.13 & 0.11 & 0.06\\
  100 & 0.32 & 5.08 & 2.98 & 0.16 & 0.10 & 0.09 & 0.05\\
  100 & 0.65 & 5.25 & 3.05 & 0.08 & 0.10 & 0.04 & 0.04\\
  200 & 0.16 & 6.85 & 3.84 & $-$0.06 & 0.11 & 0.03 & 0.04\\
  200 & 0.32 & 6.98 & 3.89 & $-$0.01 & 0.11 & 0.04 & 0.04\\
  200 & 0.65 & 7.08 & 3.94 & 0.01 & 0.09 & 0.02 & 0.04\\
  400 & 0.16 & 9.39 & 4.75 & $-$0.09 & 0.07 & $-$0.02 & 0.04\\
  400 & 0.32 & 9.41 & 4.75 & $-$0.09 & 0.06 & $-$0.01 & 0.04\\
  400 & 0.65 & 9.42 & 4.76 & $-$0.09 & 0.07 & $-$0.02 & 0.04\\
  500 & 0.16 & 9.57 & 4.77 & $-$0.15 & 0.02 & $-$0.05 & 0.02\\
  500 & 0.32 & 9.57 & 4.77 & $-$0.15 & 0.03 & $-$0.05 & 0.02\\
  500 & 0.65 & 9.58 & 4.77 & $-$0.16 & 0.02 & $-$0.05 & 0.02\\
  1000 & 0.16 & 6.52 & 3.71 & $-$0.06 & 0.06 & $-$0.01 & 0.03\\
  1000 & 0.32 & 6.52 & 3.71 & $-$0.05 & 0.06 & $-$0.01 & 0.03\\
  1000 & 0.65 & 6.52 & 3.71 & $-$0.06 & 0.05 & $-$0.01 & 0.03\\
  1500 & 0.16 & 4.65 & 3.10 & $-$0.06 & 0.05 & $-$0.03 & 0.03\\
  1500 & 0.32 & 4.65 & 3.10 & $-$0.06 & 0.06 & $-$0.03 & 0.03\\
  1500 & 0.65 & 4.65 & 3.10 & $-$0.06 & 0.05 & $-$0.03 & 0.03\\
  2000 & 0.16 & 3.95 & 2.82 & $-$0.18 & 0.07 & $-$0.07 & 0.02\\
  2000 & 0.32 & 3.95 & 2.82 & $-$0.18 & 0.05 & $-$0.07 & 0.02\\
  2000 & 0.65 & 3.95 & 2.82 & $-$0.15 & 0.06 & $-$0.06 & 0.02\\
\hline
        \end{tabular}

\end{table*}

\subsection{Global galaxy parameters and AGN component} 

\cite{STARLIGHT_TEST} (F14) made an extensive study of the
\textsc{starlight} code using 1638-zone spectra of the Sb NGC 2916,
observed with an Integral Field Unit (IFU). They performed
simulations using observed and synthetic spectra (built from
\textsc{starlight} fits to the observed data) to explore
uncertainties related to the data and the spectral synthesis
method. They added random noise and shape-changing perturbations in
both observed and synthetic spectra in order to address noise
fluctuations and continuum shape calibrations. These simulations were
then used to evaluate uncertainties in the global parameters of the stellar populations derived
by \textsc{starlight} such as the mean ages, masses, metallicities,
extinction, and star formation histories (SFHs). Despite the fact
that \textsc{starlight} solutions depend on the seed for the
random-number generator, given the pseudo-random nature of its Markov
chains, variations of this kind have a tiny effect on the derived
properties. Furthermore, synthetic unperturbed spectra produce higher
dispersions of the $\Delta$ (simulation minus original) value of each
derived quantity, compared to the observed unperturbed spectra,
attributed to the fact that observed spectra have access to a smaller
subspace of acceptable solutions than the synthetic spectra. When
introducing random noise, F14 found that the mass-weighted ages and
metallicities have broader $\Delta$ distributions ($\sigma\sim$0.15~dex)
than the luminosity-weighted counterparts ($\sigma\sim$0.1~dex),
while stellar masses are good to $\sim$0.1~dex. 
Overall, the SFHs appear less constrained than other global
galaxy parameters, being a higher-order product of the spectral
synthesis. Averaging over spatial regions comprising many zones
reduces the uncertainties.

Similar but simpler simulations were also performed by \cite{STARLIGHT} to 65 test galaxies created from the mean derived properties of fits to a sample of SDSS galaxies. The results showed that the population vectors recovered by \textsc{starlight} are subject to large uncertainties, but mean derived properties such as stellar ages and metallicities were better recovered. 

\begin{figure*}
	\begin{center}
		\hspace*{-1.0cm}\begin{tabular}{cc}
			\includegraphics[keepaspectratio=true, scale=.45]{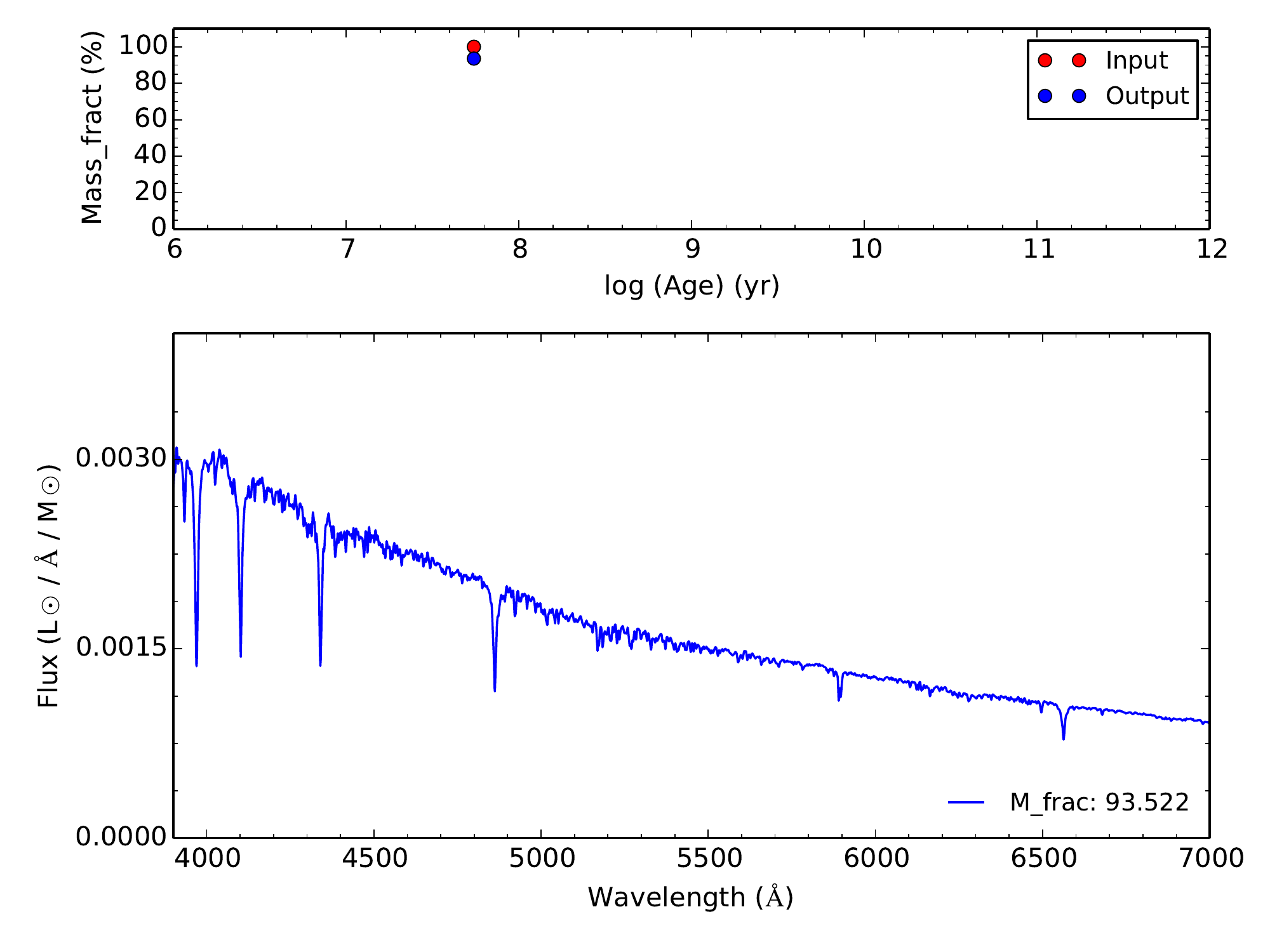} & 
			\includegraphics[keepaspectratio=true, scale=.45]{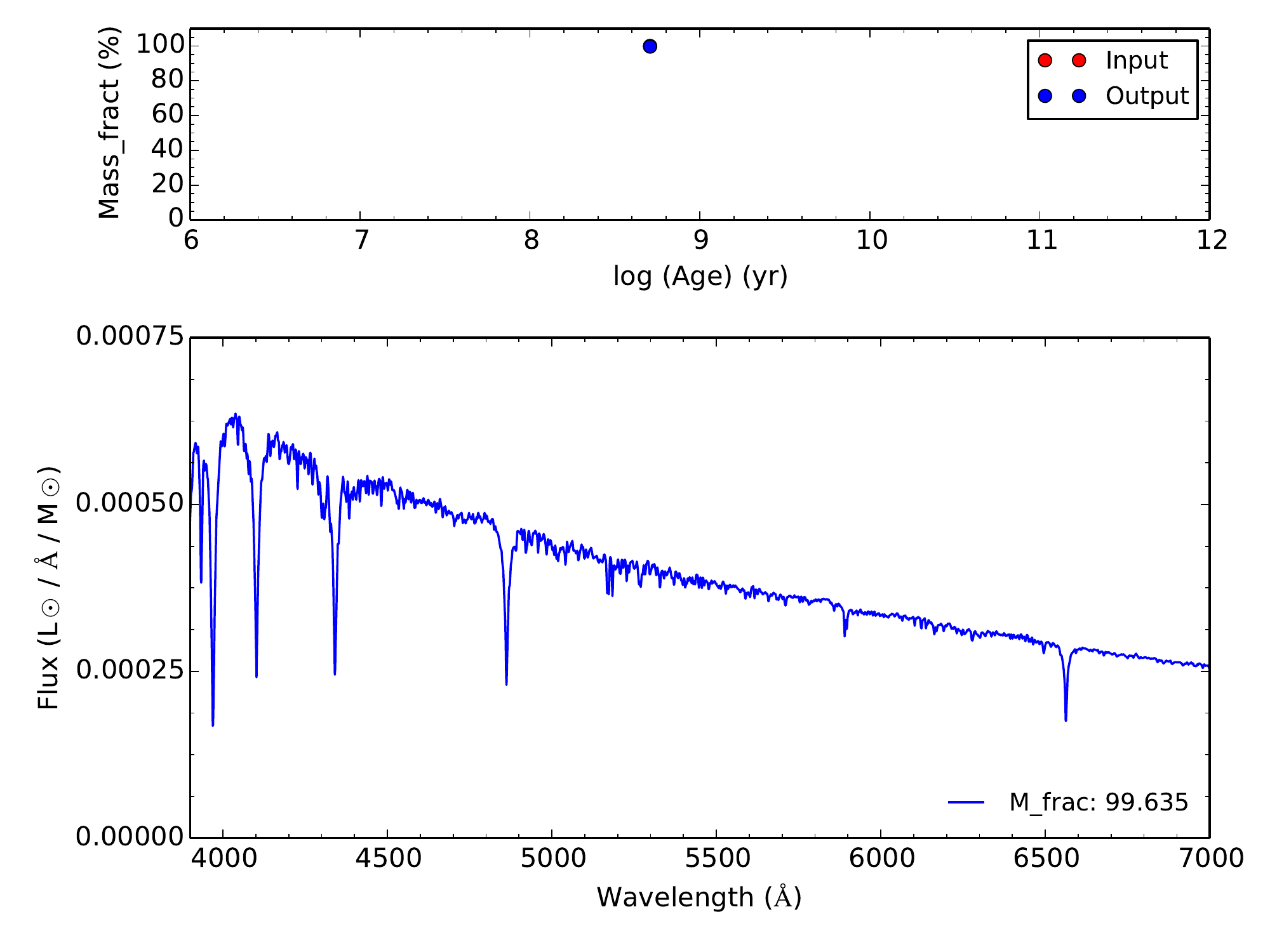} \\
			\includegraphics[keepaspectratio=true, scale=.45]{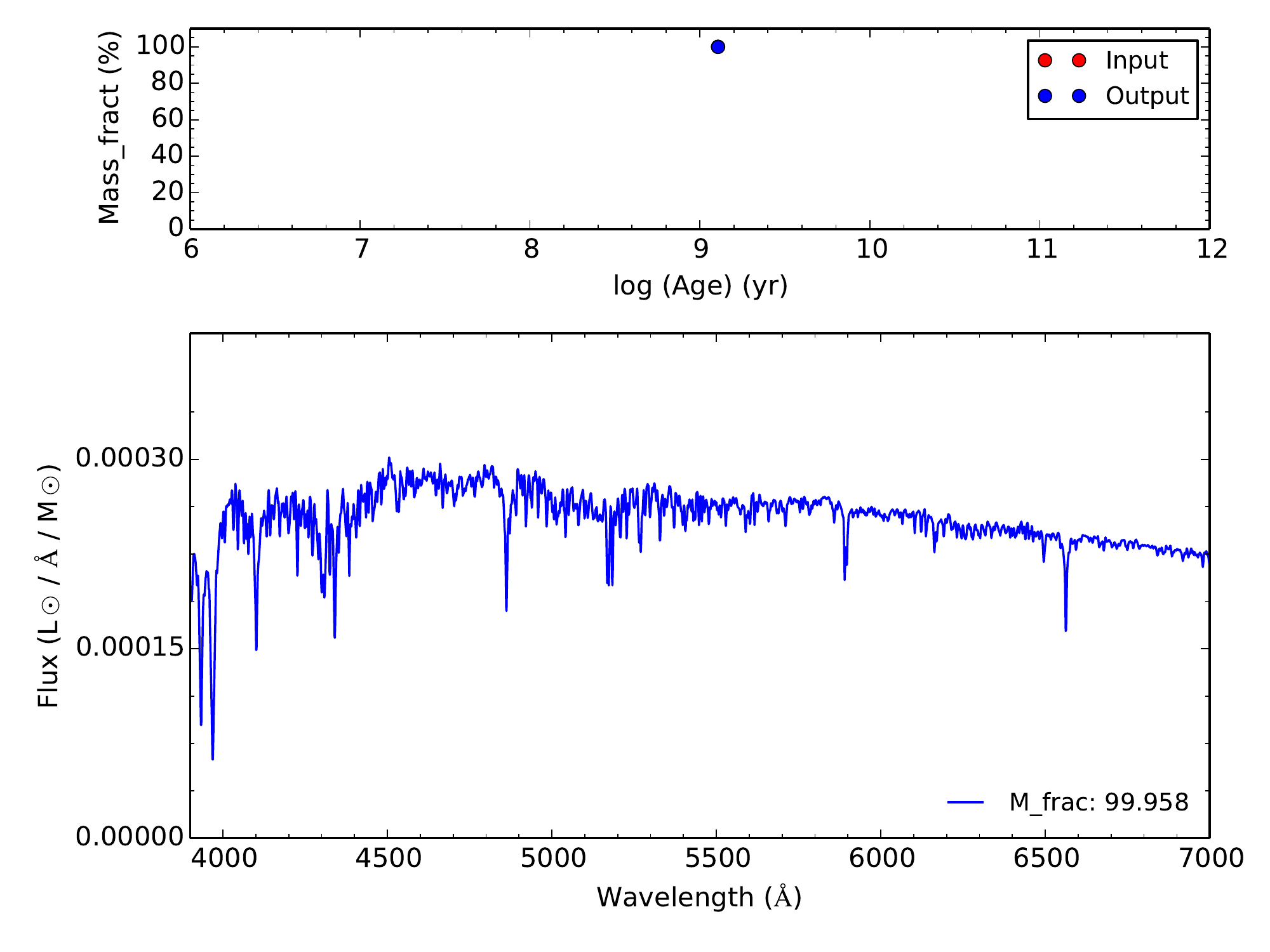} &
			\includegraphics[keepaspectratio=true, scale=.45]{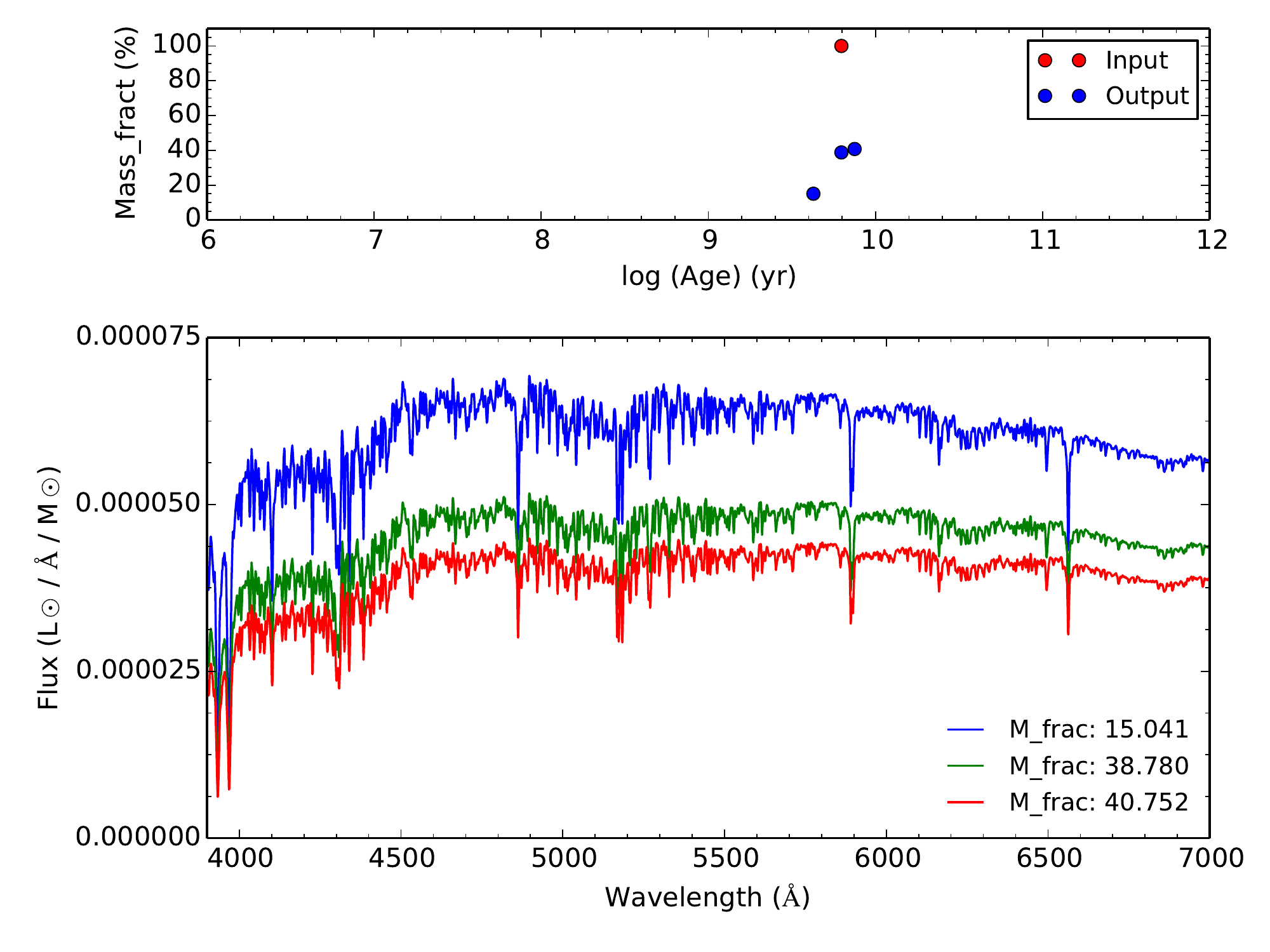} \\ 
			
		\end{tabular}
	\end{center}
	
	\caption{The top panel in all plots shows the SSP of given age and mass fraction used as input in \textsc{starlight} (red point) and the corresponding output SSP(s) (blue points) from the fit. In cases of exact match, one color is over-plotted on top of the other. The bottom panel in all figures shows the corresponding output spectrum of the SSP(s) used in the fit. The raw input SSPs from the BC03 libraries are presented.}
	\label{ST-SIMS1}				
\end{figure*}

\begin{figure*}
	\begin{center}
		\hspace*{-1.0cm}\begin{tabular}{cc}
			\includegraphics[keepaspectratio=true, scale=.45]{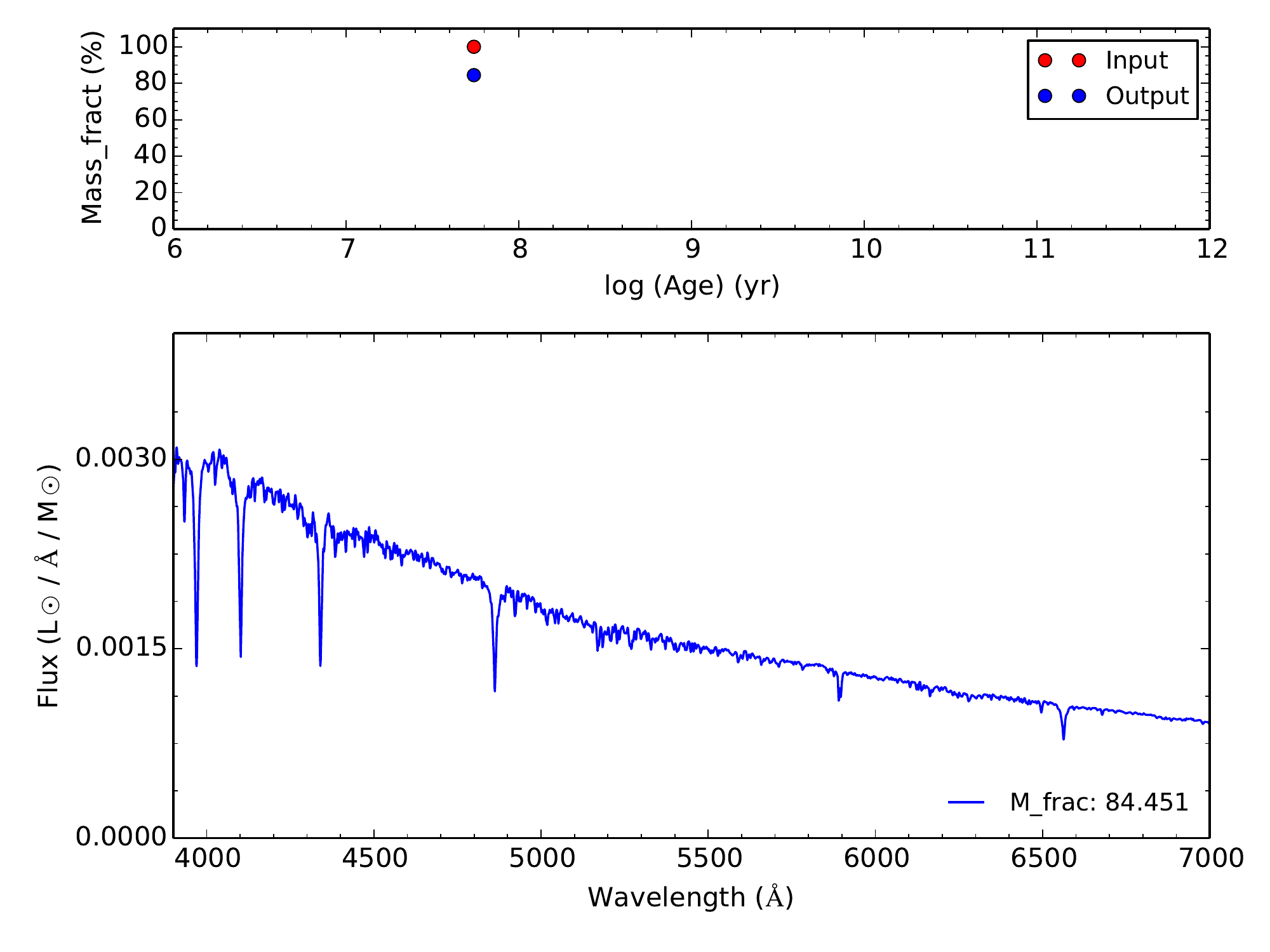} & 
			\includegraphics[keepaspectratio=true, scale=.45]{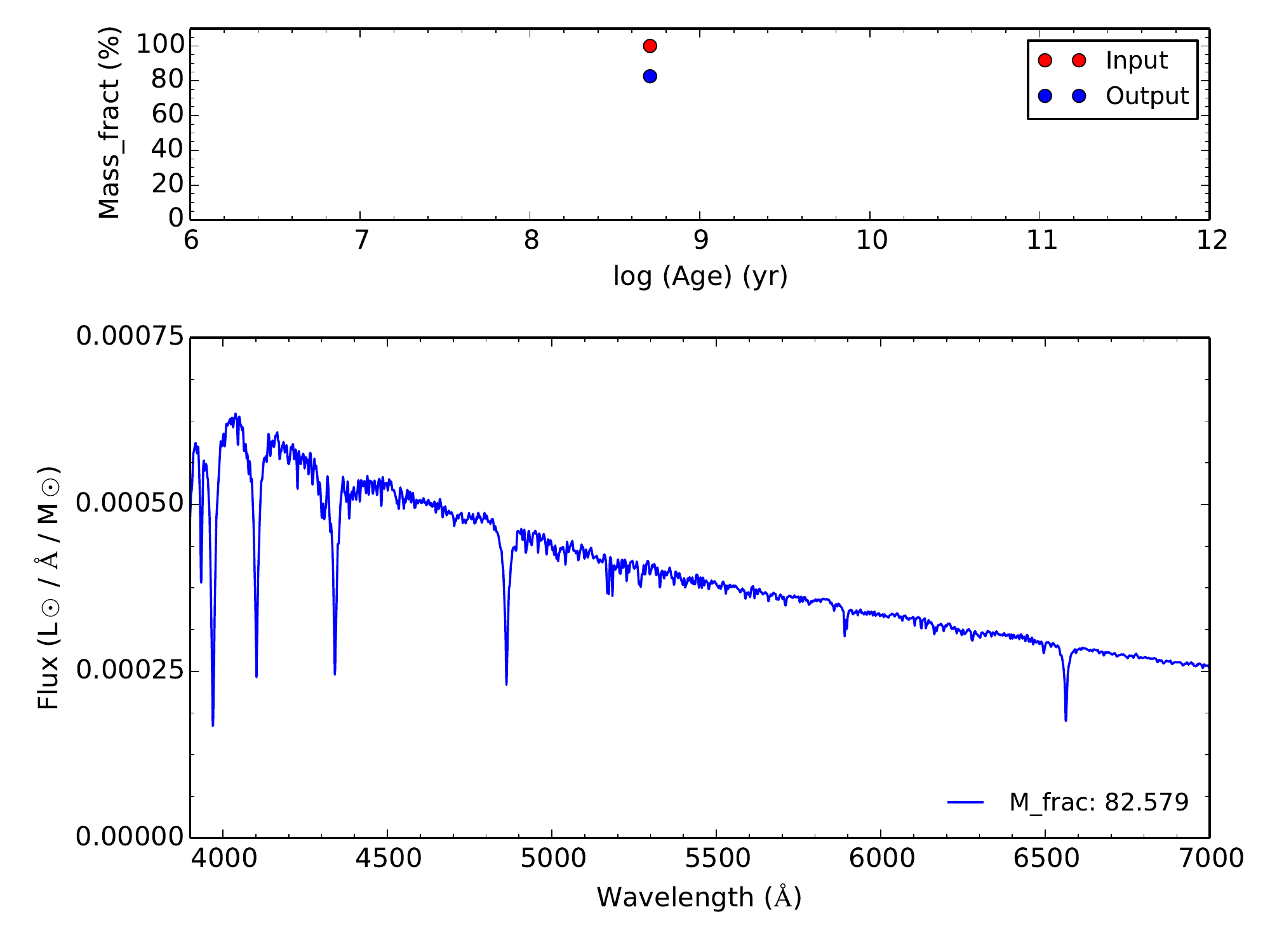} \\
			\includegraphics[keepaspectratio=true, scale=.45]{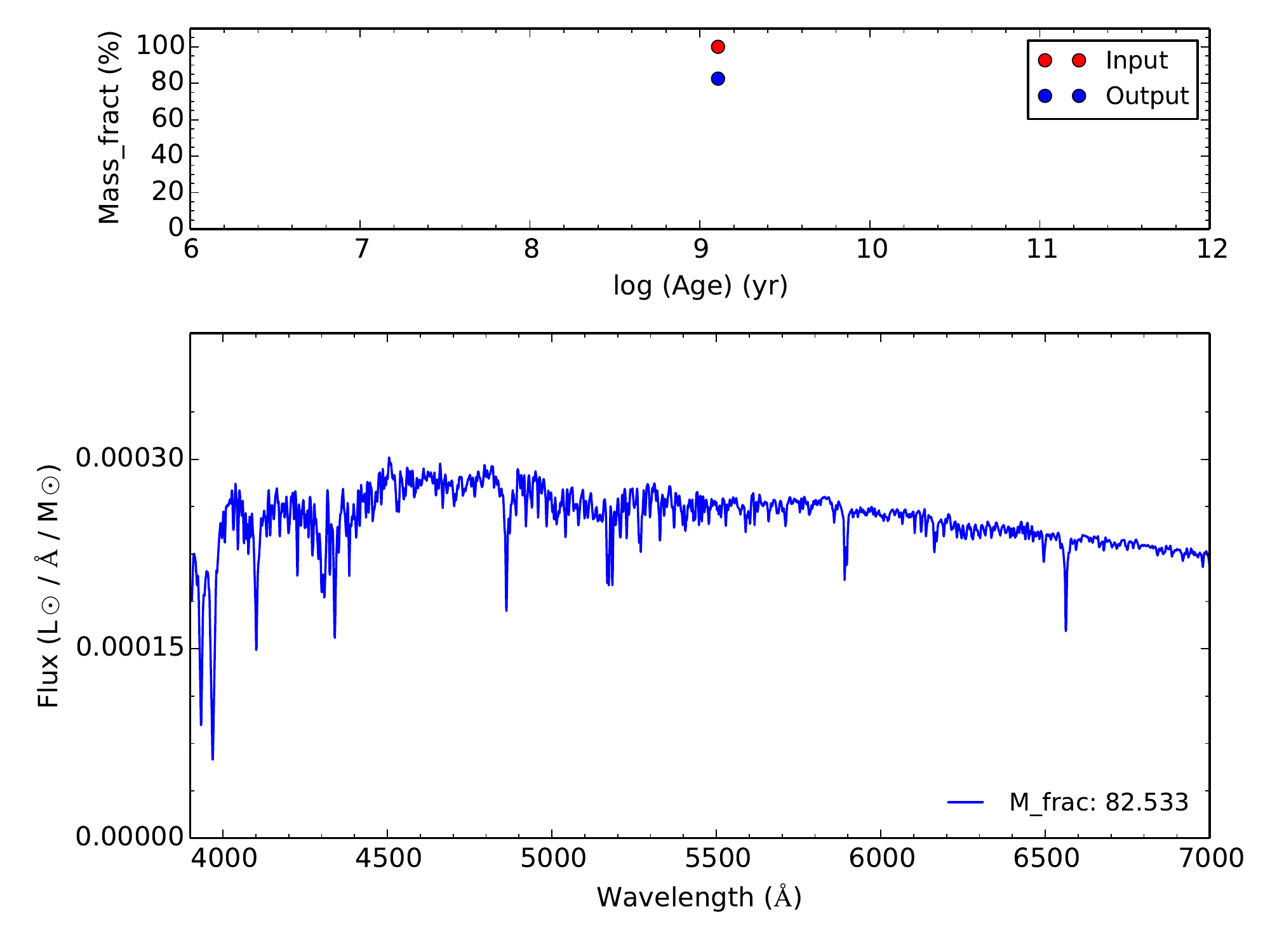} &
			\includegraphics[keepaspectratio=true, scale=.45]{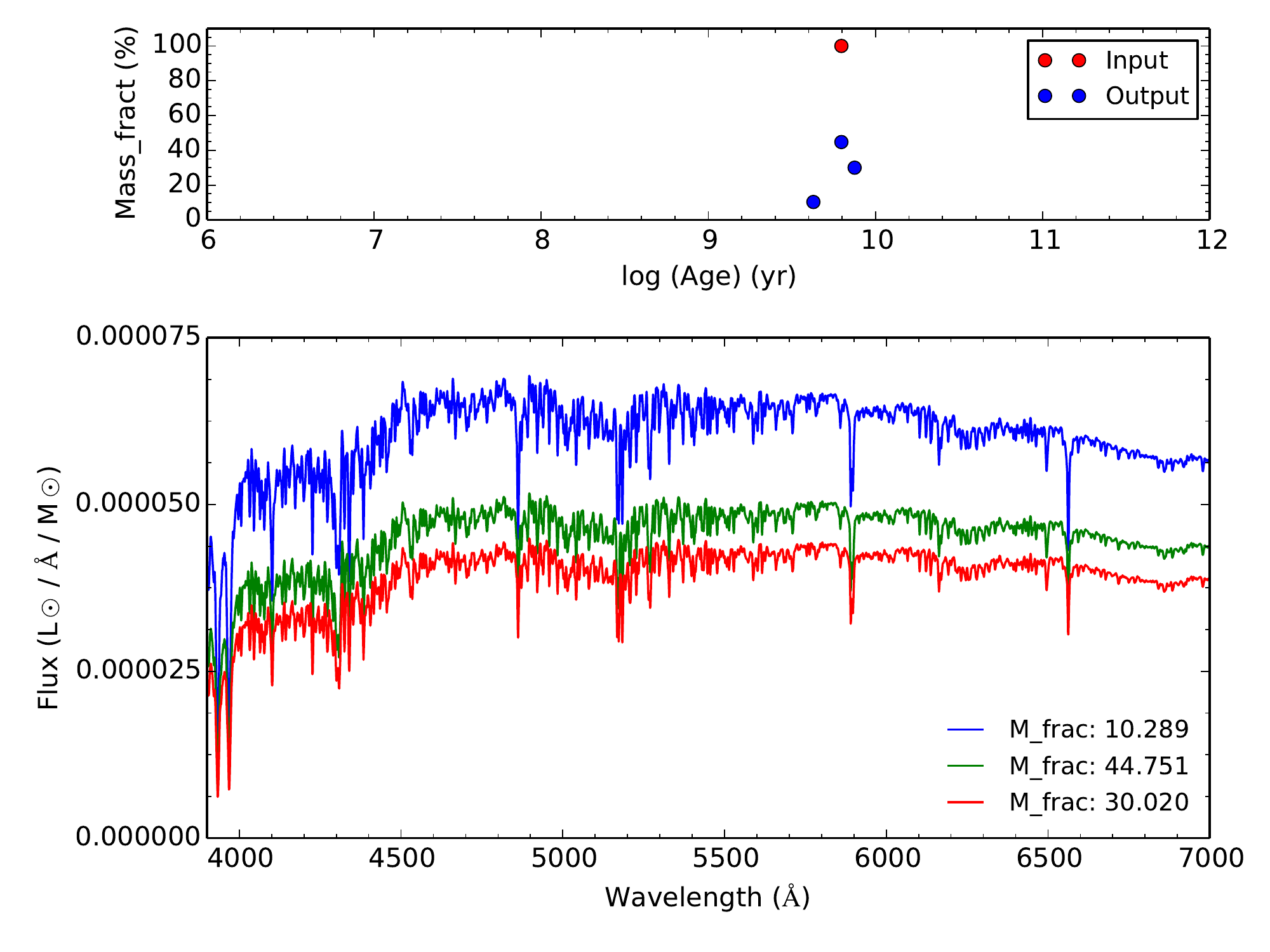} \\ 
			
		\end{tabular}
	\end{center}
	
	\caption{\textsc{starlight} code simulation study using the same SSPs as in Figure \ref{ST-SIMS1} adding Gaussian noise. The top panel in all plots shows the SSP of given age and mass fraction used as input in \textsc{starlight} (red point) and the corresponding output SSP(s) (blue points) from the fit. The bottom panel in all figures shows the corresponding output spectrum of the SSP(s) used in the fit.}
	\label{ST-SIMS2}				
\end{figure*}

\begin{figure*}
	\begin{center}
		\hspace*{-1.0cm}\begin{tabular}{cc}
			\includegraphics[keepaspectratio=true, scale=.45]{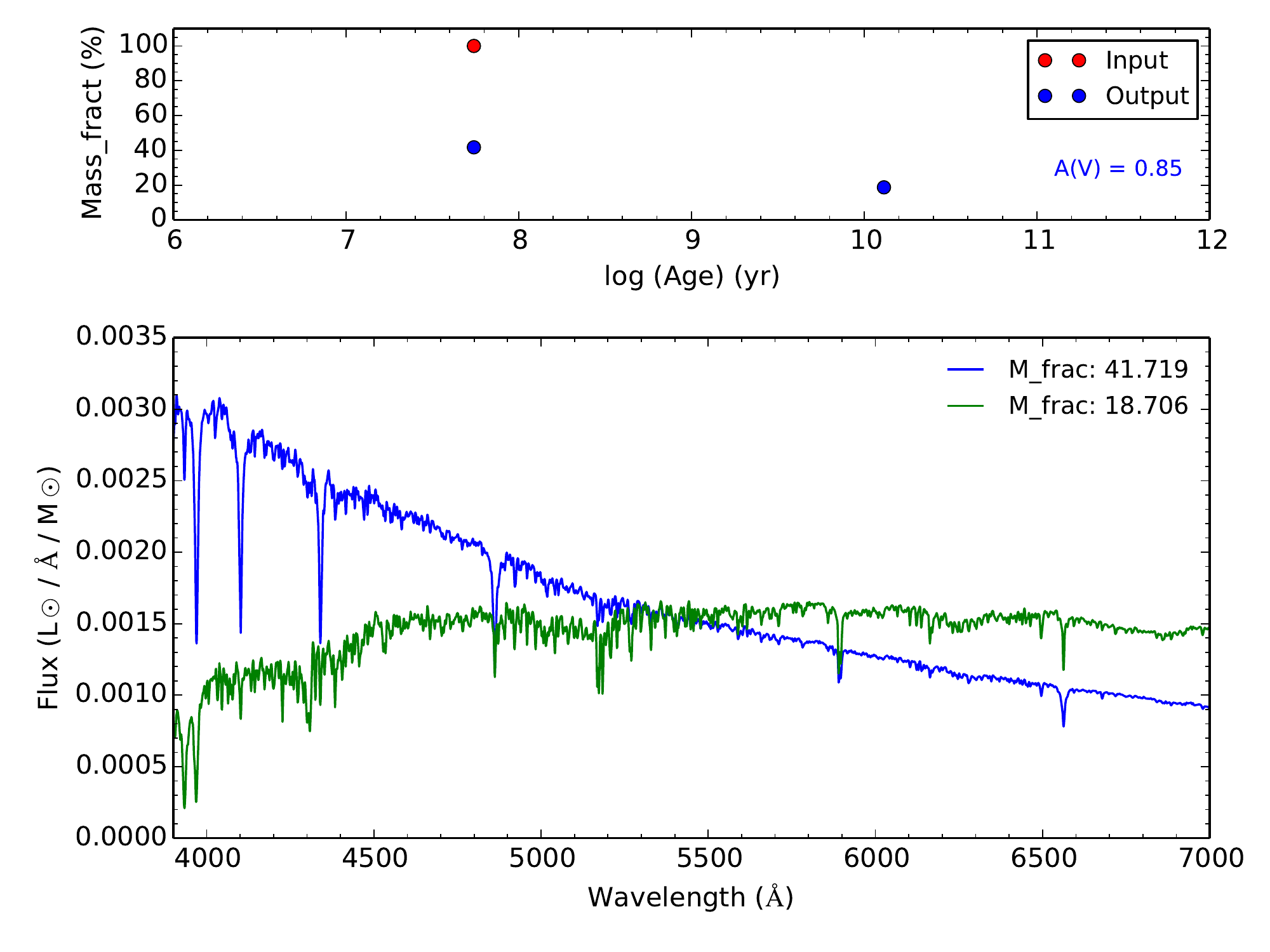} & 
			\includegraphics[keepaspectratio=true, scale=.45]{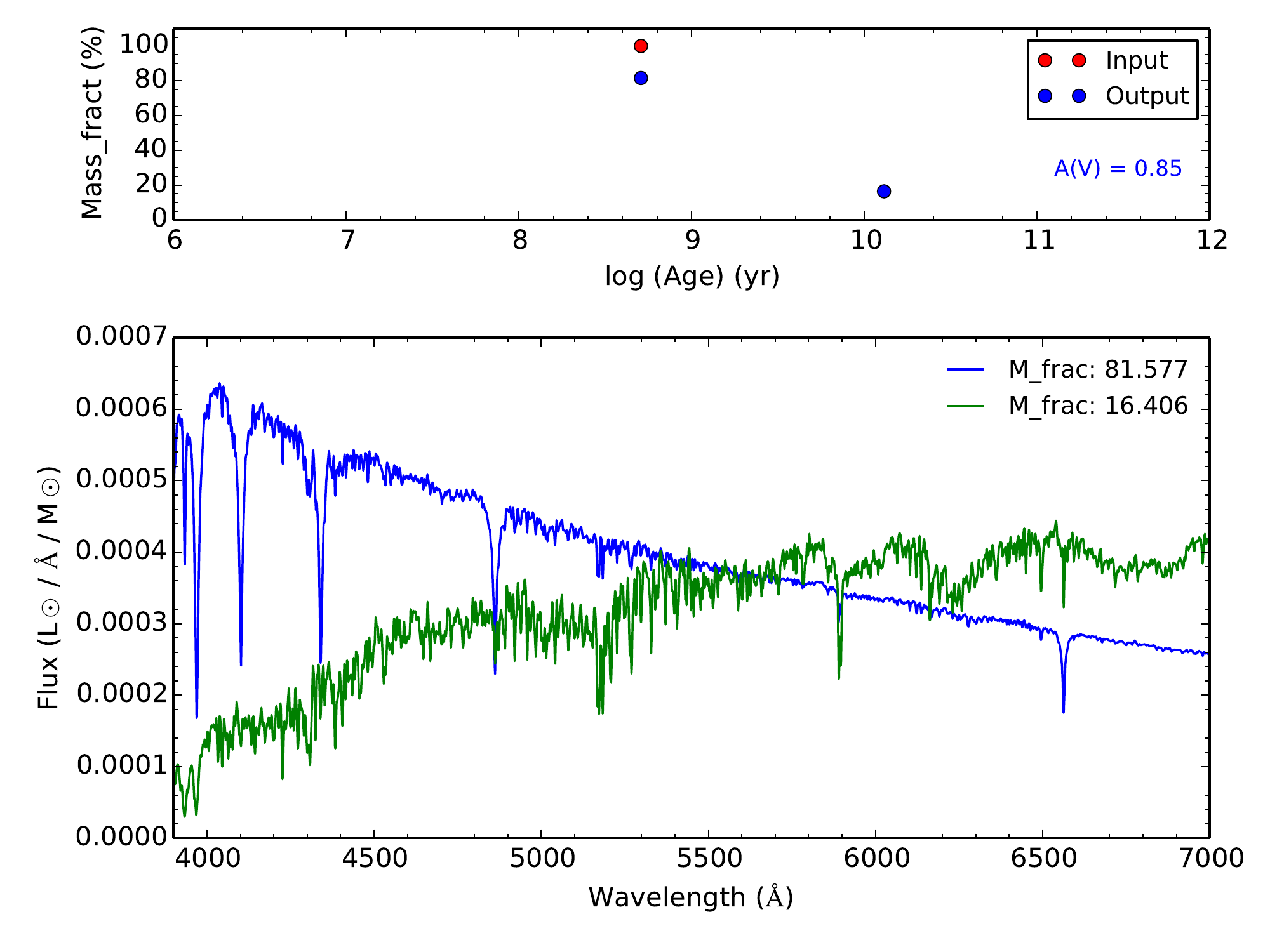} \\
			\includegraphics[keepaspectratio=true, scale=.45]{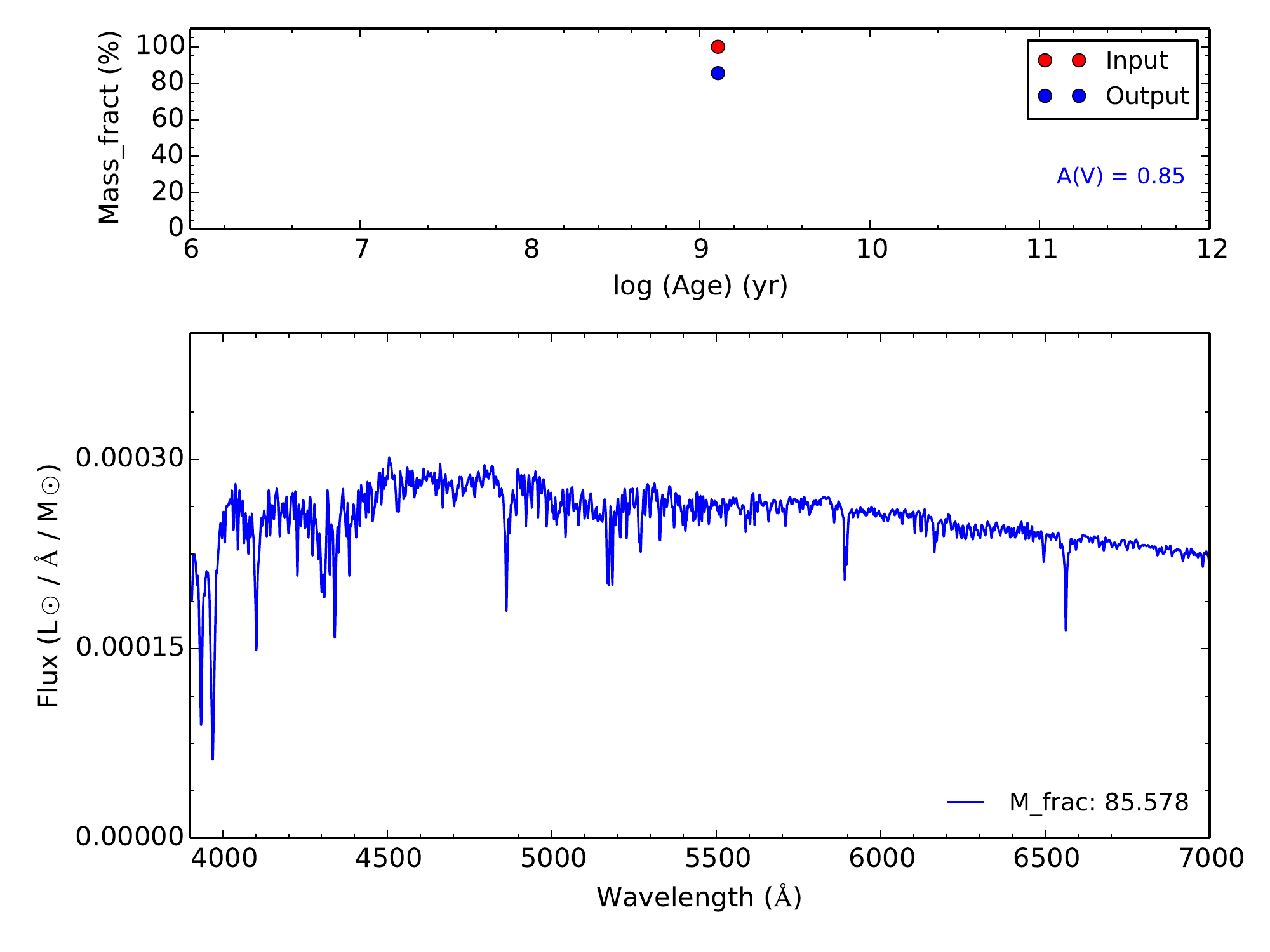} &
			\includegraphics[keepaspectratio=true, scale=.45]{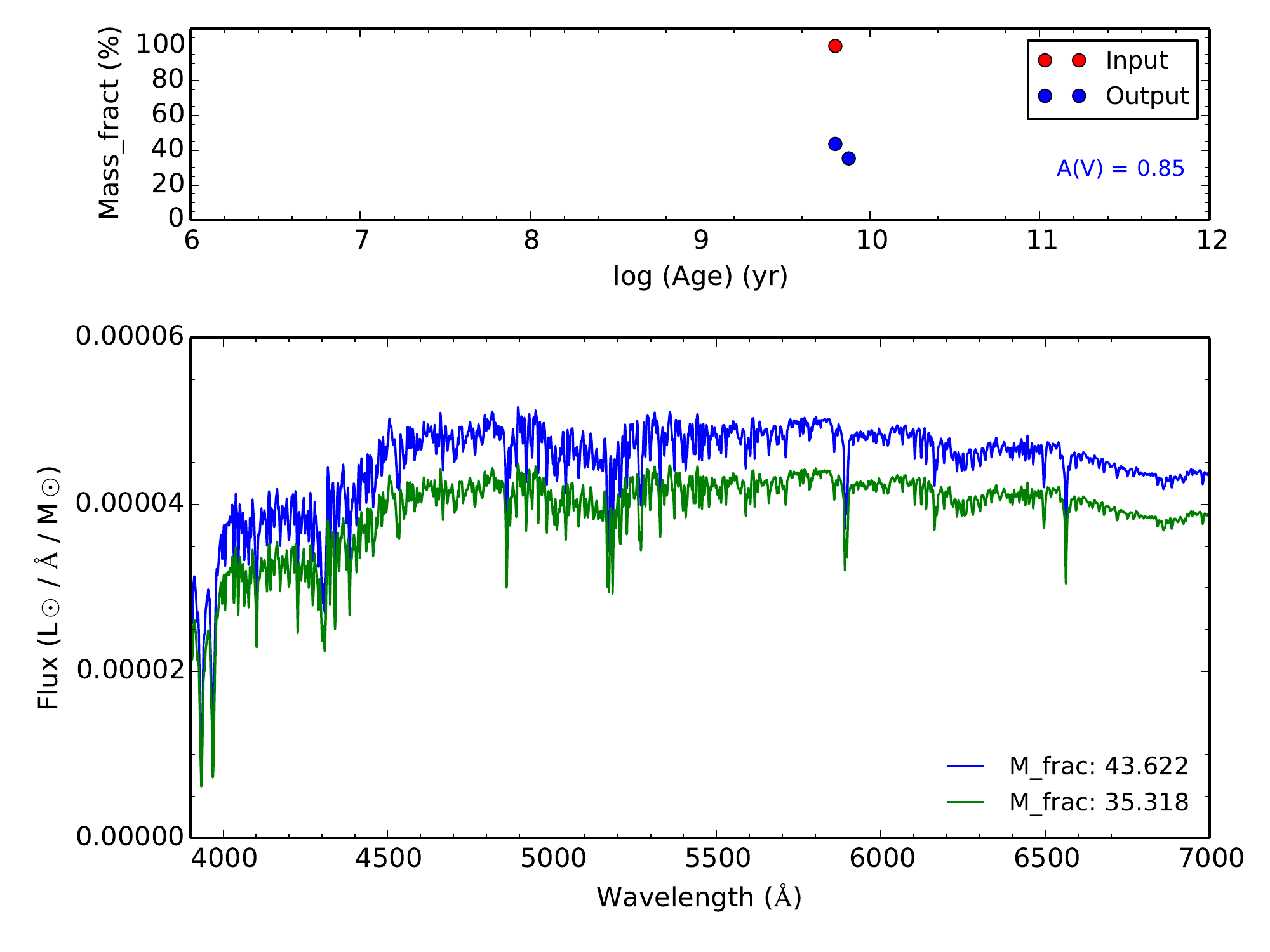} \\ 
			
		\end{tabular}
	\end{center}
	
	\caption{\textsc{starlight} code simulation study using the same SSPs as in Figure \ref{ST-SIMS1} with added extinction (A$_{V}$ = 0.85), but without Gaussian noise. The top panel in all plots shows the SSP of given age and mass fraction used as input in \textsc{starlight} (red point) and the corresponding output SSP(s) (blue points) from the fit. The bottom panel in all figures shows the corresponding output spectrum of the SSP(s) used in the fit.}
	\label{ST-SIMS3}			
\end{figure*}

\begin{figure*}
	\begin{center}
		\hspace*{-1.0cm}\begin{tabular}{cc}
			\includegraphics[keepaspectratio=true, scale=.45]{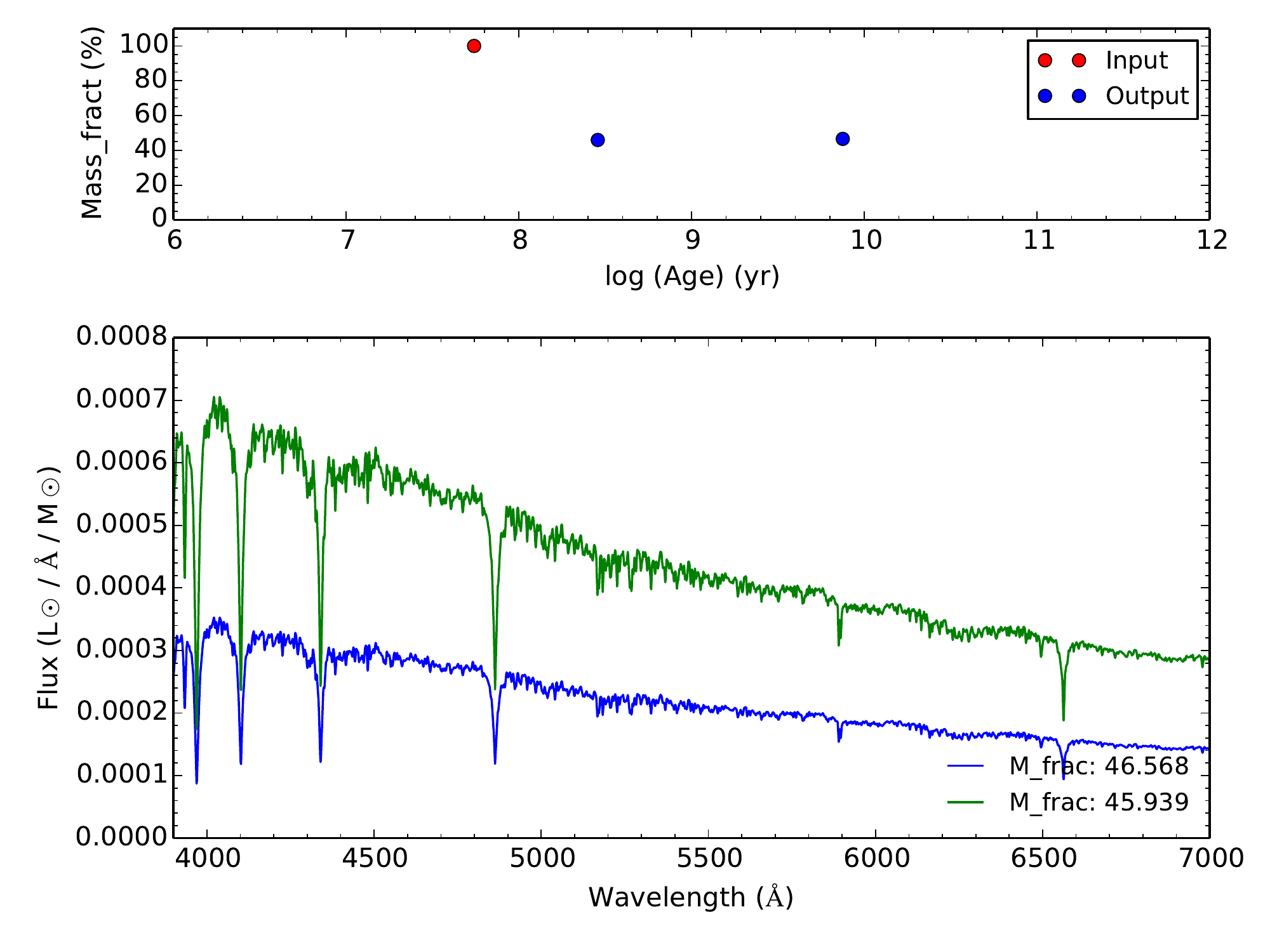} & 
			\includegraphics[keepaspectratio=true, scale=.45]{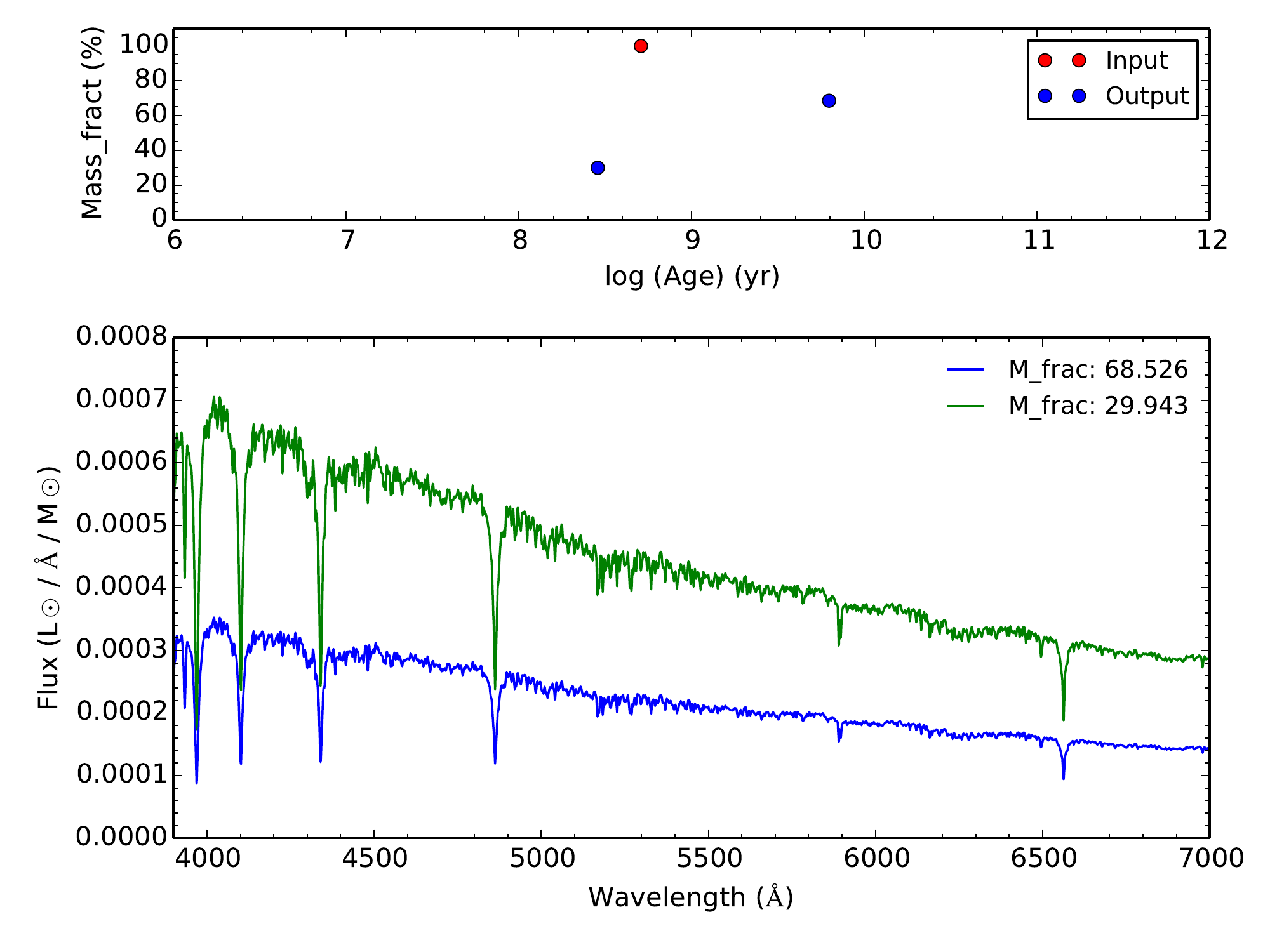} \\
			\includegraphics[keepaspectratio=true, scale=.45]{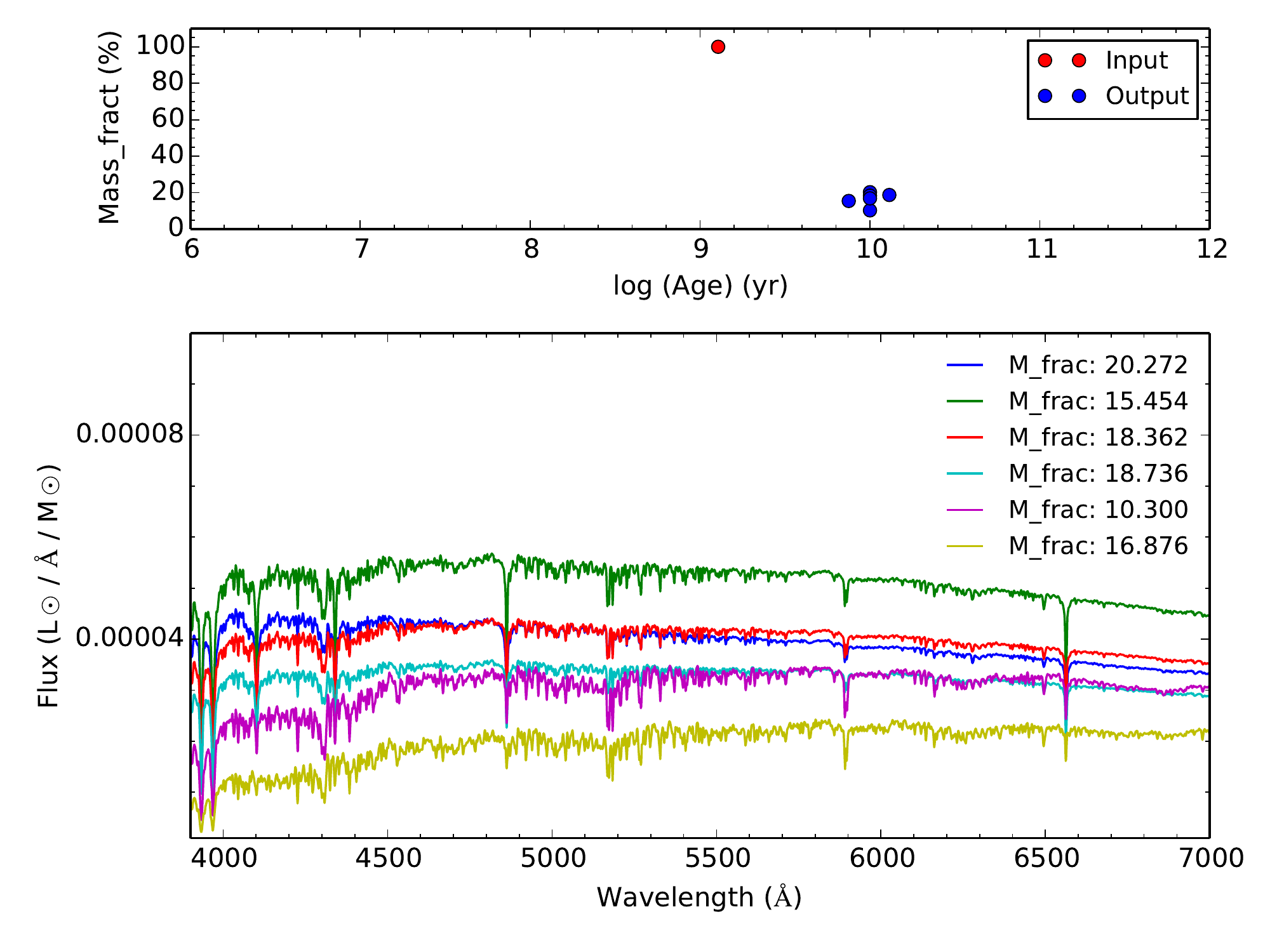} &
			\includegraphics[keepaspectratio=true, scale=.45]{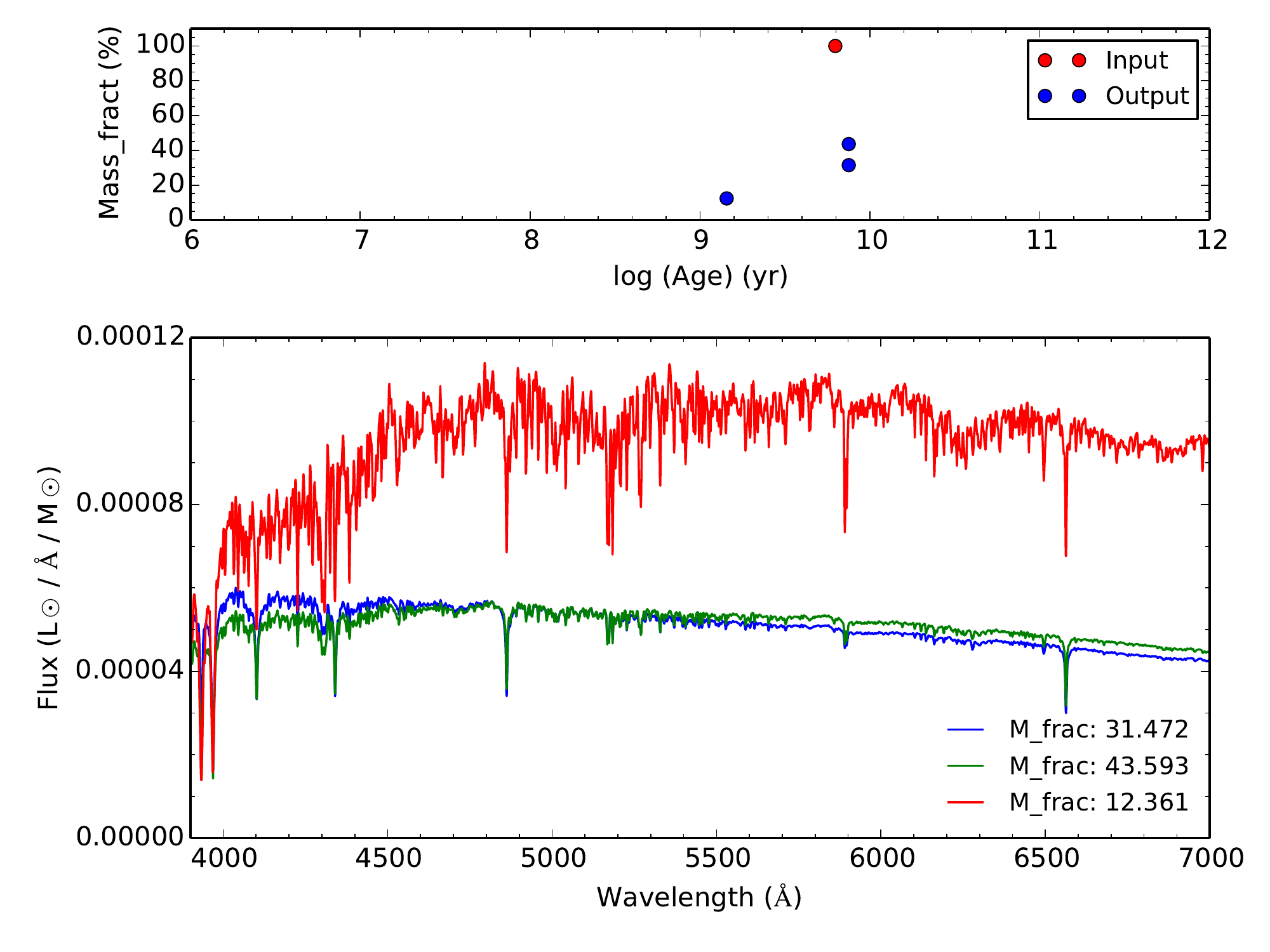} \\ 
			
		\end{tabular}
	\end{center}
	
	\caption{\textsc{starlight} code simulation study using the same SSPs as in Figure \ref{ST-SIMS1} adding an AGN component. The top panel in all plots shows the SSP of given age and mass fraction used as input in \textsc{starlight} (red point) and the corresponding output SSP(s) (blue points) from the fit. The bottom panel in all figures shows the corresponding output spectrum of the SSP(s) used in the fit.}
	\label{ST-SIMS4}			
\end{figure*}

As a supplementary analysis to those studies, we performed a set of simulations using only individual sets of SSPs or, at most, two-component composite stellar populations (CSPs) from the BC03 libraries. We used populations of different ages (5 Myr$-$10 Gyr) at a given metallicity (0.2 Z$_{\sun}$) with a 10\% uncertainty on the flux as input in \textsc{starlight}. The top panels in Figs. \ref{ST-SIMS1}, \ref{ST-SIMS2}, \ref{ST-SIMS3}, and \ref{ST-SIMS4} shows the current mass fraction against age of the input and output spectra that contributes more than 10\% to current stellar mass. \textsc{starlight} is more sensitive in identifying the actual contribution of the younger populations, while it systematically underestimates the contribution of older populations. In the latter case, while successfully recovering the actual population, it uses supplementary SSPs of similar ages to match the input spectrum. Therefore, in the simplest case of a SSP, \textsc{starlight} is able to identify both the age and metallicity of the input spectrum but not necessarily its actual contribution (100\% in the case of a SSP).

\begin{figure}
	\begin{center}
		\includegraphics[keepaspectratio=true, scale=.40]{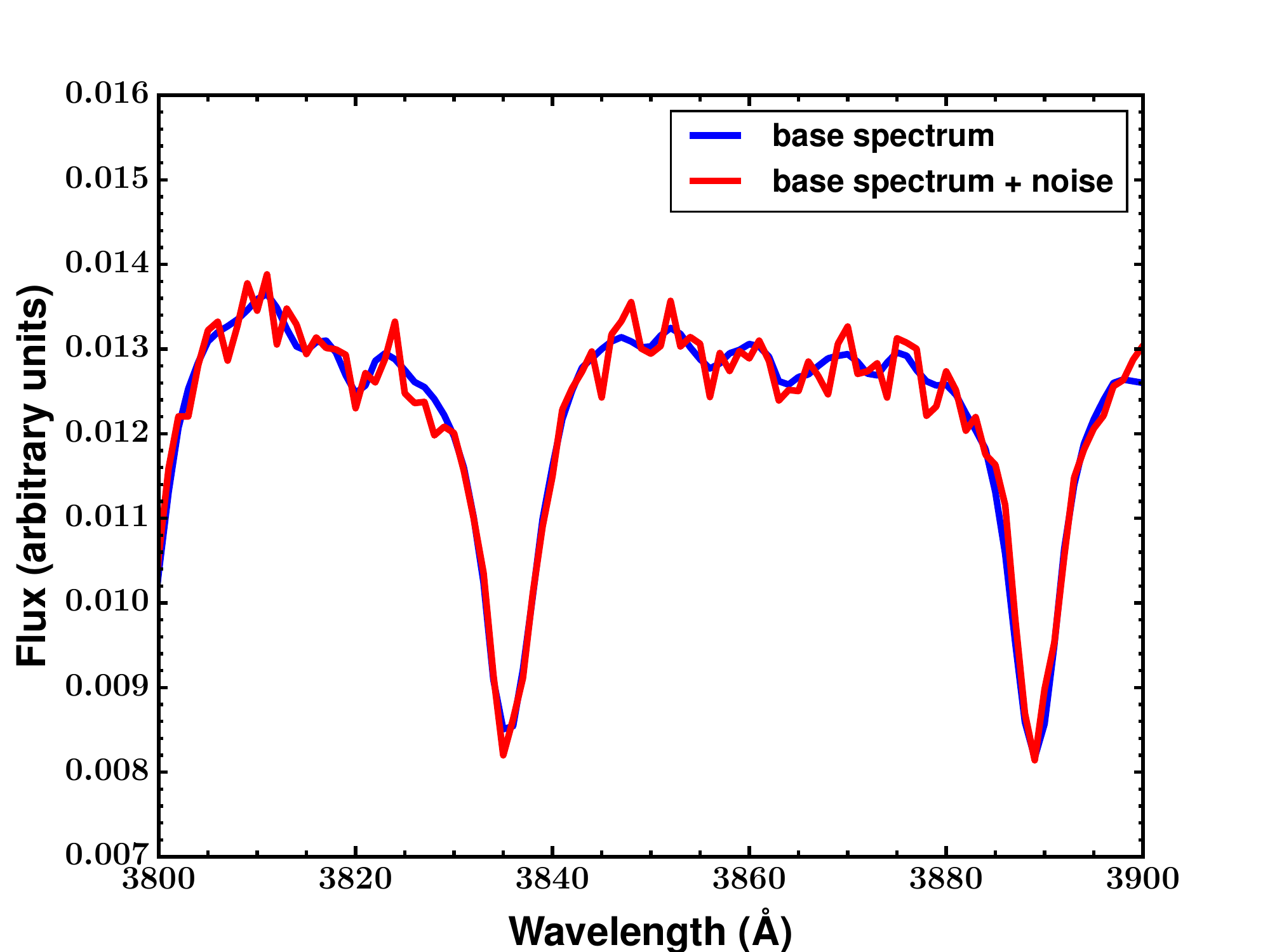}
		\caption{Example of a SSP spectrum (blue line) and the same SSP spectrum perturbed with 2\% Gaussian noise (red line)}
		\label{NOISE}		
	\end{center}
\end{figure}

To further assess \textsc{starlight}'s results, we created synthetic spectra, adding noise and extinction separately on the same SSPs used previously, in an effort to simulate an actual spectrum in a fully controlled environment. With the inclusion of Gaussian noise (2\% of the input flux, Fig. \ref{NOISE}), there is a deviation from the actual contribution to the current mass for all populations, and in some cases inclusion of extra SSPs, but the dominant component describes the correct age and metallicity (Fig. \ref{ST-SIMS2}, second panel).

Adding extinction (A$_{V}$ = 0.85) without noise to the input spectrum will force \textsc{starlight} in almost every case to split the total contribution using extra SSPs for the fit (Fig. \ref{ST-SIMS3}, third panel). While for populations between 100 Myr and 5 Gyr the highest contribution comes from the correct SSP component, in most cases that component is modeled below 50\% of the total. However, \textsc{starlight} will correctly fit the extinction of the input spectrum.

Another test we applied was to limit the spectral area \textsc{starlight} uses to fit the input spectrum by masking most spectral regions except those containing Lick indices, which are characteristic and representative of the stellar populations. In most cases \textsc{starlight} was able to attribute more accurately the actual contribution to the input spectrum or even limit the number of extra SSP used in the fit, but the overall improvement was not significant over the default configuration. 

\begin{figure}
	\begin{center}
		\includegraphics[keepaspectratio=true, scale=.40]{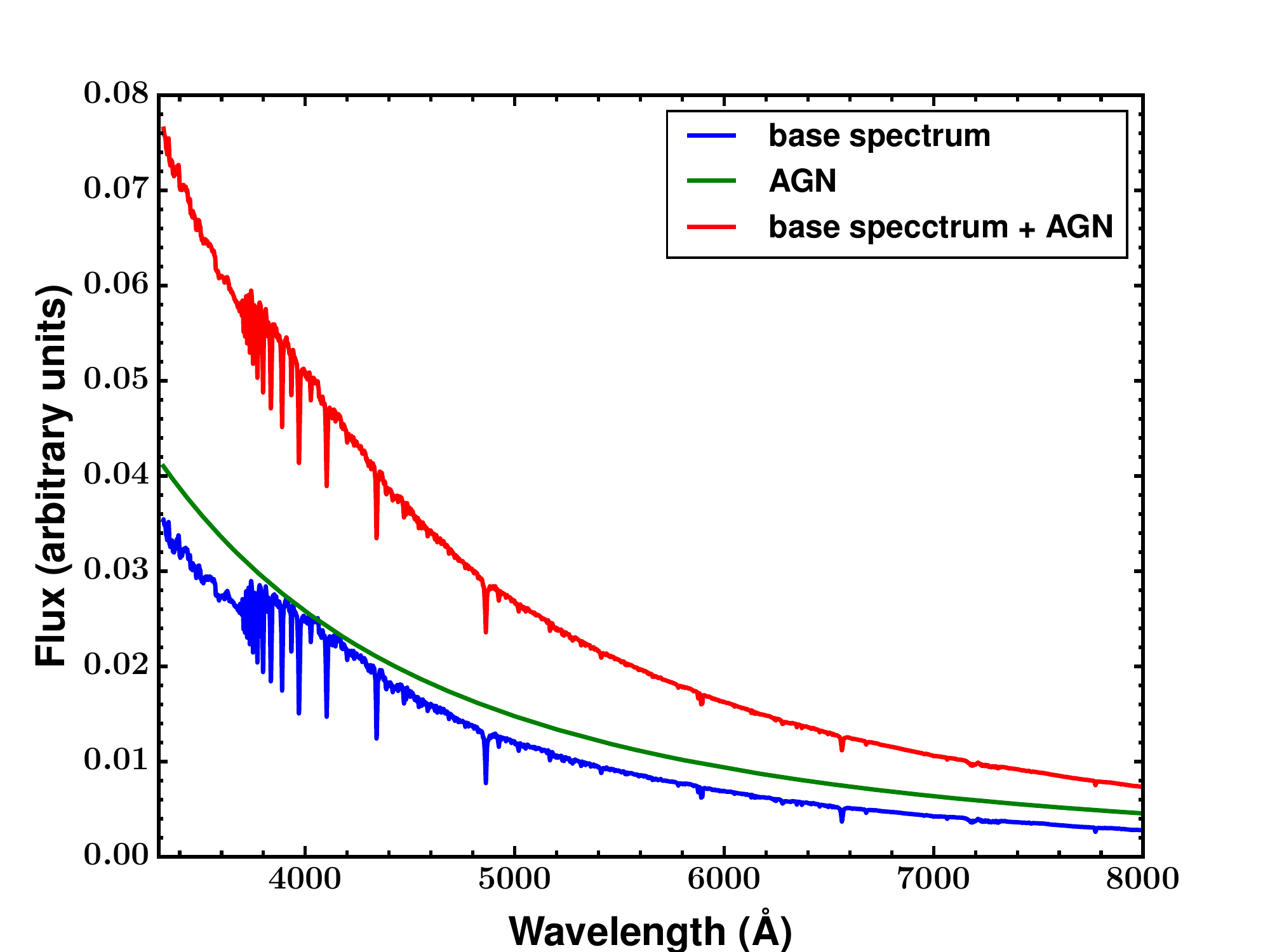}
		\caption{Example of a composite AGN spectrum (red line), composed by a SSP spectrum (green line) and an AGN power-law spectrum (blue line).}
		\label{AGN-spectrum}
	\end{center}
\end{figure}

Finally, we tested \textsc{starlight}'s sensitivity in uncovering the AGN component of a galaxy spectrum using power-law AGN templates of different indices along with the standard SSPs. We created a synthetic spectrum from an SSP adding an AGN component (Fig. \ref{AGN-spectrum}) and provided STARLIGHT with the same AGN template. Although \textsc{starlight} revealed the correct AGN component, the stellar populations used in the fit varied significantly from the input SSP used to create the synthetic spectrum (Fig. \ref{ST-SIMS4}, bottom panel). Furthermore, if more AGN templates of different power-law indices are allowed to be used in the fit, \textsc{starlight} will use most of them to reproduce the input spectrum. Additionally, by fitting a SSP without AGN component, but allowing \textsc{starlight} to use AGN templates in the fitting process (which illustrates the case of fitting an actual galaxy spectrum without a prior knowledge for the presence of an AGN or not), \textsc{starlight} has a tendency to use the AGN templates present.

As a verdict, \textsc{starlight} heads towards the right direction in uncovering the actual fitted stellar population, but even in ideal environments, absent noise and extinction, older populations are harder to uncover. The presence of an AGN is even trickier to handle, as in the absence of AGN templates the AGN continuum is attributed to and fitted with stellar populations. With the inclusion of AGN templates, \textsc{starlight} tends to over-assign an AGN contribution to the fitted spectrum even when the galaxy lacks AGN activity. In reality, an actual galactic spectrum has a much higher complexity than independent SSPs, but \textsc{starlight} remains robust in fitting and allowing the SED of the stellar component to be subtracted.

\section{Notes On Individual Galaxies} \label{Notes}

\noindent SFRS 30 ( = \textit{OJ~287}) is a blazar and is assigned a Sy classification to match the terminology adopted for the activity types in this paper. \\

\noindent SFRS 42 ( = \textit{IC~2434}) is an ambiguous case and is assigned a TO classification based on the [N\,\textsc{ii}] BPT diagram. The [S\,\textsc{ii}]~$\lambda\lambda6716, 6731$ and [O\,\textsc{i}]~$\lambda6300$  detections are considered upper limits.\\

\noindent SFRS 57 ( = \textit{CGCG~238-066}) is an ambiguous case in the BPT diagrams but is recognized as AGN in the IRAC color-color diagram and is assigned a Sy classification.\\

\noindent SFRS 61 ( = \textit{CGCG~181-068}) is an ambiguous case and is assigned a TO classification based on the [N\,\textsc{ii}] BPT diagnostic. The [S\,\textsc{ii}]~$\lambda\lambda6716, 6731$ and [O\,\textsc{i}]~$\lambda6300$ measurements are considered upper limits.\\

\noindent SFRS 80 ( = \textit{NGC~3190}) has no detectable H$\beta$ emission in the nuclear spectrum. The integrated spectrum was used to obtain activity classification (LINER).\\

\noindent SFRS 93 ( = \textit{UGC~5720}) H$\alpha$ is absent in the SDSS spectrum. The activity type (H\,\textsc{ii}) was determined solely from the \OIIIHb{} value.\\

\noindent SFRS 139 ( = \textit{NGC~3758}) is a Type-1 AGN with very broad line profiles and is automatically classified as Sy.\\

\noindent SFRS 148 ( = \textit{NGC~3822}) has no [S\,\textsc{ii}] BPT classification because the [S\,\textsc{ii}]~$\lambda\lambda6716, 6731$ doublet falls inside a telluric line region. It is assigned a Sy classification because in the [O\,\textsc{i}] BPT diagram the galaxy is located inside the Sy region, further away from the Sy/LINER demarcation line compared to the [N\,\textsc{ii}] diagnostic demarcation line.\\

\noindent SFRS 182 ( = \textit{NGC~4194}) H$\alpha$ is absent in the SDSS spectrum. The activity type (H\,\textsc{ii}) was determined solely from the \OIIIHb{} value.\\

\noindent SFRS 187 ( = \textit{NGC~4237}) is an ambiguous case, classified as TO/H\,\textsc{ii} in the N\,\textsc{ii}/S\,\textsc{ii}-BPT diagnostics. Because the [OI] line is unreliable, this galaxy is assigned an H\,\textsc{ii} classification.\\

\noindent SFRS 201 ( = \textit{NGC~4435}) is an ambiguous case and is assigned a LINER classification because in the [O\,\textsc{i}] BPT is located in the middle of the LINER region, while there is no [S\,\textsc{ii}]~$\lambda\lambda6716, 6731$ detection.\\

\noindent SFRS 204 ( = \textit{3C~273}) is a quasar and is assigned a Sy classification to match the terminology adopted for the activity types in this paper.\\

\noindent SFRS 227 ( = \textit{NGC~4689}) is assigned an H\,\textsc{ii} classification because in the [S\,\textsc{ii}] BPT diagram the galaxy is located inside the H\,\textsc{ii} region further away from the demarcation line. There is no [O\,\textsc{i}]~$\lambda6300$ detection.\\

\noindent SFRS 228 ( = \textit{NGC~4688}) is assigned an H\,\textsc{ii} classification because it is located further away from the demarcation line and into the H\,\textsc{ii} region in the [N\,\textsc{ii}] BPT diagram, while it falls right on the line in the [S\,\textsc{ii}] BPT diagnostic. There is no [O\,\textsc{i}]~$\lambda6300$ detection.\\

\noindent SFRS 233 ( = \textit{MCG~8-23-097}) is an ambiguous case and is assigned a LINER classification.\\

\noindent SFRS 239 ( = \textit{UGC~8058}) is a Type-1 AGN with very broad line profiles and is automatically classified as Sy.\\

\noindent SFRS 259 ( = \textit{NGC~5104}) is an ambiguous case and is assigned a Sy classification.\\

\noindent SFRS 261 ( = \textit{NGC~5112}) is an ambiguous case and is assigned a TO classification because it falls on the [S\,\textsc{ii}] BPT demarcation line, while in the [O\,\textsc{i}] BPT diagram the emission-line uncertainties can place the galaxy further inside the Sy region.\\

\noindent SFRS 262 ( = \textit{NGC~5123}) is assigned an H\,\textsc{ii} classification. There is no [O\,\textsc{i}]~$\lambda6300$ detection.\\

\noindent SFRS 263 ( = \textit{IRAS~13218+0552}) is a Type-1 AGN with very broad line profiles and is automatically classified as Sy.\\

\noindent SFRS 266 ( = \textit{NGC~5204}) is an ambiguous case, and is assigned an H\,\textsc{ii} classification. There is no [O\,\textsc{i}]~$\lambda6300$ detection. Visual inspection of the SDSS image shows bright blue colors, characteristic of star-forming galaxies.\\

\noindent SFRS 270 ( = \textit{IRAS~13349+2438}) is a literature verified QSO (e.g. \citealt{Lee13}) and was assigned a Sy classification.\\

\noindent SFRS 276 ( = \textit{MK~268}) is a Type-1 AGN with very broad line profiles and is automatically classified as Sy.\\

\noindent SFRS 305 ( = \textit{NGC~5515}) is an ambiguous case and is assigned a Sy classification. It is also a literature verified Sy (e.g. \citealt{Veron06})\\

\noindent SFRS 322 ( = \textit{UGC~9412}) is a Type-1 AGN with very broad line profiles and is automatically classified as Sy.\\

\noindent SFRS 331 ( = \textit{UGC~9618\_NED02}) has a TO/H\,\textsc{ii}/H\,\textsc{ii} class in the corresponding [N\,\textsc{ii}]/[S\,\textsc{ii}]/[O\,\textsc{i}] BPT diagnostics. However, because in the O\,\textsc{i} BPT diagram it falls on top of the demarcation line, it is assigned a TO classification.\\

\noindent SFRS 350 ( = \textit{UGC~10120}) has broad-line profiles and is fitted with an additional broad component along with the narrow ones when measuring its emission lines. However, it is classified as H\,\textsc{ii} in all three BPTs but is recognized as AGN in the IRAC color-color diagram and is reported as Sy in the literature (\citealt{Contini98}). Therefore it is assigned a Sy classification.\\

\end{document}